\documentclass[preprint2]{aastex}
\usepackage{graphicx}

\shorttitle{Analysis in Taurus}
\shortauthors{Huixian Li. et al.}

\slugcomment{Not to appear in Nonlearned J., 45.}

\begin{document}

\title{Outflows and Bubbles in Taurus: Star-formation Feedback Sufficient to Maintain Turbulence}

\author{Huixian Li\altaffilmark{1,2,\textbf{3}}{*}, Di Li\altaffilmark{1,2,4}{*}, Lei Qian\altaffilmark{1,2}, Duo Xu\altaffilmark{1,\textbf{3},5}, Paul F. Goldsmith\altaffilmark{6}, Alberto Noriega-Crespo\altaffilmark{7,8}, Yuefang Wu\altaffilmark{9}, Yuzhe Song\altaffilmark{5}, Rendong Nan\altaffilmark{1,2}}
\affil{
$^1$ National Astronomical Observatories, Chinese Academy of Sciences, Beijing 100012, China\\
$^2$ Key Laboratory for Radio Astronomy, Chinese Academy of Sciences, Nanjing 210008, China\\
\textbf{$^3$ University of Chinese Academy of Sciences, Beijing 100049, China}\\
$^4$ Space Science Institute, Boulder, CO, USA\\
$^5$ Nanjing University, Nanjing 210093,China\\
$^6$ Jet Propulsion Laboratory, California Institute of Technology, Pasadena, CA 91109, USA\\
$^7$ California Institute of Technology/IPAC, Pasadena, CA 91125, USA\\
$^8$ Space Telescope Science Institute/JWST, Baltimore, MD 21218, USA\\
$^9$ Department of Astronomy, Peking University, Beijing 100871, China}

\email{{*}Email: lhx@nao.cas.cn, dili@nao.cas.cn}

\hyphenpenalty=5000
\tolerance=1000

\begin{abstract}
We have identified outflows and bubbles in the Taurus molecular cloud based on the $\sim 100$ deg$^2$ Five College Radio Astronomy Observatory $^{12}$CO(1-0) and $^{13}$CO(1-0) maps and the Spitzer young stellar object catalogs. In the main 44 deg$^2$ area of Taurus we found 55 outflows, of which 31 were previously unknown. We also found 37 bubbles in the entire 100 deg$^2$ area of Taurus, all of which had not been found before. The total kinetic energy of the identified outflows is estimated to be $\bf \sim 3.9 \times 10^{45}$ erg, which is \textbf{1\%} of the cloud turbulent energy. The total kinetic energy of the detected bubbles is estimated to be $\sim 9.2 \times 10^{46}$ erg, which is 29\% of the turbulent energy of Taurus. The energy injection rate from outflows is $\bf \sim 1.3 \times 10^{33}~\rm erg\ s^{-1}$, \textbf{0.4 - 2 times} the dissipation rate of the cloud turbulence. The energy injection rate from bubbles is $\sim 6.4 \times 10^{33}$ erg s$^{-1}$, \textbf{2 - 10 times} the turbulent dissipation rate of the cloud. The gravitational binding energy of the cloud is $\bf \sim 1.5 \times 10^{48}$ {\bf erg}, \textbf{385} and 16 times the energy of outflows and bubbles, respectively. We conclude that neither outflows nor bubbles can \textbf{provide enough energy to balance the overall gravitational binding energy and the turbulent energy of Taurus. However,} in the current epoch, stellar feedback is sufficient to maintain the observed turbulence in Taurus.
\end{abstract}

\section{Introduction}
Stars during their early stage of evolution experience a phase of mass loss driven by strong stellar winds \citep{Lada85}. The stellar winds can entrain and accelerate ambient gas and inject momentum and energy into the surrounding environment, thereby significantly affect the dynamics and structure of their parent molecular clouds \citep{Narayanan08, Arce11}. Both outflows and bubbles are manifestations of strong stellar winds dispersing the surrounding gas. In general, collimated jet-like winds from young embedded protostars usually drive powerful collimated outflows, while wide-angle or spherical winds from the pre-main-sequence stars are more likely to drive less-collimated outflows or bubbles \citep{Arce11}. A bubble is a partially or fully enclosed three-dimensional structure whose projection is a partial or full ring \citep{Churchwell06}.

The kinetic energy of an outflow is very large \citep[$\rm 10^{43}$-$\rm 10^{48}$ erg;][]{Lada85, Bachiller96}, implying a substantial input of mechanical energy into its parent molecular cloud \citep{Solomon81}. Feedback from young stars has been proposed as a significant aspect of self-regulation of star formation \citep{Norman80, Franco83}. Feedback may maintain the observed turbulence in molecular clouds and it may also be responsible for stabilizing the clouds against gravitational collapse \citep{Shu87}. The impact of outflows on surrounding gas has been studied primarily in small regions such as Orion KL \citep{Kwan76}, L1551 \citep{Snell80} and GL 490 \citep{Lada81} on scales less than 10$'$. Recently there have been a few studies related to outflow feedback in nearby clouds. \citet{Arce10} undertook a complete survey of outflows in Perseus and found that outflows have an important impact on the environment immediately surrounding localized regions of active star formation, but that outflows have insufficient energy to feed the observed turbulence in the entire Perseus complex. \citet{Nakamura11Serpens} and \citet{Nakamura11Ophiuchi} studied the outflows in the $\rho$ Ophiuchi main cloud and Serpens south, respectively. Both studies concluded that outflows can power the supersonic turbulence in their parent molecular cloud but do not have enough momentum to support the entire cloud against the global gravitational contraction. \citet{Narayanan12} identified 20 outflows in the Taurus region and concluded that outflows cannot sustain the observed turbulence seen in the entire cloud. In this paper, we report a systematic and detailed search for outflows around sources from the Spitzer Space Telescope (hereinafter referred to as Spitzer) young stellar object (YSO) catalog and then estimated their impact on the entire Taurus molecular cloud.

Similar to outflows, bubbles are important morphological features in star formation process, which can give information about spherical stellar winds and physical properties of their surrounding environments \citep{Churchwell06}. Parsec-scale bubbles are usually found in massive star-forming regions \citep{Heyer92, Churchwell06, Churchwell07, Beaumont10, Deharveng10}. The conventional thought has been that high-mass stars can drive spherical winds and easily create the observed bubbles, while the spherical winds from low- and intermediate-mass stars are too weak to produce bubbles. However, \citet{Arce11} studied shells (bubbles) in Perseus, a nearby low-mass star-forming molecular cloud, and concluded that the total energy input from outflows and shells is sufficient to maintain the turbulence.

The Taurus molecular cloud is at a distance of 140 pc \citep{Torres09}. It covers an area of more than 100 deg$^2$ \citep{Ungerechts87}. Using the J=2-1 line of $^{12}$CO, 13 outflows have been found around low-mass embedded YSOs in Taurus \citep{Bontemps96}. There are 13 high velocity molecular outflows in Taurus included in the catalog of \citet{Wu04}. Using JCMT-HARP $^{12}$CO J=3-2 observations, 16 outflows have been found in L1495, a `bowl-shaped' region in the NW corner of Taurus \citep{Davis10}. Recently, 20 outflows have been identified, 8 of which were new detections with the Five College Radio Astronomy Observatory (FCRAO) $^{12}$CO J=1-0 and $^{13}$CO J=1-0 data cubes covering the entire Taurus molecular cloud \citep{Narayanan12}. The up-to-date catalog of YSOs \citep{Rebull10} from the Spitzer provides an opportunity to search for outflows and bubbles in a more comprehensive manner. Here we present a systematic and detailed search for outflows and bubbles in the vicinity of young stellar objects and estimate their impact on the overall Taurus molecular cloud.

The paper is organized as follows. In \S~\ref{Data} we describe the data used in the study. The details including searching methods, morphology and physical parameters of outflows and bubbles are presented in \S~\ref{Outflows} and \S~\ref{Bubbles}, respectively. The driving sources of outflows and bubbles, their energy feedback to the parent cloud and the related comparison between Taurus and Perseus are discussed \S~\ref{Discussion}. In \S~\ref{Conclusions} we summarize the main results.

\section{The Data}
\label{Data}
In our study we used the $^{12}$CO(1-0) and $^{13}$CO(1-0) data observed with 13.7 m FCRAO telescope \citep{Narayanan08}. We also adopted the up-to-date catalog of Spitzer YSOs, where 215 YSOs and 140 new YSO candidates in Taurus are reported \citep{Rebull10}.

\subsection{FCRAO CO Maps}
The FCRAO CO survey was taken between 2003 and 2005. The $^{12}$CO and $^{13}$CO maps are centered at $\alpha$=$ {04}{^{\rm h}}{32}{^{\rm m}}{44.6}{^{\rm s}}$, $\delta$=$ {24}{^{\circ}}25'44.6''$ (J2000) covering an area of approximately 100 $\rm deg^{2}$. The full width at half maximum (FWHM) beam width is $45''$ for $^{12}$CO and is $47''$ for $^{13}$CO. The pixel size of the resampled data is $20''$, which corresponds to 0.014 pc at a distance of 140 pc. There are 80 channels for $^{12}$CO and 76 channels for $^{13}$CO, covering approximately $-$5 to $+$14.9 km s$^{-1}$. The width of a velocity channel is 0.254 km s$^{-1}$ for $^{12}$CO and 0.266 km s$^{-1}$ for $^{13}$CO \citep{Narayanan08, Goldsmith08}.

\subsection{Spitzer MIPS Images}
The MIPS \citep[Multi-band Imaging Photometer for Spitzer;][]{Rieke04} maps were created as part of the final products from the Spitzer Legacy Taurus I and II surveys \citep{Padgett07}. The data were obtained in fast scan mode in three bands: 24, 70, and 160 $\mu$m, over an area of 44 $\rm deg^{2}$. The observations were performed in three epochs between 2005 and 2007, with an integration time of 30 s (24 $\mu$m) and 15 s (70 \& 160 $\mu$m). The maps were created using the basic calibrated data (BCDs) and coadded using the Spitzer software package MOPEX \citep[Mosaicking and Point Source Extractor;][]{Makovoz05}. Despite the fact that the data were taken with interleaved scan legs to provide optimal coverage at 70 and 160 $\mu$m, some small gaps remained, in particular at 160 $\mu$m. To mitigate this effect, the 160 $\mu$m final mosaic was created using 32 arcsec~pixels, instead of the native 16 arcsec/pixel scale. This pixel scale matches quite well with the $\sim$40 arcsec beam at 160 $\mu$m wavelength. The 24 and 70 $\mu$m maps adopted the standand 2.5 and 4 arcsec/pixel scale, respectively, to properly sample their respective 6 and 18 arcsec~beams. The maps were used successfully for photometric purposes to identify new sources in the Taurus Molecular Cloud \citep{Rebull10}.

\section{Outflows}
\label{Outflows}
We identified 55 outflows around the Spitzer YSOs in the 44 deg$^2$ area of Taurus. In total 31 of the detected outflows were previously unknown. In the following subsections, we describe the searching procedure of outflows, the morphology, physical properties, and the comparison between our findings and the known ones.

\subsection{The Search Procedures for Outflows}
\label{OutflowStep}
Instead of a blind search, we focused on seeking outflows around YSOs. The search procedure was performed with an Interactive Data Language (IDL) pipeline. We plotted spectra, position velocity diagrams (hereafter P-V diagrams) and integrated intensity maps to identify the outflows around the 355 YSOs which Spitzer identified in Taurus. Detailed steps of the search are the following.

(1) We plotted $^{12}$CO contours (hereafter contour map) overlaid on a $^{13}$CO grey-scale image around a YSO. According to the scale and velocity range of previously-detected outflows in Taurus, we chose two sizes ($10'$ and $20'$) and three sets of velocity intervals (-1 to 3.5 km s$^{-1}$, -1 to 4.5 km s$^{-1}$ and -1 to 5.5 km s$^{-1}$ for blue; 7.5 to 13 km s$^{-1}$, 8.5 to 13 km s$^{-1}$ and 9.5 to 13 km s$^{-1}$ for red) to plot the contour maps. We plotted the maps with 3 sets of velocity intervals and two scales around the 355 YSOs automatically. In total, 2130 maps were obtained. We inspected these maps to identify outflow candidates according to the morphology of the blue and red lobes. In the end, 74 candidates were selected.

(2) We plotted $^{12}$CO P-V diagrams along four directions (at position angles\footnote{The position angle is defined as the angle measured from the north clockwise to the direction along which we plotted the P-V diagram.} of $0^{\circ}$, $45^{\circ}$, $90^{\circ}$ and $135^{\circ}$) on three scales (20$'$, 40$'$ and 60$'$) around the 74 candidates. The size and high-velocity range of outflow candidates were determined roughly by checking the P-V diagrams. When the velocity bulge appears along the direction away from the central velocity, we marked it as the start of the high-velocity wing. And along the above direction, the maximum velocity corresponding to the outermost contour is the end of this high-velocity wing. We further confirmed each of the diagrams individually by visual inspection. The position range of the entire high-velocity bulge along the position axis was considered to be the rough size of the outflow. If more than one central velocity is found in the P-V diagram, it likely has multiple velocity components \citep{Wu05} and thus will be excluded from the list of outflow candidates. Therefore, 19 candidates with multiple velocity components were eliminated and the remaining 55 outflow candidates were considered to be possible outflows.

(3) Using the rough sizes and velocity ranges obtained in step (2), we plotted contour maps for the remaining 55 outflows. P-V diagrams were plotted through the midpoint of the blue and red peaks (bipolar outflow) or through the peak of the lobe in the case of monopolar outflow, at position angles spaced by $15^{\circ}$. We chose the angle with the most prominent bulge along the velocity axis to determine the velocity interval of the outflows. Then we plotted the contour map again with this velocity interval.

(4) Finally we plotted the average spectra of the blue and red lobes.
According to the morphology of P-V diagrams and contour maps, we divided the outflows into five classes. The higher is the ranking, the more likely it is that we have identified an outflow. We define a typical P-V diagram (TPV) and a representative contour map (RCM) as follows. If there is obvious high velocity gas which can be seen by the protuberance along the velocity axis on the P-V diagram and the high velocity range is not less than 1 km s$^{-1}$, then we regard the P-V diagram as a TPV. If the outermost contour of the lobe is closed and we can see a clear and unbroken lobe on the contour map, then we regard the contour map as a RCM. Table \ref{OutflowClass} shows our criteria for outflow classification with an ``$\times$'' meaning that it satisfies a certain condition. Having both $^{12}$CO TPV and $^{12}$CO RCM is required for high ranking (Class A$^{+}$), but having $^{13}$CO TPV or $^{13}$CO RCM gives a lower ranking (Class A$^{-}$ and Class B$^{-}$) because $^{13}$CO is generally optically thin in outflows and we usually found lobes of outflows with $^{12}$CO not $^{13}$CO. Having only $^{12}$CO RCM was divided into the lowest ranking (Class C$^{+}$) because the high velocity gas in P-V diagram is not obvious. The primary condition to identify an outflow is having high velocity gas which can be seen from the protuberance along the velocity axis on the P-V diagram.

\begin{table}
\begin{center}
\caption{Criteria for Outflow Classification and Outflow Distribution\label{OutflowClass}}
\begin{tabular}{lccccrr}
\tableline\tableline
Class & $^{12}$CO & $^{12}$CO & $^{13}$CO & $^{13}$CO & Outflow & Percentage\\
 & TPV & RCM & TPV & RCM & Numbers &  \\
\tableline
A$^{+}$  &  $\times$  &  $\times$  &            &             &  24  &  43.6\%  \\
A$^{-}$  &  $\times$  &  $\times$  &  $\times$  &  $\times$   &  18  &  32.7\%  \\
B$^{+}$  &  $\times$  &            &            &             &  4   &  7.3\%  \\
B$^{-}$  &  $\times$  &            &  $\times$  &             &  1   &  1.8\%   \\
C$^{+}$  &            &  $\times$  &            &             &  8   &  14.5\%  \\
\tableline
\end{tabular}
\end{center}
\end{table}

\subsection{The Results of Outflow Search}
\begin{figure*}%[htbp]
\centering %sub_figures centered
\includegraphics[width=1.1\textwidth]{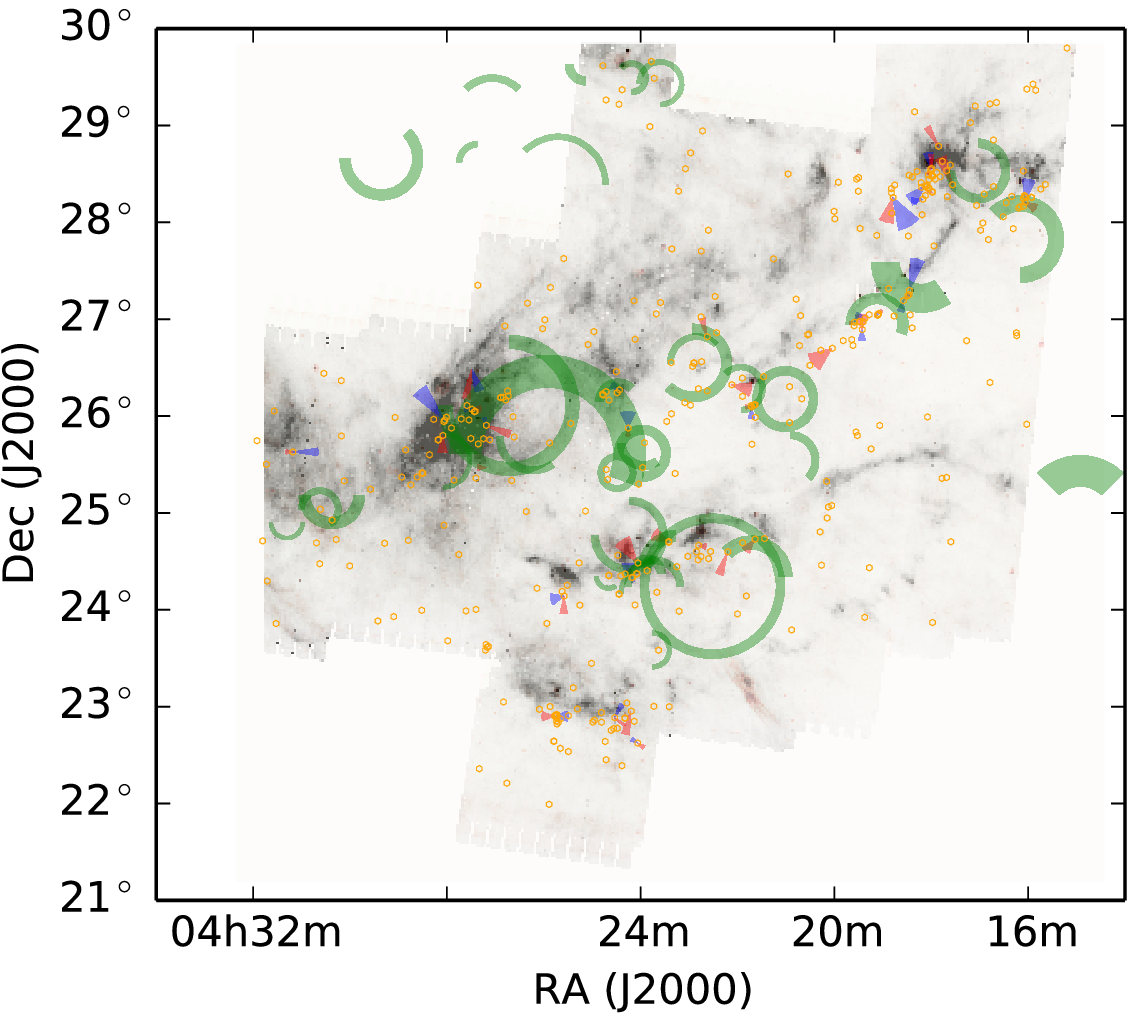}
\vspace{-5mm}

\caption{Outflows and bubbles overlaid on the Spitzer 160 $\mu$m (grey), 70 $\mu$m (green) and 24 $\mu$m (red) image from the Spitzer Legacy Taurus I and II surveys. The blue and red sectors represent blue and red lobes of outflows in Table \ref{OutflowList}, respectively. The radius of the sectors shows the scale of the outflow, while the direction of the sectors shows the projected direction of the outflow. The green rings and arcs represent the expanding and broken bubbles, respectively. The thickness and radius of the arcs and rings are the actual thickness and radius of the bubble structures. The open orange circles show the locations of YSOs, which were listed in Table 6 and Table 7 of \citet{Rebull10}.}
\label{BigMap}
\end{figure*}

Following the steps in \S~\ref{OutflowStep} we found 55 outflows in the 44 deg$^2$ area of Taurus molecular cloud. All the outflows we detected were listed in Table \ref{OutflowList}. Each outflow is referred as a ``Taurus Molecular Outflow" (TMO). We present the locations, polarities and scales of the outflows overlaid on the Spitzer MIPS Image in Fig. \ref{BigMap}. There are 31 new ones among all the detected outflows. We have thus increased the total number of known outflows by a factor of 1.3.

\begin{deluxetable}{lccllclcl}
\tabletypesize{\scriptsize}
%\rotate
\tablecaption{Outflows in Taurus
\label{OutflowList}}
\tablewidth{0pt}
\tablehead{
 \colhead{Outflow} & \colhead{R.A.} & \colhead{Dec.} & \colhead{Common} & \colhead{YSO}  & \colhead{Outflow} & \colhead{Po.\tablenotemark{b}} & \colhead{New} & \colhead{Ref.} \\
 \colhead{Name} & \colhead{(J2000)} & \colhead{(J2000)} & \colhead{Name} & \colhead{Type\tablenotemark{a}}  & \colhead{Class} & \colhead{} & \colhead{Detection\tablenotemark{c}} & \colhead{}
}
\startdata
TMO\_01	 &  04 11 59.7  &  29 42 36  & 	-                                  &  III	 &  A$^{+}$  &  MR  &  N  &  1\\
TMO\_02	 &  04 14 12.2  &  28 08 37  & 	IRAS 04113+2758 (L1495)            &  I     &  A$^{+}$  &  Bi  &  N  &  1, 2, 3\\ % L1495
TMO\_03	 &  04 14 14.5  &  28 27 58  & 	-                                  &  II	 &  A$^{+}$  &  Bi  &  Y  &  4\\ %4arcmin
TMO\_04  &  04 18 32.0  &  28 31 15  & 	-                                  &  flat  &  A$^{+}$  &  Bi  &  Y  &  4\\ %Core name:G168.69-15.47; 4arcmin
TMO\_05  &  04 19 41.4  &  27 16 07  &  IRAS 04166+2706                    &  I     &  A$^{+}$  &  MB  &  N  &  1, 2, 5, 6, 7\\
TMO\_06	 &  04 19 58.4  &  27 09 57  & 	IRAS 04169+2702                    &  I     &  A$^{+}$  &  Bi  &  N  &  1, 2, 3, 7\\
TMO\_07  &  04 21 07.9  &  27 02 20  & 	IRAS 04181+2655                    &  I     &  A$^{+}$  &  Bi  &  N  &  2, 3, 6, 7\\
TMO\_08  &  04 22 15.6  &  26 57 06  & 	FS Tau B                           &  I     &  A$^{+}$  &  Bi  &  N  &  1, 2\\
TMO\_09	 &  04 23 25.9  &  25 03 54  & 	-                                  &  II	 &  A$^{+}$  &  MR  &  Y  &  4\\ %8arcmin
TMO\_10	 &  04 24 20.9  &  26 30 51  & 	-                                  &  II	 &  A$^{+}$  &  Bi  &  Y  &  4\\ %4arcmin
TMO\_11	 &  04 24 45.0  &  27 01 44  & 	-                                  &  III	 &  A$^{+}$  &  MR  &  Y  &  4\\ %4arcmin
TMO\_12  &  04 29 30.0  &  24 39 55  &  Haro 6-10                          &  I     &  A$^{+}$  &  MR  &  N  &  1, 6, 8\\
TMO\_13  &  04 31 10.4  &  25 41 29  & 	-                                  &  -     &  A$^{+}$  &  MR  &  Y  &  4\\ %4arcmin
TMO\_14	 &  04 31 58.4  &  25 43 29  & 	-                                  &  III	 &  A$^{+}$  &  Bi  &  Y  &  4\\ %3arcmin
TMO\_15  &  04 32 14.6  &  22 37 42  &  -                                  &  flat  &  A$^{+}$  &  Bi  &  Y  &  4\\ %3arcmin
TMO\_16  &  04 32 31.7  &  24 20 02  & 	L1529                              &  II	 &  A$^{+}$  &  Bi  &  N  &  1, 6, 9, 10\\
TMO\_17  &  04 32 32.0  &  22 57 26  &  IRAS 04295+2251 (L1536)            &  I     &  A$^{+}$  &  MR  &  N  &  3, 7\\ %L1536 IRS
TMO\_18  &  04 32 43.0  &  25 52 31  & 	-                                  &  II	 &  A$^{+}$  &  Bi  &  Y  &  4\\ %6arcmin
TMO\_19	 &  04 34 15.2  &  22 50 30  &  -                                  &  II	 &  A$^{+}$  &  MR  &  Y  &  4\\ %2arcmin
TMO\_20	 &  04 37 24.8  &  27 09 19  & 	-                                  &  III	 &  A$^{+}$  &  MR  &  Y  &  4\\ %3arcmin
TMO\_21	 &  04 39 53.9  &  26 03 09  & 	L1527                              &  I     &  A$^{+}$  &  Bi  &  N  &  1, 3, 6, 7, 11, 12, 13\\
TMO\_22  &  04 41 08.2  &  25 56 07  & 	IRAS 04381+2540 (TMC-1)            &  flat  &  A$^{+}$  &  MB  &  N  &  1, 6, 7, 14\\ % IRAS 04381+2540
TMO\_23	 &  04 41 12.6  &  25 46 35  &  IRAS 04381+2540 (TMC-1)            &  I	    &  A$^{+}$  &  MR  &  N  &  1, 6, 7, 14\\ % IRAS 04381+2540
TMO\_24	 &  04 42 07.7  &  25 23 11  & 	IRAS 04390+2517 (LkH$\alpha$ 332)  &  II    &  A$^{+}$  &  MR  &  N  &  3\\
TMO\_25  &  04 18 58.1  &  28 12 23  & 	IRAS 04158+2805 (L1495)            &  flat  &  A$^{-}$  &  MB  &  Y  &  4, 7\\ %2arcmin
TMO\_26  &  04 23 18.2  &  26 41 15  & 	-                                  &  II	 &  A$^{-}$  &  MR  &  Y  &  4\\
TMO\_27  &  04 26 56.2  &  24 43 35  & 	IRAS 04239+2436 (HH 300)           &  I	    &  A$^{-}$  &  MR  &  N  &  1, 3, 6, 15\\
TMO\_28  &  04 27 02.6  &  26 05 30  & 	IRAS 04240+2559 (DG Tau)           &  I	    &  A$^{-}$  &  Bi  &  N  &  1, 16\\
TMO\_29	 &  04 27 02.8  &  25 42 22  & 	-                                  &  II	 &  A$^{-}$  &  Bi  &  Y  &  4\\ %4arcmin
TMO\_30	 &  04 27 57.3  &  26 19 18  &  IRAS 04248+2612                    &  flat  &  A$^{-}$  &  MR  &  N  &  1, 3\\
TMO\_31	 &  04 28 10.4  &  24 35 53  & 	-                                  &  flat  &  A$^{-}$  &  MR  &  Y  &  4\\
TMO\_32	 &  04 30 51.7  &  24 41 47  & 	IRAS 04278+2435 (ZZ Tau IRS)       &  flat  &  A$^{-}$  &  MR  &  N  &  1, 6, 17\\
TMO\_33  &  04 32 15.4  &  24 28 59  & 	IRAS 04292+2422 (Haro 6-13)        &  flat  &  A$^{-}$  &  Bi  &  N  &  1, 3\\
TMO\_34	 &  04 33 07.8  &  26 16 06  & 	-                                  &  III   &  A$^{-}$  &  MB  &  Y  &  4\\ %3arcmin
TMO\_35	 &  04 33 10.0  &  24 33 43  & 	-                                  &  III	 &  A$^{-}$  &  MR  &  Y  &  4\\ %3arcmin
TMO\_36  &  04 33 16.5  &  22 53 20  & 	IRAS 04302+2247                    &  I	    &  A$^{-}$  &  Bi  &  N  &  1, 3, 7\\
TMO\_37  &  04 33 34.0  &  24 21 17  & 	-                                  &  II	 &  A$^{-}$	 &  MR  &  Y  &  4\\
TMO\_38	 &  04 33 36.7  &  26 09 49  & 	-                                  &  II	 &  A$^{-}$  &  MB  &  Y  &  4\\
TMO\_39  &  04 35 57.6  &  22 53 57  & 	IRAS 04328+2248 (HP Tau)           &  II	 &  A$^{-}$	 &  Bi  &  N  &  3\\
TMO\_40	 &  04 39 11.2  &  25 27 10  & 	HH706                              &  -     &  A$^{-}$  &  MR  &  N  &  1\\
TMO\_41  &  04 39 13.8  &  25 53 20  & 	IRAS 04361+2547 (TMR-1)            &  I	    &  A$^{-}$  &  Bi  &  N  &  1, 3, 6, 7, 18\\
TMO\_42	 &  04 48 02.3  &  25 33 59  & 	Tau A 8                            &  III	 &  A$^{-}$  &  Bi  &  Y  &  4\\
TMO\_43  &  04 18 51.4  &  28 20 26  & 	HH156                              &  I	    &  B$^{+}$	 &  MB  &  Y  &  2\\%5arcmin
TMO\_44  &  04 20 21.4  &  28 13 49  & 	-                                  &  flat	 &  B$^{+}$	 &  Bi  &  Y  &  4\\%8arcmin
TMO\_45	 &  04 26 53.3  &  25 58 58  & 	-                                  &  I	    &  B$^{+}$  &  Bi  &  Y  &  4\\%3arcmin
TMO\_46	 &  04 39 56.1  &  26 28 02  & 	-                                  &  -     &  B$^{+}$  &  Bi  &  Y  &  4\\%9arcmin
TMO\_47	 &  04 35 35.3  &  24 08 19  & 	IRAS 04325+2402 (L1535)            &  I	    &  B$^{-}$  &  Bi  &  N  &  1, 3, 6, 17, 19, 20\\	
TMO\_48	 &  04 15 35.6  &  28 47 41  & 	-                                  &  I	    &  C$^{+}$  &  Bi  &  Y  &  4\\%3arcmin
TMO\_49	 &  04 17 33.7  &  28 20 46  & 	-                                  &  II	 &  C$^{+}$	 &  MR  &  Y  &  4\\%3arcmin
TMO\_50	 &  04 18 10.5  &  28 44 47  & 	-                                  &  I	    &  C$^{+}$	 &  MR  &  Y  &  4\\%6arcmin
TMO\_51  &  04 18 31.1  &  28 16 29  & 	-                                  &  II	 &  C$^{+}$  &  MR  &  Y  &  4\\%3arcmin
TMO\_52  &  04 18 31.2  &  28 26 17  & 	-                                  &  I	    &  C$^{+}$	 &  MR  &  Y  &  4\\%2arcmin
TMO\_53  &  04 18 41.3  &  28 27 25  & 	-                                  &  flat	 &  C$^{+}$	 &  MR  &  Y  &  4\\%4arcmin
TMO\_54  &  04 21 54.5  &  26 52 31  & 	-                                  &  II    &  C$^{+}$	 &  Bi  &  Y  &  4\\	
TMO\_55  &  04 29 04.9  &  26 49 07  & 	IRAS 04260+2642                    &  I     &  C$^{+}$	 &  MR  &  N  &  4\\%6arcmin
\enddata
\tablenotetext{a}{The YSO classification from \citet{Rebull10}. ``flat" represents a flat-spectrum YSO, which is intermediate between Class I and II.}
\tablenotetext{b}{The polarity of the outflow. ``Bi" represents bipolar outflow, ``MB" and ``MR" indicate blue and red monopolar outflow, respectively.}
\tablenotetext{c}{The column represents whether the outflow is detected for the first time in our study. Y = new, N = has reported in previous work.}
\tablerefs{(1) Narayanan et al. (2012); (2) Davis et al. (2010); (3) Moriarty-Schieven et al. (1992); (4) this paper; (5) Tafalla et al. (2004); (6) Wu et al. (2004); (7) Bontemps et al. (1996); (8) Stojimirovi\'{c}, Narayanan \& Snell (2007); (9) Lichten (1982); (10) Goldsmith et al. (1984); (11) Tamura et al. (1996); (12) Hogerheijde et al. (1998); (13) Zhou, Evans \& Wang (1996); (14) Chandler et al. (1996); (15) Arce \& Goodman (2001); (16) Mitchell et al. (1994); (17) Heyer, Snell \& Goldsmith (1987); (18) Terebey et al. (1990); (19) Myers et al. (1988); (20) Wu, Zhou, \& Evans II (1992).}
\end{deluxetable}

Table~\ref{OutflowClass} lists the numbers and percentages in the five classes of outflows. We can see Class A$^{+}$ and class A$^{-}$ account for 76.3\% of all the detected outflows. These two types can be considered as the ``most probable" outflows in our study. Table \ref{OutflowClassNum} lists the numbers of previously known and newly detected outflows in different classes. We found more new outflows of Class A$^{+}$ and class A$^{-}$, which account for 64.5\% of all the newly detected outflows. That is, most of the new outflows we found are likely true outflows.

\begin{table}
\begin{center}
\caption{The Numbers of Outflows in Different Classes\label{OutflowClassNum}}
\begin{tabular}{lcc}
\tableline\tableline
Type & Previously & Newly\\
     & Known & Detected\\
\tableline
A$^{+}$  &  13  &  11  \\
A$^{-}$  &  9   &  9   \\
B$^{+}$  &  0   &  4   \\
B$^{-}$  &  1   &  0   \\
C$^{+}$  &  1   &  7   \\
\tableline
\end{tabular}
\end{center}
\end{table}

Table~\ref{OutflowNumYSO} lists the outflow numbers and percentages according to the types of their driving sources. Class I accounts for 36.4\%, which is the largest proportion of all the YSOs driving outflow. The outflows driven by the Class I YSOs are closer to the YSOs and have more collimated bi-polar morphology. Compared with the Class I, Class III YSOs drive a small proportion (12.7\%) of outflows, which tend to be farther from the YSOs. This indicates that the outflows from Class III YSOs are more evolved than those from Class I YSOs. We also found three outflows (TMO\_13, TMO\_40, TMO\_46) without YSOs, indicating that they are possibly Class 0 objects. Among the three outflows, TMO\_13 and TMO\_40 are newly found in our study, while TMO\_46 has been reported in \citet{Narayanan12}.

\begin{table}
\begin{center}
\caption{Number of Outflows Around Different Types of YSOs \label{OutflowNumYSO}}
\begin{tabular}{lrr}
\tableline\tableline
YSO & Number of & Percentage\\
Type & Outflows &  \\
\tableline
I       &  20  &  36.4\%  \\
Flat    &  10  &  18.2\%  \\
II      &  15  &  27.3\%  \\
III     &  7   &  12.7\%  \\
No YSO  &  3   &  5.4\%   \\
\tableline
\end{tabular}
\end{center}
\end{table}

\subsection{Morphology of Outflows}
We found 25 bipolar, 22 monopolar redshifted and 6 monopolar blueshifted outflows. Bipolar and redshifted outflows account for the vast majority of outflows in the Taurus molecular cloud. This is consistent with the results of \citet{Narayanan12}. Fig. \ref{sample-figure01} - Fig. \ref{sample-figure55} show the $\rm {}^{12}CO$ integrated intensity map, $\rm {}^{12}CO$ P-V diagram and average spectrum for each outflow. For Class A$^{-}$ and B$^{-}$ outflows, we also plotted the $\rm {}^{13}CO$ integrated intensity maps and $\rm {}^{13}CO$ P-V diagrams.

\subsection{Comparison with Previously Found Outflows}
Using the FCRAO large-scale survey data \citep{Narayanan08} and the latest YSO catalog from \citet{Rebull10} we were able to identify the previously known outflows, obtain more complete morphology, and find additional new outflows. The YSO catalog is also convenient for identifying the driving sources of the outflows. Comparing with the previous works, we confirmed more driving sources of outflows.

L1527 (TMO\_21) is a typical outflow in Taurus \citep{Narayanan12, Hogerheijde98, Zhou96}. The P-V diagram and contour map in our work are very similar to those in \citet{Hogerheijde98} and \citet{Narayanan12}. TMO\_08 (SST 042215.6+265706) and FS Tau B in \citet{Narayanan12} are the same outflow with the same location. They have the same structure, which can be seen in our Fig. \ref{sample-figure08} and Figure 15 in \citet{Narayanan12}. In addition, TMO\_30 (SST 042757.3+261918), TMO\_32 (SST 043051.7+244147), TMO\_33 (SST 043215.4+242859) and TMO\_41 (SST 043913.8+255320) also have the same morphology as IRAS 04248+2612, ZZ Tau IRS, IRAS 04292+2422 and IRAS 04361+2547 in \citet{Narayanan12}, respectively. These confirm the general consistency between the two works in terms of strong and extended outflows.

For TMO\_02 (SST 041412.2+280837), we obtained a good bipolar structure shown in the upper left panel of Fig. \ref{sample-figure01}, while \citet{Narayanan08} considered this outflow (IRAS 04113+2758) to be redshifted only. \citet{Davis10} did not identify the driving source of this outflow (named by W-CO-flow1), while we determined that the YSO SST 041412.2+280837 is driving the outflow. \citet{Moriarty-Schieven92} only presented the central spectrum of IRAS 04390+2517 and IRAS 04328+2248, while we illustrated the two outflows (TMO\_24 and TMO\_39) more clearly through contour maps and P-V diagrams.

The morphology of TMO\_07 (SST 042107.9+270220) shown in Fig. \ref{sample-figure07} is similar to that of J04210795+2702204 in \citet{Davis10}. This outflow was also reported by \citet{Moriarty-Schieven92}, \citet{Bontemps96} and \citet{Wu04}. However, \citet{Narayanan12} did not find it with the same FCRAO survey data. \citet{Mitchell94} reported that IRAS 04240+2559 was a monopolar redshifted outflow with $^{12}$CO (3-2) line. But at this location we found the well-defined bipolar outflow as shown in Fig. \ref{sample-figure28}.

As for L1529, \citet{Lichten82} presented high-velocity $^{12}$CO wings observed by antenna No. 2 of the Caltech 10.4 m array, but \citet{Goldsmith84} did not find any high-velocity gas in observations at FCRAO. We identified a bipolar outflow named TMO\_16 (SST 043231.7+242002) and demonstrated the result of \citet{Lichten82} with the FCRAO data. At the position of IRAS 04295+2251 \citet{Moriarty-Schieven92} showed line wings while \citet{Bontemps96} found no outflow. We found a red monopolar outflow as shown in Fig. \ref{sample-figure17}.

We have found 31 new outflows which are labeled ``Y" in the eighth column of Table \ref{OutflowList}. Two of these new outflows were not identified as outflows in the literature. \citet{Bontemps96} considered IRAS 04158+2805 (L1495) but did not find any sign of outflow activity in the $^{12}$CO (2-1) transition at the location of TMO\_025 (SST 041858.1+281223). \citet{Davis10} had some doubt about CO flow of CoKU Tau-1 when analyzing the $^{12}$CO (3-2) emission, while we found TMO\_043 (SST 041851.4+282026) at this site. The rest of the new outflows have not been reported in the literature and are identified as outflows for the first time. All of the new outflows are of small angular extent, less than 10$'$. They were missed in previous searches perhaps because of their small sizes.

\subsection{Physical Parameters of Outflows}
To study the effects of outflows on their environment we calculated their masses, momenta, kinetic energy and energy deposition rates. The total column density of the outflowing gas is
\begin{equation}
N_{\rm tot}(^{12}{\rm CO})=\frac{3k^2T_{\rm ex}}{4\pi^3\mu^2_{\rm d}h\nu^2 \exp(-h\nu/kT_{\rm ex})}\int T_{\rm s} dv,
\label{equ:DerivedTotalColumnDensity}
\end{equation}
where $k=1.38\times10^{-16}~{\rm erg/K}$, $h=6.626\times10^{-27}$ erg s, $\mu_{\rm d}=0.112\times10^{-18}~{\rm esu}$, $\nu=115.2712\times10^9~{\rm Hz}$ and $T_{\rm s}$ is the observed source antenna temperature with proper correction for antenna efficiency. We assumed an excitation temperature of 25 K. The excitation temperature assumed in the literature \citep{Zhou96, Tamura96, Ohashi97a, Ohashi97b, Davis10, Narayanan12} ranges from 11 K to 50 K. The lowest temperature will decrease the mass estimate by a factor of 3 and the highest temperature will increase the mass estimate by a factor of 2.2. The detailed derivations regarding the physical parameters of the outflows are given in appendix A.

\citet{Arce01} described three major issues that can cause uncertainties in the calculation of an outflow's parameters, namely, the inclination, opacity, and blending. Our prescriptions are the following. (1) We defined the inclination angle of the outflow as the angle between the long axis of the outflow and the line of sight. Since the outflows with \textbf{small} inclination angle (especially when \textbf{the outflow is perpendicular to the plane of the sky}) are hard to detect, our outflow searching is biased to those with \textbf{large} inclination angle. If the inclination angle $\theta$ is randomly distributed, the average value is given by
\begin{equation}
\langle \theta\rangle=\int^{\pi/2}_{0}\theta\sin\theta{\rm d}\theta.
\label{equ:InclinationAngle}
\end{equation}
From the above formula we got the average inclination angle of \textbf{57.3$^\circ$}, which differs from the usually used median value of 45$^\circ$. Then the velocity and the dynamic age, $t_{\rm dyn}$, should be scaled up by a factor of \textbf{1.9} and \textbf{0.64}, respectively. (2) Using the $\rm {}^{12}CO$ and $\rm {}^{13}CO$ data we can correct for the opacity in the $^{12}$CO line when $\rm {}^{12}CO$ emission of an outflow is optically thick. The algorithm for the opacity correction is described in appendix A. (3) We probably missed some low-velocity outflowing gas, which blended into the ambient gas, when we conservatively determined the emission only from outflows. Previous studies \citep{Margulis85, Arce10, Narayanan12} showed that neglecting this gas results in the underestimate of the outflow mass almost by a factor of 2.

Table \ref{OutflowParameters} gives the length, mass, momentum, kinetic energy, dynamical timescale and the luminosity of the outflows in Taurus.

\begin{deluxetable}{llcccccccc}
\tabletypesize{\scriptsize}
%\rotate
\tablecaption{Physical Parameters of Outflows
\label{OutflowParameters}}
\tablewidth{0pt}
\tablehead{
 \colhead{Outflow} & \colhead{Lobe} & \colhead{$v_{\rm avg}$\tablenotemark{a}}  & \colhead{Area\tablenotemark{b}} & \colhead{Length} & \colhead{Mass} & \colhead{Momentum}  & \colhead{Energy} & \colhead{$t_{\rm dyn}$} & \colhead{$L_{\rm flow}$} \\
 \colhead{Name} & \colhead{} & \colhead{($\rm km\ s^{-1}$)} & \colhead{(arcmin)} & \colhead{(pc)} & \colhead{($M_{\odot}$)} & \colhead{($M_{\odot}\ km\ s^{-1}$)} & \colhead{($10^{43}$ erg)} & \colhead{($10^{5}$ yr)} & \colhead{($10^{30}\rm erg\ s^{-1}$)}
}

\startdata
TMO\_01 & Blueshifted & -   &  -             & -    & -     & -      &  -    & -   & -    \\
        & Redshifted  & 2.5 & 11 $\times$ 25 & 1.11 & 0.083 & 0.205  &  0.51 & 4.4 & 0.37 \\
TMO\_02 & Blueshifted & 3.4 &  8 $\times$ 12 & 0.58 & 0.168 & 0.570  &  1.92 & 1.7 & 3.63 \\
        & Redshifted  & 3.6 &  6 $\times$ 5  & 0.31 & 0.061 & 0.220  &  0.79 & 0.8 & 2.97 \\
TMO\_03 & Blueshifted & 3.1 &  6 $\times$ 9  & 0.41 & 0.050 & 0.156  &  0.49 & 1.3 & 1.19 \\
        & Redshifted  & 2.4 & 11 $\times$ 2  & 0.46 & 0.013 & 0.030  &  0.07 & 1.9 & 0.12 \\
TMO\_04 & Blueshifted & 2.6 &  6 $\times$ 8  & 0.41 & 0.026 & 0.069  &  0.18 & 1.5 & 0.36 \\
        & Redshifted  & 2.5 &  3 $\times$ 8  & 0.33 & 0.005 & 0.013  &  0.03 & 1.3 & 0.08 \\
TMO\_05 & Blueshifted & 4.6 &  9 $\times$ 18 & 0.83 & 0.078 & 0.355  &  1.61 & 1.8 & 2.89 \\
        & Redshifted  & -   &  -             & -    & -     & -      &  -    & -   & -    \\
TMO\_06 & Blueshifted & 2.8 &  5 $\times$ 5  & 0.29 & 0.038 & 0.109  &  0.31 & 1.0 & 0.96 \\
        & Redshifted  & 3.9 &  3 $\times$ 5  & 0.21 & 0.010 & 0.038  &  0.15 & 0.5 & 0.88 \\
TMO\_07 & Blueshifted & 2.1 &  7 $\times$ 7  & 0.38 & 0.012 & 0.025  &  0.05 & 1.8 & 0.09 \\
        & Redshifted  & 2.7 &  2 $\times$ 2  & 0.13 & 0.002 & 0.005  &  0.01 & 0.5 & 0.08 \\
TMO\_08 & Blueshifted & 3.1 &  5 $\times$ 4  & 0.26 & 0.013 & 0.039  &  0.12 & 0.8 & 0.48 \\
        & Redshifted  & 2.8 &  2 $\times$ 2  & 0.12 & 0.001 & 0.003  &  0.01 & 0.4 & 0.07 \\
TMO\_09 & Blueshifted & -   &  -             & -    & -     & -      &  -    & -   & -    \\
        & Redshifted  & 1.8 & 17 $\times$ 18 & 1.01 & 0.311 & 0.565  &  1.02 & 5.5 & 0.59 \\
TMO\_10 & Blueshifted & 2.3 &  3 $\times$ 7  & 0.29 & 0.010 & 0.023  &  0.05 & 1.3 & 0.13 \\
        & Redshifted  & 2.9 &  7 $\times$ 3  & 0.30 & 0.016 & 0.047  &  0.14 & 1.0 & 0.43 \\
TMO\_11 & Blueshifted & -   &  -             & -    & -     & -      &  -    & -   & -    \\
        & Redshifted  & 2.6 &  7 $\times$ 7  & 0.40 & 0.020 & 0.052  &  0.14 & 1.5 & 0.29 \\
TMO\_12 & Blueshifted & -   &  -             & -    & -     & -      &  -    & -   & -    \\
        & Redshifted  & 2.5 &  4 $\times$ 3  & 0.22 & 0.014 & 0.036  &  0.09 & 0.8 & 0.33 \\
TMO\_13 & Blueshifted & -   &  -             & -    & -     & -      &  -    & -   & -    \\
        & Redshifted  & 2.2 &  7 $\times$ 4  & 0.34 & 0.053 & 0.115  &  0.25 & 1.5 & 0.51 \\
TMO\_14 & Blueshifted & 2.6 &  5 $\times$ 5  & 0.29 & 0.019 & 0.050  &  0.13 & 1.1 & 0.38 \\
        & Redshifted  & 3.2 &  7 $\times$ 4  & 0.31 & 0.031 & 0.099  &  0.32 & 0.9 & 1.05 \\
TMO\_15 & Blueshifted & 1.9 &  3 $\times$ 3  & 0.18 & 0.013 & 0.024  &  0.05 & 0.9 & 0.16 \\
        & Redshifted  & 1.5 &  3 $\times$ 4  & 0.20 & 0.010 & 0.016  &  0.02 & 1.3 & 0.06 \\
TMO\_16 & Blueshifted & 2.3 &  2 $\times$ 2  & 0.11 & 0.005 & 0.012  &  0.03 & 0.4 & 0.21 \\
        & Redshifted  & 2.0 &  4 $\times$ 4  & 0.23 & 0.023 & 0.046  &  0.09 & 1.1 & 0.25 \\
TMO\_17 & Blueshifted & -   &  -             & -    & -     & -      &  -    & -   & -    \\
        & Redshifted  & 1.7 &  5 $\times$ 14 & 0.60 & 0.042 & 0.072  &  0.12 & 3.5 & 0.11 \\
TMO\_18 & Blueshifted & 2.0 & 10 $\times$ 3  & 0.42 & 0.012 & 0.024  &  0.05 & 2.1 & 0.07 \\
        & Redshifted  & 1.9 &  3 $\times$ 4  & 0.20 & 0.007 & 0.013  &  0.02 & 1.0 & 0.08 \\
TMO\_19 & Blueshifted & -   &  -             & -    & -     & -      &  -    & -   & -    \\
        & Redshifted  & 2.2 &  4 $\times$ 4  & 0.23 & 0.008 & 0.017  &  0.04 & 1.0 & 0.11 \\
TMO\_20 & Blueshifted & -   &  -             & -    & -     & -      &  -    & -   & -    \\
        & Redshifted  & 4.0 &  7 $\times$ 3  & 0.31 & 0.017 & 0.067  &  0.26 & 0.8 & 1.08 \\
TMO\_21 & Blueshifted & 3.1 &  4 $\times$ 2  & 0.17 & 0.004 & 0.012  &  0.04 & 0.5 & 0.22 \\
        & Redshifted  & 2.9 &  5 $\times$ 3  & 0.22 & 0.011 & 0.032  &  0.09 & 0.7 & 0.40 \\
TMO\_22 & Blueshifted & 1.6 & 10 $\times$ 22 & 0.97 & 0.226 & 0.351  &  0.54 & 6.1 & 0.28 \\
        & Redshifted  & -   &  -             & -    & -     & -      &  -    & -   & -    \\
TMO\_23 & Blueshifted & -   &  -             & -    & -     & -      &  -    & -   & -    \\
        & Redshifted  & 3.7 &  3 $\times$ 6  & 0.27 & 0.019 & 0.072  &  0.26 & 0.7 & 1.18 \\
TMO\_24 & Blueshifted & -   &  -             & -    & -     & -      &  -    & -   & -    \\
        & Redshifted  & 3.2 &  4 $\times$ 6  & 0.29 & 0.015 & 0.047  &  0.15 & 0.9 & 0.52 \\
TMO\_25 & Blueshifted & 2.4 &  5 $\times$ 5  & 0.31 & 0.267 & 0.640  &  1.52 & 1.2 & 3.87 \\
        & Redshifted  & -   &  -             & -    & -     & -      &  -    & -   & -    \\
TMO\_26 & Blueshifted & -   &  -             & -    & -     & -      &  -    & -   & -    \\
        & Redshifted  & 2.4 & 13 $\times$ 8  & 0.60 & 0.369 & 0.880  &  2.09 & 2.5 & 2.68 \\
TMO\_27 & Blueshifted & -   &  -             & -    & -     & -      &  -    & -   & -    \\
        & Redshifted  & 2.4 &  6 $\times$ 7  & 0.37 & 0.074 & 0.174  &  0.41 & 1.5 & 0.86 \\
TMO\_28 & Blueshifted & 2.2 &  4 $\times$ 6  & 0.30 & 0.170 & 0.371  &  0.80 & 1.3 & 1.91 \\
        & Redshifted  & 3.9 &  3 $\times$ 2  & 0.15 & 0.031 & 0.122  &  0.47 & 0.4 & 3.80 \\
TMO\_29 & Blueshifted & 1.9 &  9 $\times$ 4  & 0.40 & 0.264 & 0.492  &  0.91 & 2.1 & 1.37 \\
        & Redshifted  & 2.1 &  6 $\times$ 5  & 0.32 & 0.076 & 0.161  &  0.34 & 1.5 & 0.74 \\
TMO\_30 & Blueshifted & -   &  -             & -    & -     & -      &  -    & -   & -    \\
        & Redshifted  & 2.4 &  9 $\times$ 9  & 0.49 & 0.286 & 0.673  &  1.57 & 2.1 & 2.43 \\
TMO\_31 & Blueshifted & -   &  -             & -    & -     & -      &  -    & -   & -    \\
        & Redshifted  & 2.1 &  5 $\times$ 14 & 0.62 & 0.277 & 0.587  &  1.24 & 2.9 & 1.37 \\
TMO\_32 & Blueshifted & -   &  -             & -    & -     & -      &  -    & -   & -    \\
        & Redshifted  & 3.3 &  9 $\times$ 6  & 0.42 & 0.131 & 0.429  &  1.39 & 1.2 & 3.53 \\
TMO\_33 & Blueshifted & 2.1 &  8 $\times$ 5  & 0.38 & 0.120 & 0.249  &  0.51 & 1.8 & 0.90 \\
        & Redshifted  & 3.7 & 11 $\times$ 13 & 0.68 & 0.896 & 3.299  & 12.08 & 1.8 & 1.174\\
TMO\_34 & Blueshifted & 3.2 &  5 $\times$ 3  & 0.23 & 0.061 & 0.198  &  0.63 & 0.7 & 2.86 \\
        & Redshifted  & -   &  -             & -    & -     & -      &  -    & -   & -    \\
TMO\_35 & Blueshifted & -   &  -             & -    & -     & -      &  -    & -   & -    \\
        & Redshifted  & 3.2 &  6 $\times$ 4  & 0.28 & 0.049 & 0.156  &  0.49 & 0.9 & 1.79 \\
TMO\_36 & Blueshifted & 2.6 &  9 $\times$ 14 & 0.65 & 0.484 & 1.241  &  3.17 & 2.5 & 4.03 \\
        & Redshifted  & 1.9 &  8 $\times$ 6  & 0.42 & 0.092 & 0.177  &  0.34 & 2.1 & 0.51 \\
TMO\_37 & Blueshifted & -   &  -             & -    & -     & -      &  -    & -   & -    \\
        & Redshifted  & 2.1 & 15 $\times$ 19 & 0.99 & 0.348 & 0.737  &  1.55 & 4.6 & 1.08 \\
TMO\_38 & Blueshifted & 2.3 & 13 $\times$ 8  & 0.61 & 0.377 & 0.860  &  1.95 & 2.6 & 2.35 \\
        & Redshifted  & -   &  -             & -    & -     & -      &  -    & -   & -    \\
TMO\_39 & Blueshifted & 2.3 &  7 $\times$ 5  & 0.35 & 0.136 & 0.313  &  0.72 & 1.5 & 1.51 \\
        & Redshifted  & 2.2 &  6 $\times$ 5  & 0.32 & 0.076 & 0.167  &  0.37 & 1.4 & 0.81 \\
TMO\_40 & Blueshifted & -   &  -             & -    & -     & -      &  -    & -   & -    \\
        & Redshifted  & 4.7 &  4 $\times$ 4  & 0.25 & 0.129 & 0.607  &  2.84 & 0.5 & 7.683\\
TMO\_41 & Blueshifted & 3.8 &  3 $\times$ 4  & 0.23 & 0.028 & 0.104  &  0.39 & 0.6 & 2.09 \\
        & Redshifted  & 3.9 &  5 $\times$ 15 & 0.63 & 0.317 & 1.226  &  4.72 & 1.6 & 9.401\\
TMO\_42 & Blueshifted & 3.2 &  5 $\times$ 15 & 0.65 & 0.105 & 0.338  &  1.08 & 2.0 & 1.72 \\
        & Redshifted  & 2.3 &  3 $\times$ 3  & 0.17 & 0.004 & 0.010  &  0.02 & 0.7 & 0.10 \\
TMO\_43 & Blueshifted & 2.7 &  9 $\times$ 9  & 0.51 & 0.034 & 0.092  &  0.24 & 1.9 & 0.41 \\
        & Redshifted  & -   &  -             & -    & -     & -      &  -    & -   & -    \\
TMO\_44 & Blueshifted & 2.4 & 15 $\times$ 14 & 0.84 & 0.148 & 0.359  &  0.87 & 3.4 & 0.81 \\
        & Redshifted  & 1.4 &  9 $\times$ 13 & 0.63 & 0.070 & 0.097  &  0.13 & 4.4 & 0.10 \\
TMO\_45 & Blueshifted & 2.9 &  2 $\times$ 4  & 0.19 & 0.013 & 0.037  &  0.11 & 0.6 & 0.53 \\
        & Redshifted  & 2.1 &  3 $\times$ 2  & 0.14 & 0.006 & 0.013  &  0.03 & 0.6 & 0.14 \\
TMO\_46 & Blueshifted & 2.8 &  6 $\times$ 13 & 0.58 & 0.050 & 0.141  &  0.39 & 2.0 & 0.61 \\
        & Redshifted  & 2.9 &  5 $\times$ 18 & 0.76 & 0.123 & 0.362  &  1.06 & 2.5 & 1.34 \\
TMO\_47 & Blueshifted & 1.5 &  8 $\times$ 3  & 0.36 & 0.193 & 0.282  &  0.41 & 2.4 & 0.53 \\
        & Redshifted  & 1.9 &  5 $\times$ 10 & 0.45 & 0.304 & 0.583  &  1.11 & 2.3 & 1.53 \\
TMO\_48 & Blueshifted & 5.2 &  2 $\times$ 3  & 0.15 & 0.002 & 0.012  &  0.06 & 0.3 & 0.71 \\
        & Redshifted  & 3.4 &  2 $\times$ 2  & 0.11 & 0.001 & 0.004  &  0.02 & 0.3 & 0.16 \\
TMO\_49 & Blueshifted & -   &  -             & -    & -     & -      &  -    & -   & -    \\
        & Redshifted  & 4.4 &  7 $\times$ 17 & 0.76 & 0.042 & 0.182  &  0.79 & 1.7 & 1.47 \\
TMO\_50 & Blueshifted & -   &  -             & -    & -     & -      &  -    & -   & -    \\
        & Redshifted  & 2.2 &  4 $\times$ 13 & 0.55 & 0.020 & 0.044  &  0.10 & 2.4 & 0.12 \\
TMO\_51 & Blueshifted & -   &  -             & -    & -     & -      &  -    & -   & -    \\
        & Redshifted  & 2.9 &  2 $\times$ 2  & 0.10 & 0.001 & 0.003  &  0.01 & 0.4 & 0.08 \\
TMO\_52 & Blueshifted & -   &  -             & -    & -     & -      &  -    & -   & -    \\
        & Redshifted  & 2.3 &  2 $\times$ 3  & 0.16 & 0.002 & 0.004  &  0.01 & 0.7 & 0.04 \\
TMO\_53 & Blueshifted & -   &  -             & -    & -     & -      &  -    & -   & -    \\
        & Redshifted  & 2.4 &  2 $\times$ 10 & 0.41 & 0.005 & 0.011  &  0.03 & 1.7 & 0.05 \\
TMO\_54 & Blueshifted & 1.9 &  5 $\times$ 5  & 0.30 & 0.013 & 0.025  &  0.05 & 1.5 & 0.10 \\
        & Redshifted  & 1.8 &  8 $\times$ 5  & 0.38 & 0.034 & 0.061  &  0.11 & 2.1 & 0.17 \\
TMO\_55 & Blueshifted & -   &  -             & -    & -     & -      &  -    & -   & -    \\
        & Redshifted  & 2.0 &  4 $\times$ 11 & 0.49 & 0.011 & 0.023  &  0.05 & 2.4 & 0.06 \\
\enddata
\tablenotetext{a}{The average velocity of the outflow relative to the cloud systemic velocity.}
\tablenotetext{b}{The extent along right ascension $\times$ the extent along declination.}
\end{deluxetable}

\begin{figure*}%[htbp]
\centering %sub_figures centered
\includegraphics[width=0.40\textwidth]{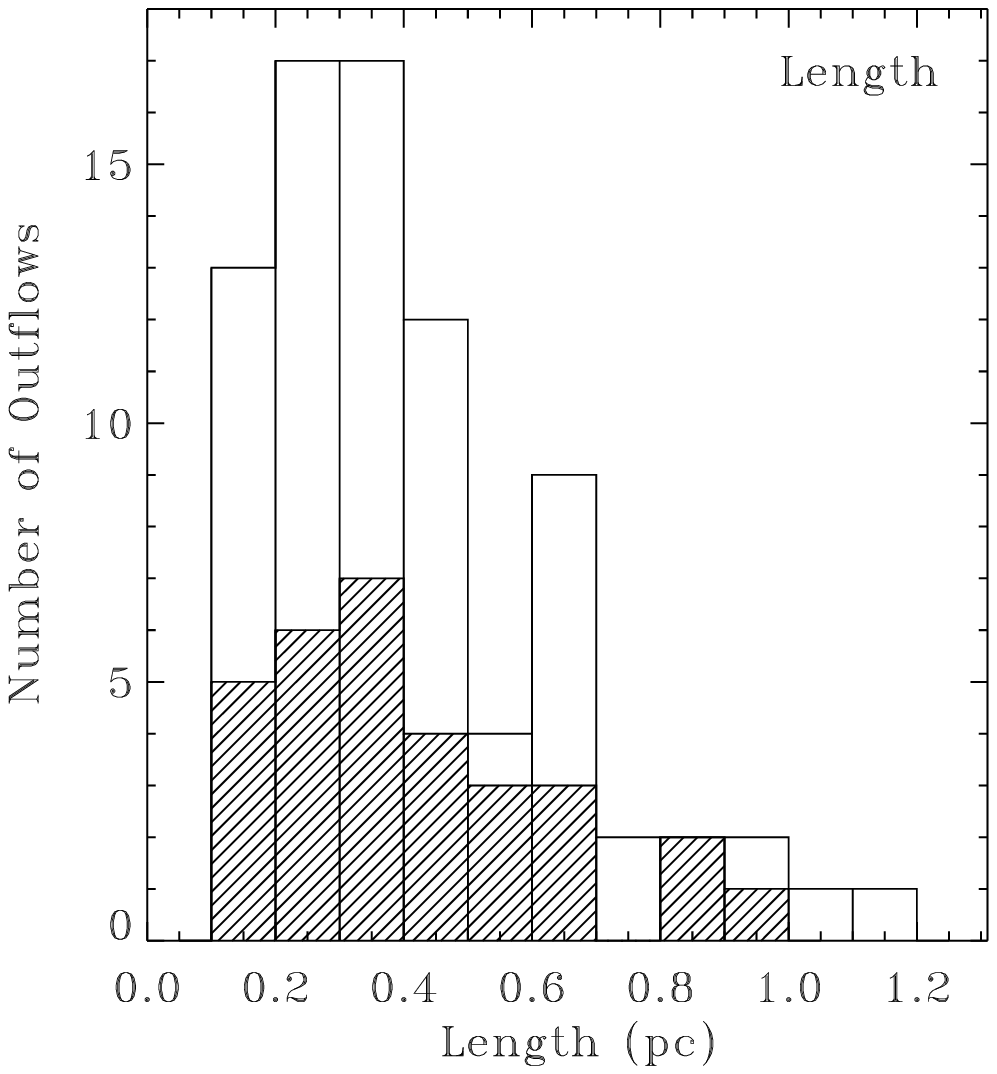}\hspace{-2mm}
\includegraphics[width=0.40\textwidth]{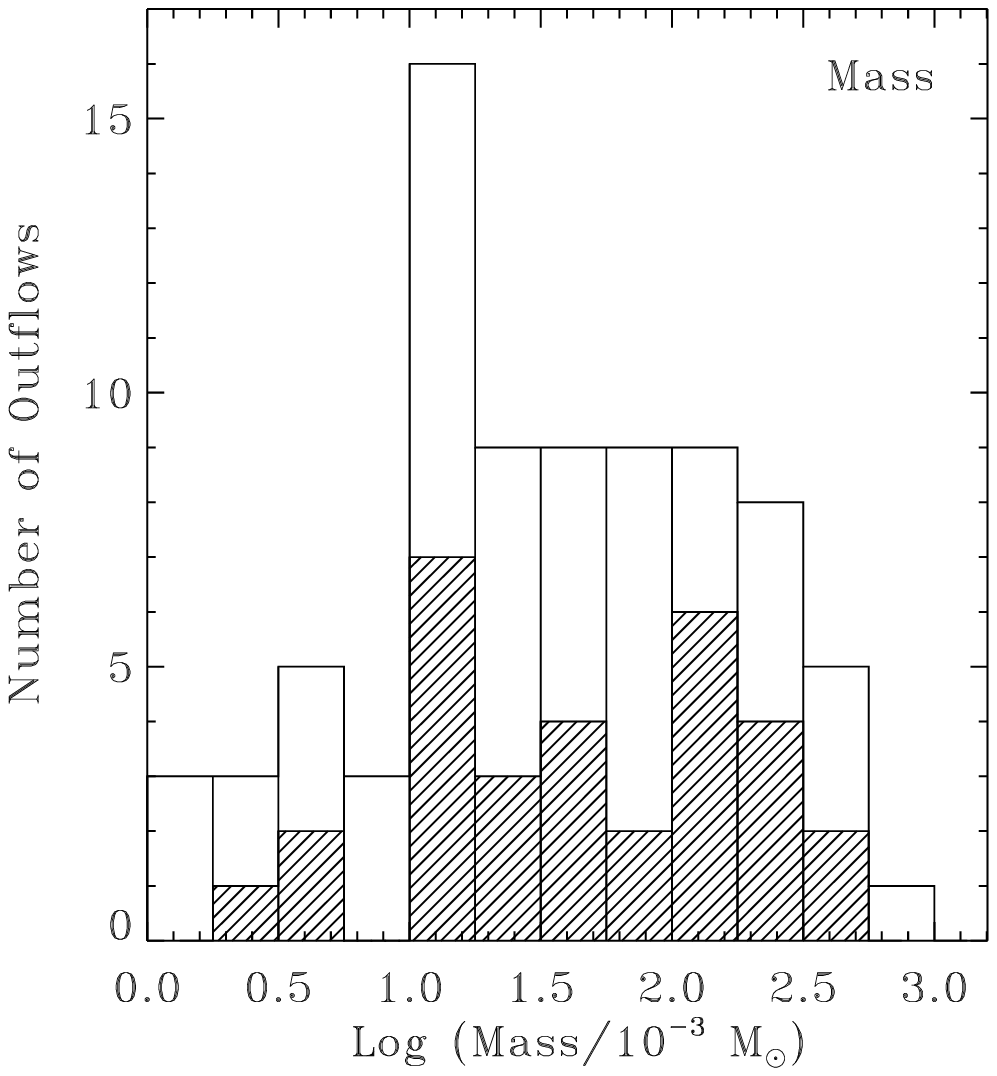}

\vspace{1mm}

\includegraphics[width=0.40\textwidth]{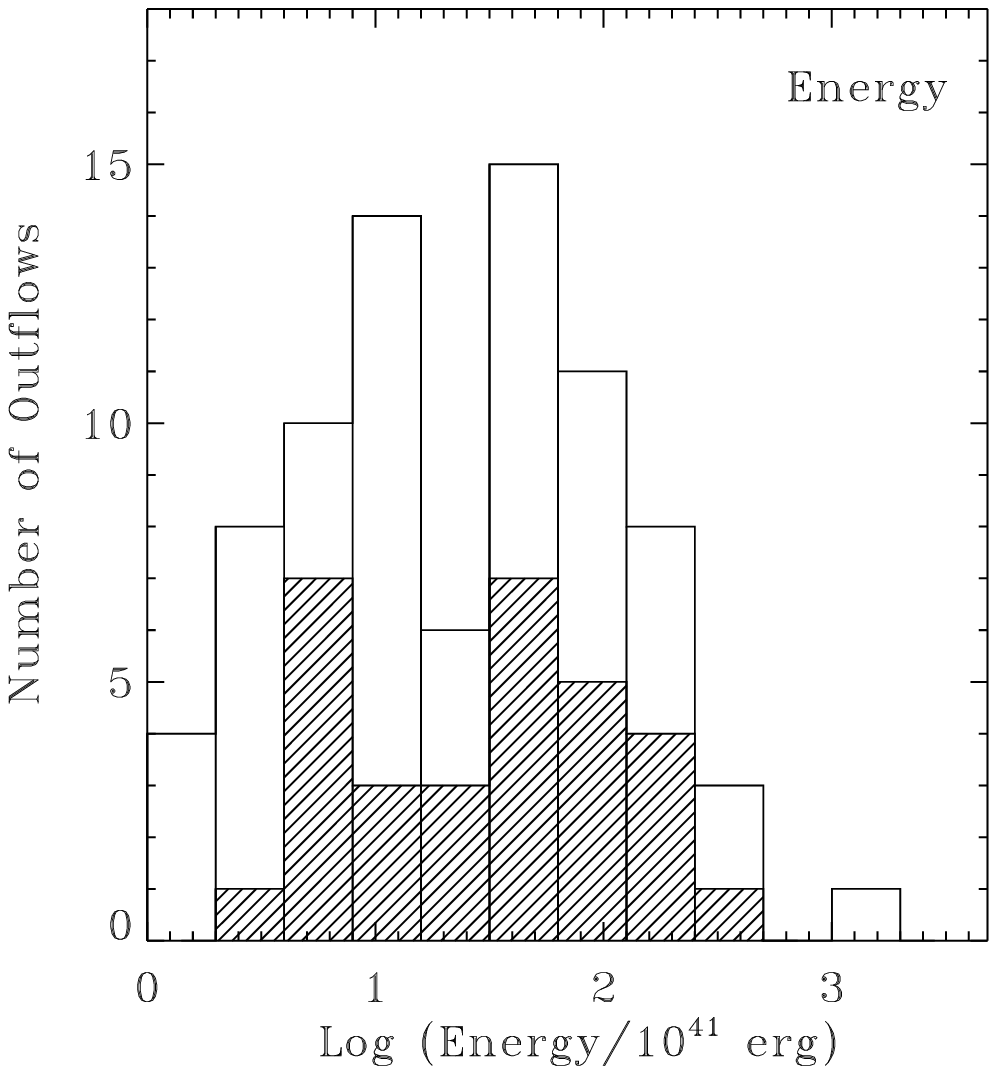}\hspace{-2mm}
\includegraphics[width=0.40\textwidth]{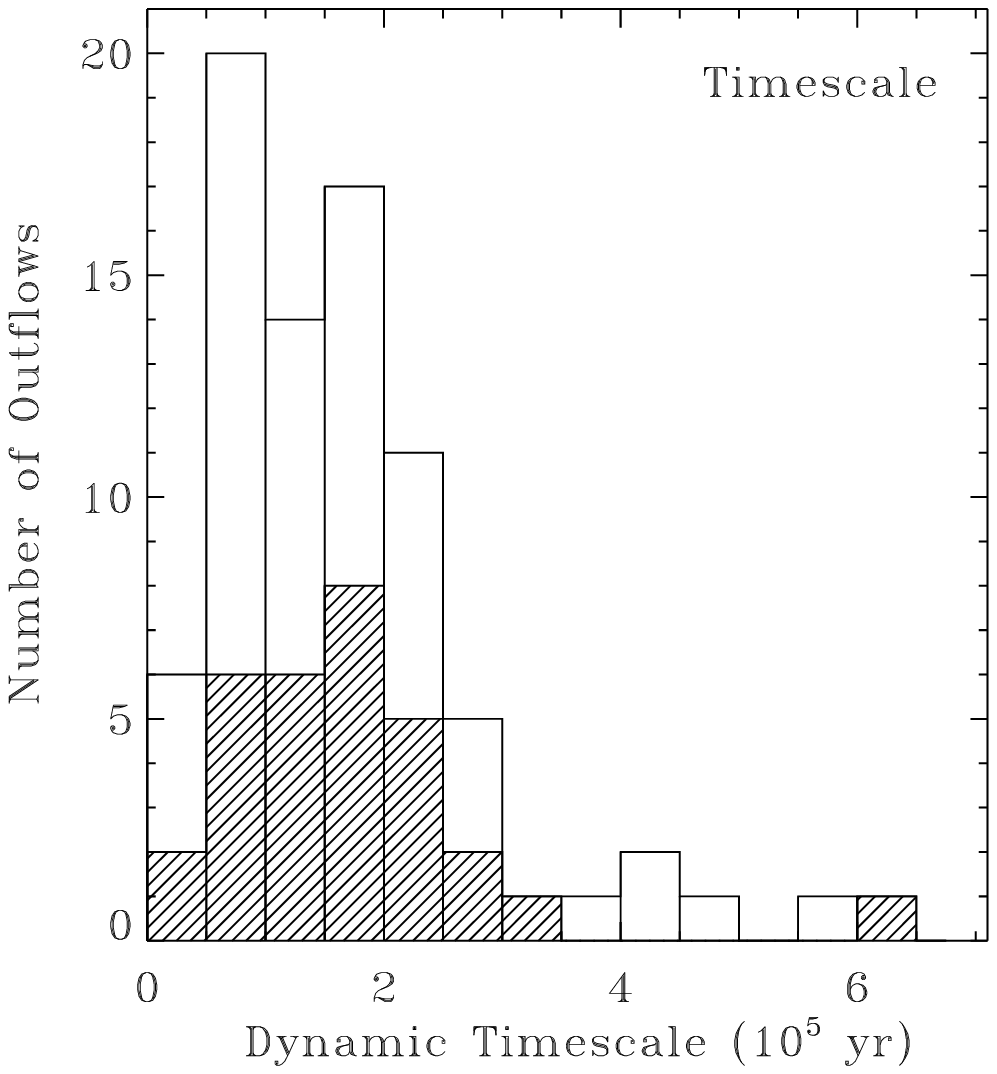}

\caption{Histograms of the distributions of outflow parameters. The shaded histograms represent the blue lobes and the open histograms represent the red lobes.}
\label{outflow-histogram}
\end{figure*}

The distributions of length, mass, energy and dynamical timescale of outflows are shown in Fig. \ref{outflow-histogram}. The extents of outflows are in the range of 0.1 - 1.11 pc. 79\% of outflows are smaller than 0.6 pc. The mass of 54\% of outflows is between $0.01~M_{\odot}$ and $0.1~M_{\odot}$. The outflows with mass lower than $0.01~M_{\odot}$ and higher than $0.1~M_{\odot}$ account for 17\% and 29\% of the total, respectively. The energy of 48\% of outflows is in the range $10^{42}$ - $10^{43}$ erg. The outflows with energy lower than $10^{42}$ erg and higher than $10^{43}$ erg account for 31\% and 21\%, respectively. The dynamical timescales of outflows are between $0.3 \times 10^{5}$ yr and $6.1 \times 10^{5}$ yr. 85\% of outflows have dynamical timescale shorter than $2.5 \times 10^{5}$ yr.

The mass, momentum, energy and luminosity in Table \ref{OutflowParameters} are only lower limits because we did not take into account the inclination and blending correction in the calculation. The mass should be multiplied by a factor of 2 due to blending. Assuming the average inclination angle of outflows is \textbf{57.3$^\circ$}, the velocity and the dynamic age should be scaled up by a factor of \textbf{1.9} and \textbf{0.64}, respectively. Combining the correction factors due to blending and inclination, the momentum, the kinetic energy and luminosity of outflows should be multiplied by a factor of \textbf{3.8, 6.8 and 11}, respectively. After correction, the total mass, momentum, energy, and luminosity of all outflows found in Taurus are approximately $15.4~M_{\odot}$, $\bf 77~M_{\odot}~km~s^{-1}$, $\bf 3.9 \times 10^{45}$ \textbf{erg}, and $\bf 1.3 \times 10^{33}~\rm erg~s^{-1}$, respectively. The totals of the previously known outflows are about $8.6~M_{\odot}$, $\bf 47~M_{\odot}~km~s^{-1}$, $\bf 2.6 \times 10^{45}$ \textbf{erg}, and $\bf 9.4 \times 10^{32}~\rm erg~s^{-1}$, respectively. We found 1.8 times more outflowing mass, \textbf{1.6} times more momentum and 1.5 times more energy from outflows injecting into the Taurus molecular cloud than previously study. A high spatial dynamic range and systematic spectral line survey with good angular resolution is clearly necessary for obtaining a more complete picture of the influence of outflows on their parent cloud.

\section{Bubbles}
\label{Bubbles}
Following the method of identifying bubbles in \citet{Arce11} we have identified 37 bubbles in the $\sim 100$ deg$^2$ region of Taurus. The procedures for bubble searching, the morphology and physical parameters of bubbles are described in the following sections.

\subsection{The Procedures of Searching for Bubbles}
We undertook a blind search for bubbles using the FCRAO $ {^{13}}{}$CO data cube. The integrated intensity map, P-V diagram and channel maps of each bubble were examined. The detailed steps of the search were as follows.

(1) We first searched for circular or arc-like (hereafter bubble-like) structures in $ {^{13}}{}$CO data cube channel by channel through visual inspection. If there is a bubble-like structure in at least three contiguous channels, we considered it as a bubble candidate. The approximate central position and radius of each candidate were recorded for further analysis. We also marked the channels where the bubble-like structure appears. With the marked channels we obtained the expanding velocity interval of a bubble.

(2) We plotted $ {^{13}}{}$CO contour maps around the central position of the bubble candidates with the expanding velocity intervals.

(3) We plotted P-V diagrams in $ {^{13}}{}$CO through the central position of each candidate of every $15^{\circ}$ in position angle. We chose the one with the most obvious circular or ``V" structure to show in the figures. The circular or ``V" structure in the P-V diagram is described in the expanding bubble model \citep{Arce11}.

(4) We plotted the $ {^{13}}{}$CO channel maps of each candidate to look over the variation of radius with velocity.

(5) Finally, we fitted a Gaussian profile to the azimuthally averaged profile of $ {^{13}}{}$CO intensity of each candidate in the channel where the bubble morphology is most like a ring or arc. The radius of a bubble was obtained from the peak position of the fitted profile.

The contour map, P-V diagram, channel maps and Gaussian fitted curves helped us not only to analyze the morphology but also to determine the confidence level of a bubble. The bubble candidates were classified into 6 categories according to the characteristic of the above four types of plots. The criteria for bubble classification, as well as the numbers and ratios of bubbles in different classes are illustrated in Table~\ref{BubbleClass}. In this table ``$\times$" means that it meets a certain condition. For each type of plot the condition is as follows.

(a) There is an obvious bubble-like structure in the contour map.

(b) The P-V diagram has an obvious circular or ``V" structure.

(c) There is an obvious bubble-like structure in the channel map and the radius of bubble is increasing or decreasing with channel.

(d) The average intensity distribution can be fitted with a Gaussian profile.

Meeting all the above four items is required for a high ranking (Class A). If the plots only meet (b) and (c), we then divided the bubbles into the lower ranking (Class B1) because only the expanding velocity is detected but there is no obvious bubble-like structure and good Gaussian fitted profile. Then B2, B3 and B4 are in descending order of ranking. A candidate bubble only meeting (a) is assigned the lowest ranking (Class C) because the gas with expanding velocity is not obvious.

\begin{deluxetable}{lccccrr}
\tabletypesize{\scriptsize}
%\rotate
\tablecaption{Criteria for Bubble Classification and Bubble Distribution
\label{BubbleClass}}
\tablewidth{0pt}
\tablehead{
 \colhead{Bubble} & \colhead{Contour} & \colhead{P-V}  & \colhead{Channel} & \colhead{Fitting} & \colhead{Bubble} & \colhead{Percentage} \\
 \colhead{Class} & \colhead{Map} & \colhead{Diagram} & \colhead{Maps} & \colhead{Curve} & \colhead{Numbers} & \colhead{}}

\startdata
A   &  $\times$  &  $\times$  &  $\times$  &  $\times$  &  13  &  35.2\%\\
B1  &            &  $\times$  &  $\times$  &            &  6   &  16.2\% \\
B2  &  $\times$  &            &  $\times$  &            &  4   &  10.8\% \\
B3  &  $\times$  &  $\times$  &            &            &  4   &  10.8\% \\
B4  &            &            &  $\times$  &            &  4   &  10.8\% \\
C   &  $\times$  &            &            &            &  6   &  16.2\% \\
\enddata
\end{deluxetable}

Following the above procedures we found 37 bubbles in the entire 100 deg$^2$ area of Taurus molecular cloud. Each bubble is referred as a ``Taurus Molecular Bubble'' (TMB). The positions and classifications are listed in Table \ref{bubble}. The numbers and percentage of each class are illustrated in Table \ref{BubbleClass}.

\subsection{Morphology of Bubbles}
The $ {^{13}}{}$CO integrated intensity maps, P-V diagrams, Gaussian fitting profiles and channel maps for the bubbles are presented in Fig. \ref{Bubble_fig1.} - Fig. \ref{Bubble_fig37.}. If the morphology of a contour map is a closed ring, we then considered it to be an expanding bubble. If the ring on the contour map is incomplete, we then called it a broken bubble. There are 3 expanding bubbles (TMB\_07, TMB\_10 and TMB\_24) and 34 broken bubbles among all the bubbles in Taurus.

\subsection{Physical Parameters of Bubbles}

To examine the impact of bubbles on the host cloud we calculated the mass, momentum, kinetic energy, dynamical timescale and energy deposition rate of the bubbles.
Assuming the $^{13}$CO(1-0) emission of the bubble is optically thin, the total column density is derived as follows:
\begin{equation}
N_{\rm tot}(^{13}{\rm CO})=\frac{3k^2T_{\rm ex}f_{\tau}}{4\pi^3\mu^2_dh\nu^2 \exp(-h\nu/kT_{\rm ex})}\int T_{\rm s} dv,
\label{equ:BubbleDerivedTotalColumnDensity}
\end{equation}
where $k=1.38\times10^{-16}~{\rm erg/K}$, $h=6.626\times10^{-27}$ erg s, $\mu_{\rm d}=0.112\times10^{-18}~{\rm esu}$ and $\nu=110.2014\times10^9~{\rm Hz}$. The excitation temperature, $T_{\rm ex}$ is assumed to be 25 K. $T_{\rm s}$ is the observed source antenna temperature with proper correction for antenna efficiency. The optical depth correction factor, $f_{\tau}$, is estimated from the following formulae \citep{Qian12}.
\begin{equation}
f_{\tau}=\frac{\int \tau(^{13}{\rm CO}) dv}{\int [1-e^{-\tau(^{13}{\rm CO})}] dv},
\label{equ:ftau}
\end{equation}
where $\tau(^{13}{\rm CO})$ is the opacity of the $^{13}$CO transition. Assuming equal excitation temperatures for $^{13}$CO and $^{12}$CO, we can get
\begin{equation}
\frac{T(^{12}{\rm CO})}{T(^{13}{\rm CO})}=\frac{1-e^{-\tau(^{12}{\rm CO})}}{1-e^{-\tau(^{13}{\rm CO})}},
\label{equ:BrightnessRatio}
\end{equation}
where $T(^{12}{\rm CO})$ and $T(^{13}{\rm CO})$ are the brightness temperature of $^{12}$CO and $^{13}$CO, respectively. $\tau(^{12}{\rm CO})$ is the opacity of the $^{12}$CO transition. Assuming the $^{12}$CO emission from the bubbles is optically thick ($\tau(^{12}{\rm CO})\gg 1$), the opacity of $^{13}$CO can be obtained from
\begin{equation}
\tau(^{13}{\rm CO})=-\ln\left(1-\frac{T(^{13}{\rm CO})}{T(^{12}{\rm CO})}\right).
\label{equ:Tao13}
\end{equation}
%
%The derivation of Eq.(\ref{equ:BubbleDerivedTotalColumnDensity}) is very similar to Eq.(\ref{equ:DerivedTotalColumnDensity}) in appendix A.
With the column density and area we can obtain the bubble mass. Using the bubble mass and expansion velocity we can then get the momentum and kinetic energy of the bubble using $P_{\mathrm{bubble}} =  M_{\mathrm{bubble}} V_{\mathrm{exp}}$ and $E_{\mathrm{bubble}} = \frac{1}{2} M_{\mathrm{bubble}} V_{\mathrm{exp}}^2$, respectively. The kinetic timescale of bubble can be calculated as $t_{\rm {kinetic}} = R/V_{\rm {exp}}$, where $R$ is the radius and $V_{\rm {exp}}$ is the expansion velocity of the bubble. The bubble energy injection rate, $L_{\rm {bubble}}$, can be estimated as $L_{\rm {bubble}} = E_{\mathrm{bubble}}/t_{\rm {kinetic}}$.

The physical parameters of all bubbles are listed in Table \ref{bubble}. The momentum and kinetic energy are lower limits mainly because of the underestimate of the minimum expansion velocity. The total mass, momentum, energy and energy injection rate of all detected bubbles in Taurus molecular cloud are about $1704~M_{\odot}$, $3780~M_{\odot}~km~s^{-1}$, $9.2~\times10^{46}$ erg and $6.4 \times 10^{33}~\rm erg~s^{-1}$, respectively.

\clearpage
\begin{deluxetable}{cccccccrrrccr}
\tabletypesize{\scriptsize}
\rotate
\tablecaption{BUBBLES IN TAURUS
\label{bubble}}
\tablewidth{0pt}
\tablehead{
 \colhead{Bubble} & \colhead{R.A.} & \colhead{Dec.} & \colhead{Bubble} & \colhead{YSO} & \colhead{Radius} & \colhead{V$_{\rm exp}$} & \colhead{Mass} & \colhead{Momentum} & \colhead{Energy}  & \colhead{$t_{\rm kinetic}$} & \colhead{$L_{\rm bubble}$} & \colhead{$\rm {\dot m_{w}}$\tablenotemark{a}} \\
 \colhead{Name} & \colhead{(J2000)} & \colhead{(J2000)} & \colhead{Class} & \colhead{} & \colhead{(pc)} & \colhead{(km s$^{-1}$)} & \colhead{(M$_{\sun}$)} & \colhead{(M$_{\sun}$ km s$^{-1}$)} & \colhead{($10^{45}$ erg)} & \colhead{$10^{6}$ yr} & \colhead{($10^{32}$ erg s$^{-1}$)} & \colhead{($10^{-8}$ M$_{\sun}$~yr$^{-1}$)}
}
\startdata
    TMB\_01 & 04 12 08 & 24 53 33  & A  & N  & 0.98 & 1.3  & 25  & 31    & 0.39   & 0.8   & 0.17  & 15.5 \\
    TMB\_02 & 04 14 28 & 27 45 53  & B1 & Y  & 0.60 & 1.3  & 6   & 7     & 0.09   & 0.5   & 0.06  & 3.5  \\
    TMB\_03 & 04 16 20 & 28 28 53  & A  & Y  & 0.76 & 1.8  & 10  & 18    & 0.33   & 0.4   & 0.25  & 9.0  \\
    TMB\_04 & 04 19 05 & 27 33 33  & A  & Y  & 0.62 & 1.3  & 7   & 9     & 0.12   & 0.5   & 0.08  & 4.5  \\
    TMB\_05 & 04 21 12 & 26 55 33  & B3 & Y  & 0.56 & 1.8  & 4   & 8     & 0.14   & 0.3   & 0.14  & 4.0  \\
    TMB\_06 & 04 25 17 & 25 32 13  & B1 & N  & 0.77 & 1.3  & 10  & 12    & 0.16   & 0.6   & 0.08  & 6.0  \\
    TMB\_07 & 04 25 29 & 26 10 13  & B1 & N  & 1.12 & 2.3  & 102 & 234   & 5.31   & 0.5   & 3.51  & 117.0\\
    TMB\_08 & 04 27 07 & 24 20 13  & A  & N  & 0.70 & 2.3  & 18  & 42    & 0.95   & 0.3   & 1.00  & 21.0 \\
    TMB\_09 & 04 27 31 & 26 16 53  & B4 & N  & 0.49 & 1.0  & 4   & 4     & 0.04   & 0.5   & 0.03  & 2.0  \\
    TMB\_10 & 04 28 52 & 24 14 33  & A  & Y  & 1.58 & 1.5  & 213 & 325   & 4.92   & 1.0   & 1.53  & 162.5\\
    TMB\_11 & 04 29 32 & 26 32 33  & A  & Y  & 0.84 & 2.0  & 41  & 83    & 1.68   & 0.4   & 1.31  & 41.5 \\
    TMB\_12 & 04 29 44 & 26 32 53  & B4 & Y  & 0.70 & 2.5  & 23  & 58    & 1.47   & 0.3   & 1.72  & 29.0 \\
    TMB\_13 & 04 30 31 & 24 26 13  & A  & Y  & 0.73 & 2.3  & 18  & 41    & 0.93   & 0.3   & 0.94  & 20.5 \\
    TMB\_14 & 04 31 14 & 29 25 53  & B2 & Y  & 0.84 & 2.0  & 10  & 20    & 0.40   & 0.4   & 0.31  & 10.0 \\
    TMB\_15 & 04 31 30 & 24 14 33  & A  & Y  & 0.84 & 1.3  & 34  & 43    & 0.54   & 0.6   & 0.27  & 21.5 \\
    TMB\_16 & 04 31 32 & 24 09 53  & B3 & N  & 0.70 & 1.8  & 22  & 39    & 0.69   & 0.4   & 0.57  & 19.5 \\
    TMB\_17 & 04 31 35 & 23 35 13  & B4 & N  & 0.42 & 1.0  & 2   & 2     & 0.02   & 0.4   & 0.02  & 1.0  \\
    TMB\_18 & 04 31 50 & 24 22 13  & C  & N  & 0.56 & 2.0  & 14  & 29    & 0.59   & 0.3   & 0.69  & 14.5 \\
    TMB\_19 & 04 31 59 & 25 43 13  & B1 & Y  & 0.70 & 2.0  & 16  & 32    & 0.64   & 0.3   & 0.60  & 16.0 \\
    TMB\_20 & 04 32 03 & 25 36 53  & B2 & N  & 0.42 & 1.3  & 4   & 5     & 0.07   & 0.3   & 0.07  & 2.5  \\
    TMB\_21 & 04 32 37 & 29 29 13  & A  & N  & 0.62 & 2.3  & 6   & 14    & 0.31   & 0.3   & 0.37  & 7.0  \\
    TMB\_22 & 04 32 39 & 24 46 13  & A  & Y  & 1.06 & 1.3  & 28  & 36    & 0.45   & 0.8   & 0.17  & 18.0 \\
    TMB\_23 & 04 33 10 & 26 08 53  & B4 & N  & 0.28 & 1.3  & 2   & 2     & 0.03   & 0.2   & 0.04  & 1.0  \\
    TMB\_24 & 04 33 13 & 25 24 53  & C  & N  & 0.49 & 1.3  & 5   & 7     & 0.09   & 0.4   & 0.07  & 3.5  \\
    TMB\_25 & 04 33 34 & 24 20 53  & A  & Y  & 0.28 & 3.0  & 5   & 15    & 0.46   & 0.1   & 1.61  & 7.5  \\
    TMB\_26 & 04 34 47 & 29 37 13  & B3 & N  & 0.46 & 2.3  & 3   & 6     & 0.14   & 0.2   & 0.23  & 3.0  \\
    TMB\_27 & 04 36 02 & 28 23 13  & C  & Y  & 1.40 & 2.5  & 64  & 161   & 4.08   & 0.5   & 2.39  & 80.5 \\
    TMB\_28 & 04 36 23 & 25 36 33  & B2 & Y  & 1.40 & 3.3  & 386 & 1275  & 41.84  & 0.4   & 31.93 & 637.5\\
    TMB\_29 & 04 37 04 & 25 46 33  & C  & Y  & 0.84 & 1.5  & 20  & 30    & 0.46   & 0.5   & 0.27  & 15.0 \\
    TMB\_30 & 04 38 11 & 26 05 53  & B3 & Y  & 1.90 & 1.8  & 340 & 604   & 10.68  & 1.0   & 3.23  & 302.0\\
    TMB\_31 & 04 39 11 & 29 05 13  & B2 & N  & 1.26 & 1.5  & 74  & 113   & 1.71   & 0.8   & 0.67  & 56.5 \\
    TMB\_32 & 04 39 48 & 28 35 33  & C  & N  & 0.70 & 2.0  & 7   & 14    & 0.27   & 0.3   & 0.26  & 7.0  \\
    TMB\_33 & 04 41 10 & 25 31 13  & C  & N  & 0.70 & 1.0  & 10  & 10    & 0.10   & 0.7   & 0.05  & 5.0  \\
    TMB\_34 & 04 44 20 & 28 36 53  & B1 & N  & 1.26 & 2.8  & 143 & 399   & 11.08  & 0.4   & 7.95  & 199.5\\
    TMB\_35 & 04 46 12 & 25 07 33  & A  & N  & 0.62 & 1.5  & 8   & 13    & 0.19   & 0.4   & 0.15  & 6.5  \\
    TMB\_36 & 04 46 43 & 24 59 13  & B1 & Y  & 0.56 & 1.8  & 12  & 21    & 0.38   & 0.3   & 0.39  & 10.5 \\
    TMB\_37 & 04 48 12 & 24 50 33  & A  & N  & 0.63 & 2.3  & 8   & 18    & 0.41   & 0.3   & 0.48  & 9.0  \\
\enddata
\tablenotetext{a}{Estimate of minimum stellar wind mass loss rate needed to drive the bubbles.}
\end{deluxetable}

\begin{figure*}%[htbp]
\centering %sub_figures centered
\includegraphics[width=0.40\textwidth]{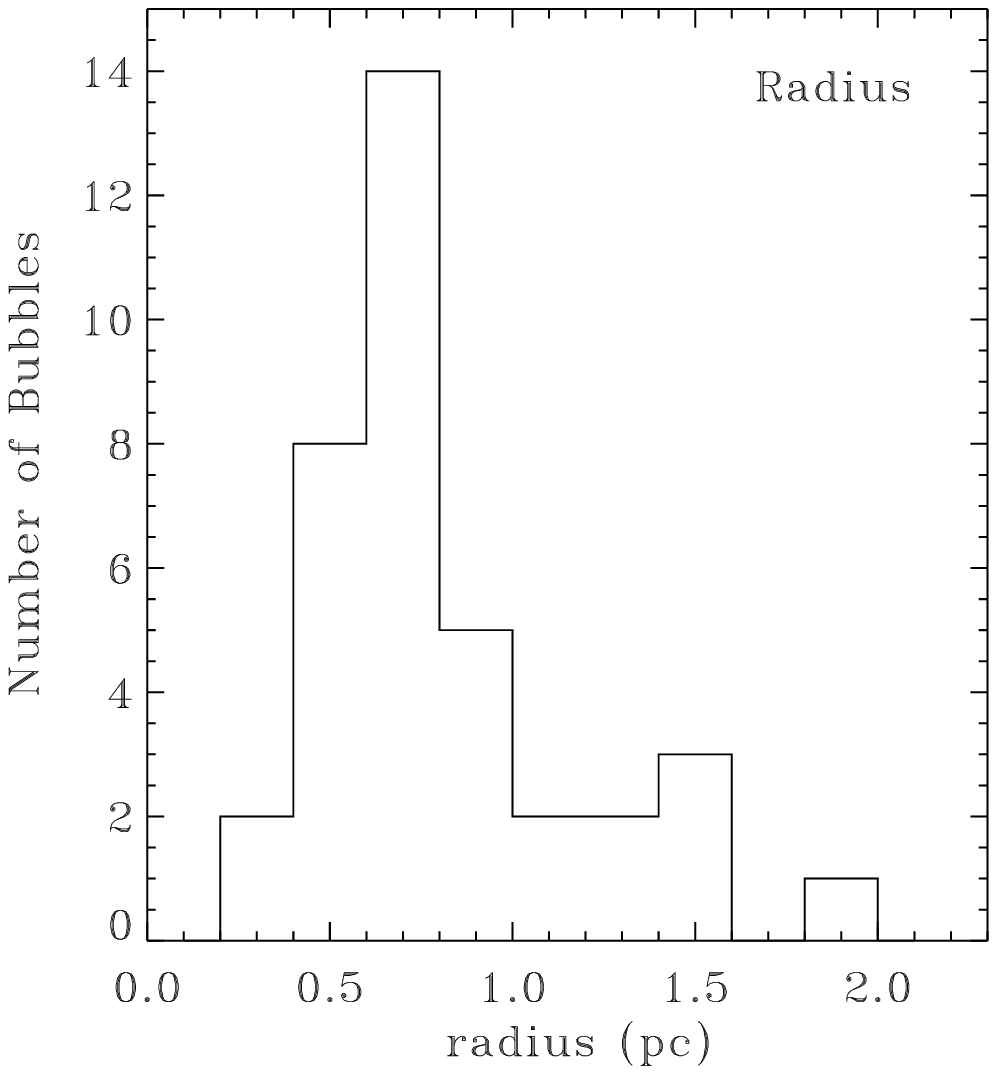}\hspace{-2mm}
\includegraphics[width=0.40\textwidth]{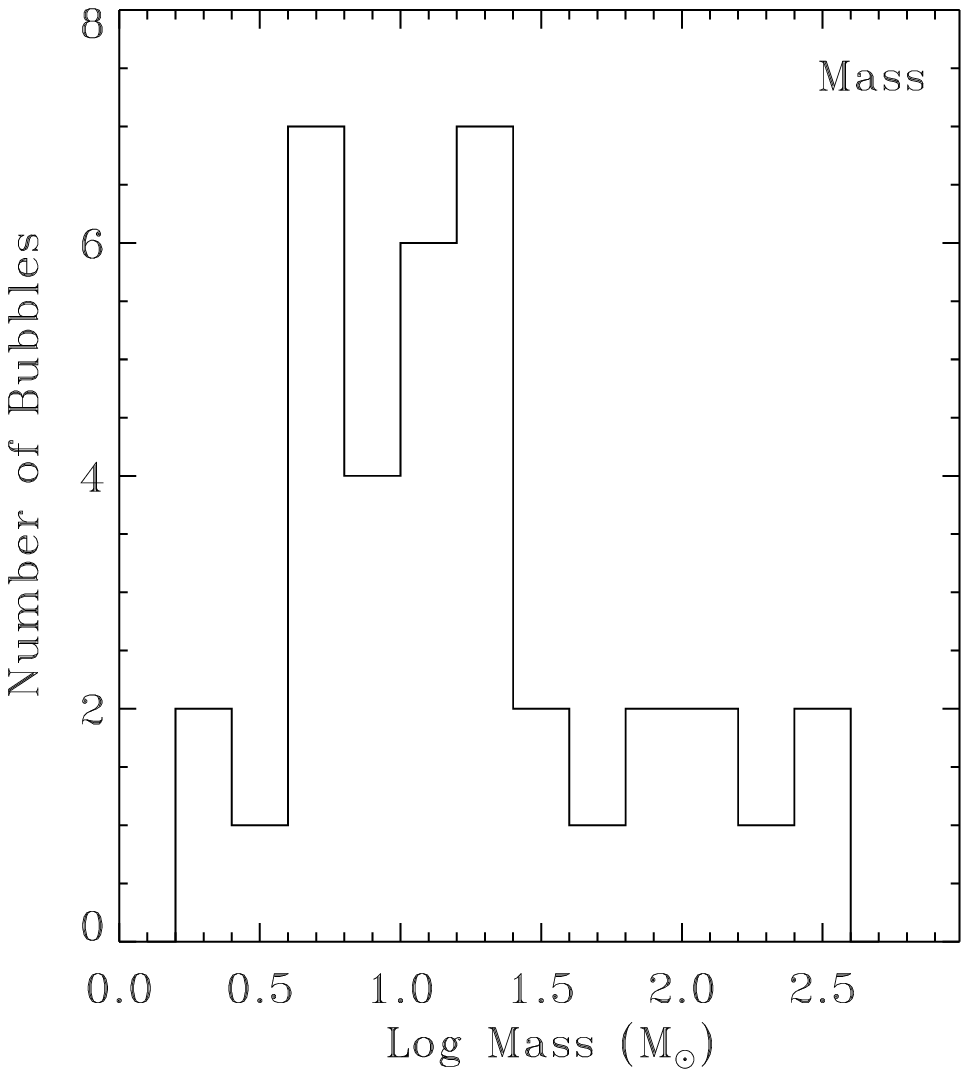}

\vspace{1mm}

\includegraphics[width=0.40\textwidth]{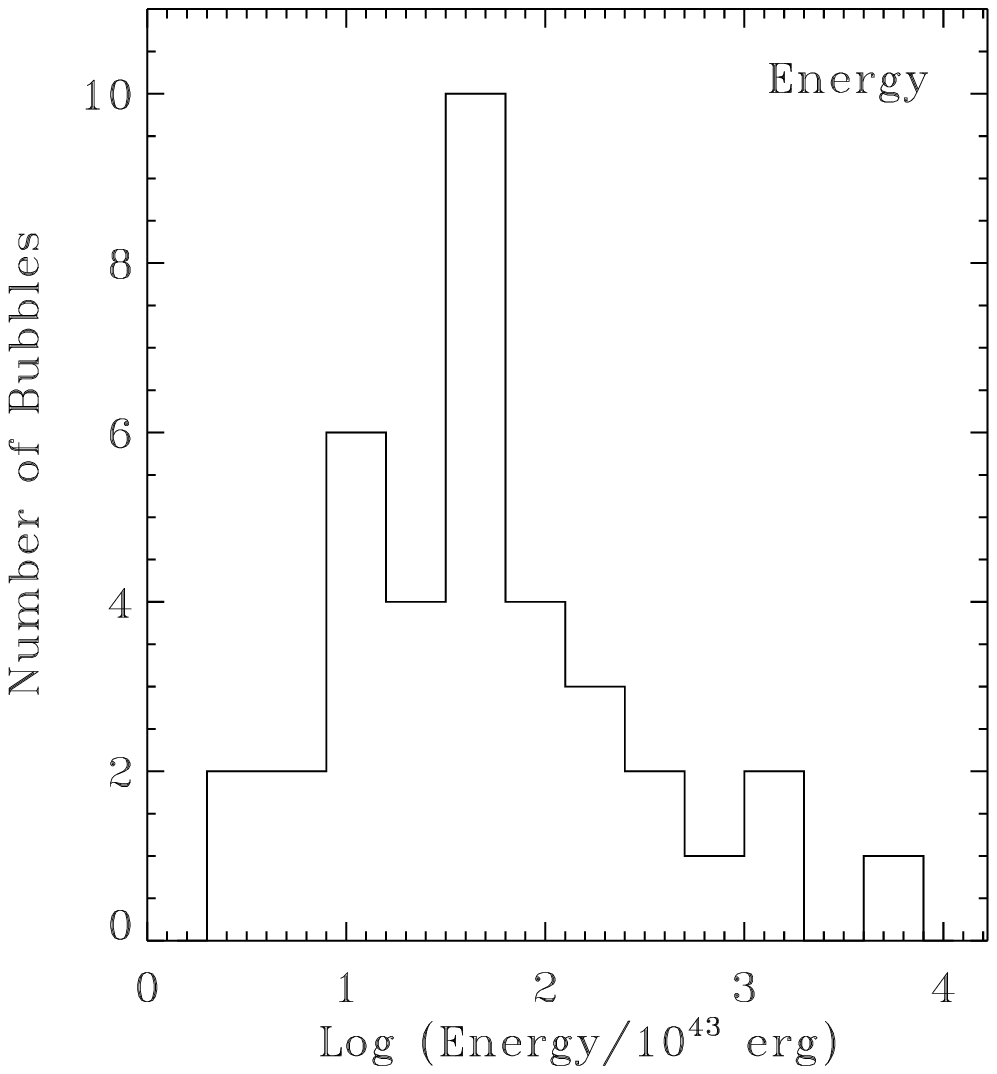}\hspace{-2mm}
\includegraphics[width=0.40\textwidth]{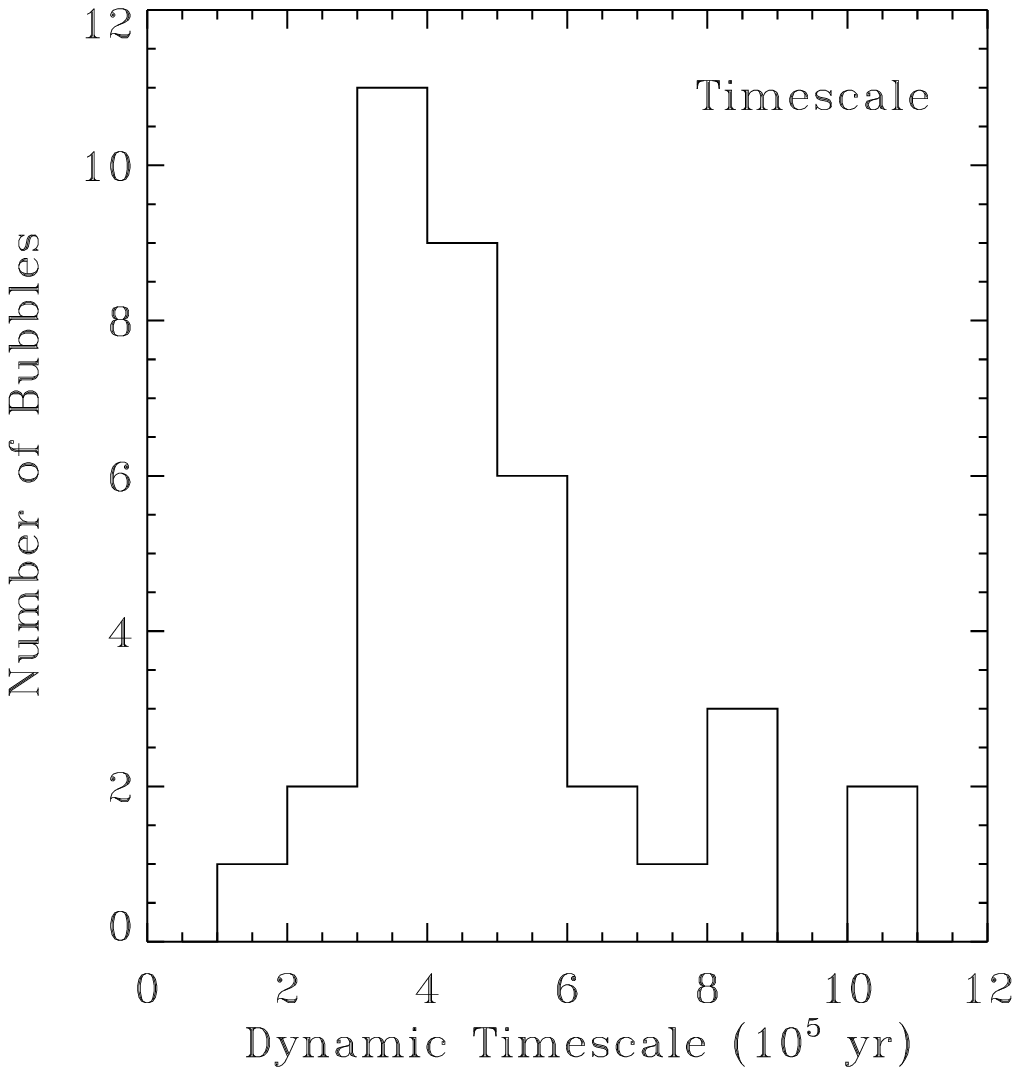}

\caption{Histograms of the distributions of bubble parameters.}
\label{bubble-histogram}
\end{figure*}

The distribution of radius, mass, energy and dynamical timescale of bubbles are shown in Fig. \ref{bubble-histogram}. The radius of bubbles is in the range 0.28 - 1.9 pc. 78\% of the bubbles are smaller than 1 pc. The mass of 65\% of the bubbles is between $5~M_{\odot}$ and $50~M_{\odot}$. The bubbles with mass lower than $5~M_{\odot}$ and higher than $50~M_{\odot}$ account for 16\% and 19\%, respectively. The highest bubble mass is $386~M_{\odot}$. The energy of 60\% of bubbles is in the range $10^{44}$ - $10^{45}$ erg. The bubbles with energy lower than $10^{44}$ erg and higher than $10^{45}$ erg account for 16\% and 24\%, respectively. The dynamical timescales of bubbles are between $10^{5}$ yr and $10^{6}$ yr. Almost 95\% of bubbles are younger than $10^{6}$ yr. Compared to the outflows, the bubbles have about \textbf{110} times larger mass and \textbf{24} times higher energy. The extents of bubbles are larger than outflows, which can be seen from Fig. \ref{yso-distribute}. The dynamical timescales of bubbles are longer than that of outflows.

\section{Analysis and Discussion}
\label{Discussion}
\subsection{Polarity of Outflows in Taurus}
Among the 55 outflows we found that bipolar, monopolar redshifted and monopolar blueshifted outflows account for 45\%, 44\% and 11\%, respectively. There are more red lobes than blue ones , which can be seen from the histograms in Fig. \ref{outflow-histogram}. The occurrence of more red lobes may result from the fact that Taurus is thin \citep{Qian2015}. Red lobes tend to be smaller and younger. The total mass and energy of red lobes is similar to blue lobes on average, which can be seen from the upper right panel and lower left panel of Fig. \ref{outflow-histogram}.

\subsection{The Driving Sources of Outflows and Bubbles in Taurus}
\begin{figure*}%[htbp]
\centering %sub_figures centered
\includegraphics[width=0.8\textwidth]{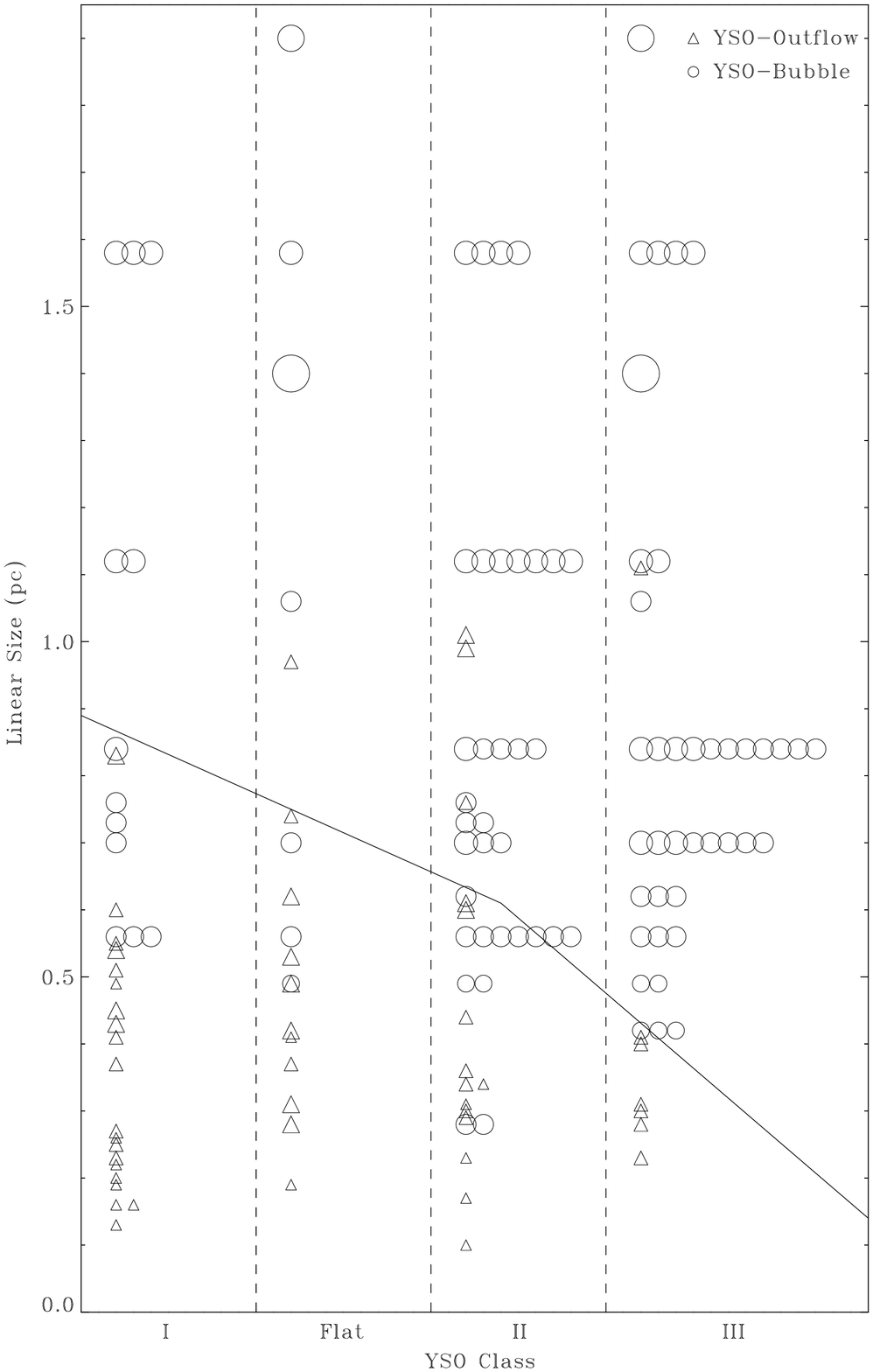}
\vspace{-5mm}

\caption{Distribution of Outflows and Bubbles. The open upward triangles represent the ``outflow-driving YSOs'' and the open circles represent the ``bubble-driving YSOs''. The black solid line roughly divides the outflows and bubbles. The size of symbols is proportional to the energy of outflows and bubbles.}
\label{yso-distribute}
\end{figure*}

\begin{figure*}%[htbp]
\centering %sub_figures centered
\hspace{6mm}\includegraphics[width=0.40\textwidth]{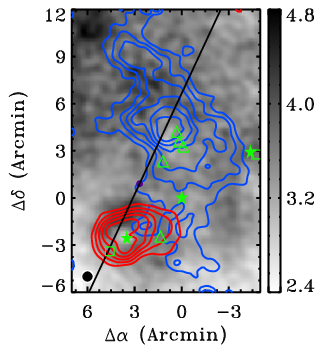}\hspace{-15mm}
\includegraphics[width=0.40\textwidth]{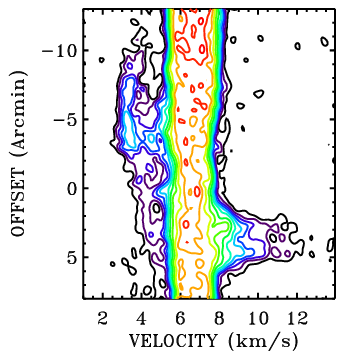}

\vspace{-5mm}

\includegraphics[width=0.40\textwidth]{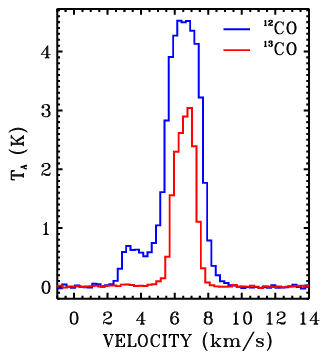}\hspace{-9mm}
\includegraphics[width=0.40\textwidth]{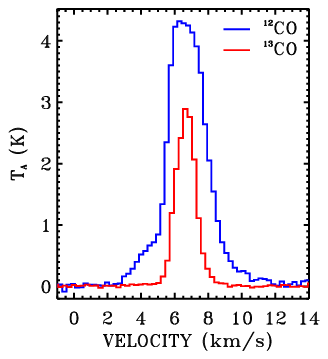}

\caption{TMO\_02 (SST 041412.2+280837): \textbf{upper left}, $\rm {}^{12}CO$ integrated intensity map overlaid on the $\rm {}^{13}CO$ grey-scale image, integrated over 2 to 5 km s$^{-1}$ for the blue lobe and 8 to 12.5 km s$^{-1}$ for the red lobe with $\rm {}^{12}CO$, and integrated over 5.5 to 7.5 km s$^{-1}$ for the $\rm {}^{13}CO$ grey-scale image. The blue and red contour levels are 40\%, 50\%,..., 90\% and 30\%, 45\%,..., 90\% of their peak value, respectively. In this panel and all subsequent upper left panels including those of bubbles, the green filled stars, if present, show the location of Class I YSOs in Taurus, the green open squares show the location of flat-spectrum YSOs, the green open triangles show the location of Class II YSOs and the green filled squares show the location of Class III YSOs. All the above YSOs were listed in Table 6 and Table 7 of \citep{Rebull10}. The black solid line represents a cut for the P-V diagram shown in the upper right panel, and the amaranth filled circle indicates the origin on the Y-axis of the upper right panel. The filled circle in the lower left corner shows the beam. \textbf{upper right}, P-V diagram of $\rm {}^{12}CO$, through the slice shown in the upper left panel at a position angle of $25^{\circ}$. Contour levels are 0.4 to 1.6 K by 0.3 K, 2 to 5 K by 0.5 K. \textbf{lower left}, average spectra of $\rm {}^{12}CO$ emission (blue lines) and $\rm {}^{13}CO$ emission (red lines) towards the blueshifted lobe shown in the upper left panel. \textbf{lower right}, average spectra of $\rm {}^{12}CO$ emission (blue lines) and $\rm {}^{13}CO$ emission (red lines) towards the redshifted lobe shown in the upper left panel.}
\label{sample-figure01}
\end{figure*}

The outflows are driven by four types of YSOs. From Table \ref{OutflowNumYSO} we can see Class I, Flat, Class II and Class III account for 36.4\%, 18.2\%, 27.3\% and 12.7\% of all the driving sources, respectively. Fig. \ref{yso-distribute} shows the distribution of different classes of YSOs driving outflows (hereafter outflow-driving YSO) and YSOs inside the bubbles (hereafter bubble-driving YSO). The rough dividing line shows that there are more outflow-driving YSOs in Class I, Flat and Class II while few outflow-driving YSOs in Class III, which indicates that outflows are more likely appear in the earlier stage (Class I) than in the later phase (Class III) of star formation. There are more bubble-driving YSOs of Class II and Class III while there are few bubble-driving YSOs of Class I and Flat, implying that the bubble structures are more likely to occur in the later stage of star formation. From the size of the symbols we can see that the larger outflows and bubbles are, the higher energy they have.

\subsection{The Feedback of Outflows and Bubbles in Taurus}
With the complete sample of outflows and bubbles we can estimate the overall impact of dynamical structures on the Taurus molecular cloud. We investigated whether the outflows and bubbles have enough energy to potentially unbind the entire Taurus molecular cloud or drive the turbulence in the cloud.

\subsubsection{\bf The Energy of Outflows and Bubbles Cannot Balance the Gravitational Binding Energy of the Entire Cloud}
Using a total mass of $\bf 1.5 \times 10^{4}~M_{\odot}$ \citep{Pineda10} and an effective radius of 13.8 pc \citep{Narayanan12} for the 100 deg$^2$ region of Taurus, we calculated the magnitude of the gravitational binding energy $GM_{\rm cloud}^2/R_{\rm cloud}$ to be $\bf \sim 1.5 \times 10^{48}$ {\bf erg}. The total kinetic energy of outflows from the 44 deg$^2$ region of Taurus is $\bf 3.9 \times 10^{45}$ erg, much less than the gravitational binding energy. Given that we searched for outflows around YSOs only in the Spitzer 44 deg$^2$ survey region not the overall area of Taurus, we may well have missed some outflows. Most of the gas of Taurus is centered on the Spitzer 44 deg$^2$ survey region, which can be seen from Fig. 2 of \citet{Goldsmith08}. There are not many YSOs outside the Spitzer coverage in Taurus and those YSOs are generally clustered, which can be seen from Figure 1 of \citet{Rebull11}. So there should be few outflows outside the Spitzer coverage in Taurus and the outflows we found around YSOs in the 44 deg$^2$ area account for the majority of outflows in Taurus. Similarly, the total kinetic energy of the detected bubbles in the 100 deg$^2$ Taurus region is $9.2 \times 10^{46}$ erg, which also cannot balance the gravitational potential energy of the entire cloud.

\subsubsection{\bf Turbulent Energy is Greater than the Energy of Outflows and Bubbles}
\label{TurbulentEnergy}
\textbf{The turbulent energy of the Taurus molecular cloud is given approximately by}
\begin{equation}
\bf E_{\rm turb} = {{1}\over{2}}M_{\rm cloud}\sigma_{\rm 3d}^2,
\label{equ:EnergyTurb}
\end{equation}
\textbf{where $\bf \sigma_{\rm 3d}$ is the three dimensional turbulent velocity dispersion, which can be calculated by}
\begin{equation}
\bf \sigma_{\rm 3d} = {{\sqrt{3}}\over{2\sqrt{2\ {\rm ln\ 2}}}}\Delta v_{\rm FWHM}.
\label{equ:sigma3d}
\end{equation}
\textbf{Here $\bf \Delta v_{\rm FWHM}$ = 2 km s$\bf ^{-1}$ is the one dimensional FWHM velocity dispersion based on typical $\bf ^{13}$CO spectra in Taurus \citep{Narayanan12}. Then we get $\bf \sigma_{\rm 3d} = 1.47$ km s$\bf ^{-1}$. The total mass of the 100 deg$\bf ^2$ region of Taurus is $\bf M_{\rm cloud} = 1.5 \times 10^{4}~M_{\odot}$ \citep{Pineda10}. Using Eq. (\ref{equ:EnergyTurb}) we obtain the turbulent energy of the Taurus to be $\bf 3.2 \times 10^{47}$ erg.}
The energy of all detected outflows (\textbf{$3.9 \times 10^{45}$ erg}) is about two orders of magnitude less than the turbulent energy of the cloud. The lower limit of the total energy of the bubbles ($9.2 \times 10^{46}$ erg) is 29\% of the turbulent energy. \textbf{We conclude that the total energy of outflows and bubbles cannot balance the turbulence in Taurus.}

\subsubsection{\bf Turbulent Dissipation Rate is Comparable to Bubble Energy Injection Rate}
We also estimated the total outflow energy rate (outflow luminosity, $L_{\rm flow}$) and the total bubble energy rate (bubble luminosity, $L_{\rm bubble}$) with the energy rate needed to maintain the turbulence (turbulent energy dissipation, $L_{\rm turb}$). \textbf{The luminosity of outflows is $\bf 1.3 \times 10^{33}$ erg s$\bf ^{-1}$ after inclination and blending correction. Summing up the luminosity in Table \ref{OutflowParameters} we get the energy injection rate of bubbles to be $\bf 6.4 \times 10^{33}$ erg s$\bf ^{-1}$.}

\textbf{The turbulent dissipation rate can be calculated as}
\begin{equation}
\bf L_{\rm turb} = {{E_{\rm turb}}\over{t_{\rm diss}}},
\label{equ:L_turb}
\end{equation}
\textbf{where $\bf t_{\rm diss}$ is the turbulent dissipation time. We estimate the turbulent dissipation time through two methods based on numerical simulations.}

\textbf{First, the turbulent dissipation time of the cloud is given by \citep{McKee07}}
\begin{equation}
\bf t_{\rm diss} \sim 0.5{{d}\over{\sigma_{\rm 1d}}},
\label{equ:t_diss_McKee}
\end{equation}
\textbf{where $d = 27.6$ pc is the cloud diameter and $\sigma_{\rm 1d}$ is the one dimensional turbulent velocity dispersion along the line of sight,}
\begin{equation}
\bf \sigma_{\rm 1d} = {{\Delta v_{\rm FWHM}}\over{2\sqrt{2\ {\rm ln\ 2}}}}.
\label{equ:sigma_1d}
\end{equation}
\textbf{Here $\bf \Delta v_{\rm FWHM}$ = 2 km s$\bf ^{-1}$ is the same as that in \S~\ref{TurbulentEnergy}. Then we get $\bf \sigma_{\rm 1d} = 0.85$ km s$^{-1}$. Combining Eq. (\ref{equ:t_diss_McKee}) and Eq. (\ref{equ:sigma_1d}) we obtain the turbulent dissipation time, $\bf t_{\rm diss} = 1.6 \times 10^{7}$ yr, which is about 6 times larger than the result ($2.7 \times 10^{6}$ yr) in \citet{Narayanan12}. Then using Eq. (\ref{equ:L_turb}) we get the turbulent dissipation rate to be $6.6 \times 10^{32}$ erg s$\bf ^{-1}$, which is 51\% of the luminosity of outflows and only 10\% of the energy injection rate of bubbles.}

\textbf{Second, we follow \citet{MacLow99} to calculate the dissipation time of the cloud by}
\begin{equation}
\bf t_{\rm diss} \sim ({{3.9\kappa}\over {M_{\rm rms}}})t_{\rm ff},
\label{equ:t_diss}
\end{equation}
\textbf{where $\bf t_{\rm ff}$ is the free-fall timescale, $\bf M_{\rm rms}$ is the Mach number of the turbulence, and $\bf \kappa$ is the ratio of the driving length to the Jean's length of the cloud,}
\begin{equation}
\bf \kappa = {{\lambda_d}\over{\lambda_J}}.
\label{equ:kappa}
\end{equation}

\textbf{Using our extensive data sets we get $\bf \lambda_d = $ 0.53 pc, which is the average size of outflows and bubbles we found in Taurus. The Jean's length is defined as}
\begin{equation}
\bf \lambda_J = c_s(\pi/G\rho_{\rm reg})^{1/2},
\label{equ:lambda_J}
\end{equation}
\textbf{where $c_s$ is the sound speed \citep{MacLow99}. For an ideal gas}
\begin{equation}
\bf c_s = (3kT/\mu)^{1/2},
\label{equ:c_s}
\end{equation}
\textbf{where $\bf k$ is the Boltzmann's constant, $\bf T$ = 10 K is the temperature of the Taurus molecular cloud, and $\bf \mu$ = 2.72 is the mean molecular weight \citep{Brunt10}. Then we get the sound speed, $\bf c_s$ = 0.3 km $\bf s^{-1}$. $\bf \rho_{\rm reg}$ is the representative volume density of the region where dissipation takes place. We estimate the volume density to be $\bf \sim$ 1500 cm$\bf ^{-3}$. Using Eq. (\ref{equ:lambda_J}) we obtain the Jean's length of the region, $\bf \lambda_J = 0.83$ pc, which is about 4 times larger than that ($\bf \lambda_J \sim 0.2$ pc) in Perseus \citep{Arce10}. And then using Eq. (\ref{equ:kappa}) we obtain $\bf \kappa = 0.64$, which is different from the assumption ($\bf \kappa \sim $ 1) by \citet{Arce10} and \citet{Narayanan12}.}

\textbf{The Mach number of the turbulence can be calculated by \citep{MacLow99}}
\begin{equation}
\bf M_{\rm rms} = {{\sigma_{\rm 3d}}\over{c_{\rm s}}}.
\label{equ:M_rms}
\end{equation}
\textbf{Using Eq. (\ref{equ:sigma3d}), Eq. (\ref{equ:c_s}) and Eq. (\ref{equ:M_rms}) we get the Mach number, $\bf M_{\rm rms}$ = 5, which is different from the assumption ($\bf M_{\rm rms} = $ 10) by \citet{Arce10} and \citet{Narayanan12}.}

\textbf{The free-fall timescale of the cloud,}
\begin{equation}
\bf t_{\rm ff} = (3\pi/32G\rho_{\rm cloud})^{1/2},
\label{equ:t_ff}
\end{equation}
\textbf{where}
\begin{equation}
\bf \rho_{\rm cloud} = {{3M_{\rm cloud}}\over{4\pi R_{\rm cloud}^3}}
\label{equ:rho_cloud}
\end{equation}
\textbf{is the average volume density of the cloud. Then we get the free-fall timescale, $\bf t_{\rm ff} = 7 \times 10^{6}$ yr.}

\textbf{Using the formulas from Eq. (\ref{equ:t_diss}) to Eq. (\ref{equ:rho_cloud}) we obtain the turbulent dissipation time to be $3.5 \times 10^{6}$ yr. Then we get the turbulent dissipation rate to be $3.1 \times 10^{33}$ erg s$\bf ^{-1}$, which is about 2.4 times larger than the luminosity of outflows but 48\% of the energy injection rate of bubbles. The turbulent dissipation rate we obtained is close to that ($3.8 \times 10^{33}$ erg s$\bf ^{-1}$) of \citet{Narayanan12}. In Table~\ref{DissipationRate} we list the parameters related to the dissipation rate we get from the above two methods and compare them with the results of \citet{Narayanan12}.}

\begin{table}
\begin{center}
\caption{The Parameters Related to the Dissipation Rate\label{DissipationRate}}
\begin{tabular}{lcccc}
\tableline\tableline
Method & $\kappa$ & $M_{\rm rms}$ & $t_{\rm diss}$ & $L_{\rm turb}$ \\
       &          &               & ($10^{6}$ yr)  & ($10^{33}$ erg s$^{-1}$) \\
\tableline
MO07\tablenotemark{a}   &  -       &  -    &  $16  $  &  $0.7 $\\
ML99\tablenotemark{b}   &  0.64    &  5    &  $3.5 $  &  $3.1  $\\
Na12\tablenotemark{c}   &  1       &  10   &  $2.7 $  &  $3.8  $\\
\tableline
\end{tabular}
\end{center}
\tablenotemark{a}{The result we obtain using the method given by \citet{McKee07}.}
\tablenotemark{b}{The result we obtain using the method given by \citet{MacLow99}.}
\tablenotemark{c}{The result of \citet{Narayanan12} using the method given by \citet{MacLow99}.}
\end{table}

\textbf{Both methods invoke numerical simulations to calibrate the numerical factors in addition to essentially dimensional arguments. The main difference of the two methods is the scale of the region where dissipation takes place. \citet{McKee07} adopted the dimension of the entire cloud, while when we use the method given by \citet{MacLow99} the scale is the average size of outflows and bubbles. None of the simulations so far implements the physics (excitation, radiative transfer, etc.) necessary for actually modeling dissipation. Thus we should treat the calculations above with caution and take them as dimensional and order-of magnitude estimates.}

\textbf{Comparing the energy injection rate of outflows and bubble with the turbulent dissipation rate, we conclude that in the current episode of star formation in Taurus, both outflows and bubbles can sustain the currently observed turbulence in Taurus.}

\subsection{Protostellar Winds Can Drive Bubbles to Sustain Turbulence in Taurus}
\label{StellarWind}
Protostellar winds will inject energy into the cloud and may help sustain turbulence \citep{Nakamura07}. The winds can clear the gas surrounding the young star and form a bubble structure \citep{Arce11}. To assess whether the winds can drive bubbles in Taurus we compared the wind energy injection rate into the cloud ($\dot{E}_{\mathrm{w}}$) with the total energy injection rate from bubbles of the cloud. Following \citet{Arce11}, we estimated the wind energy injection rate using Equation 3.7 from \citet{McKee89}:
\begin{equation}
\bf \dot{E}_{w} = \frac{1}{2}(\dot{M}_{\mathrm{w}} v_{\mathrm{w}})\sigma_{\rm 3d},
\label{equ:WindEnergyInjection}
\end{equation}
where $v_{\mathrm{w}}$ is the wind velocity, which is generally assumed to be close to the star escape velocity. For the low- and intermediate-mass stars the escape velocity is about $1 - 4 \times 10^2$ km s$^{-1}$ \citep{Lamers99}. Similar to \citet{Arce11} we assumed $v_{\mathrm{w}} \sim 200$ km s$^{-1}$. The total mass loss rate from the protostellar winds is given by $\dot{M}_{\mathrm{w}}$, which can be estimated by the sum of the wind mass loss rate for each bubble ($\dot{m}_{\mathrm{w}}$). The wind mass loss rate required to produce the bubbles is roughly estimated by Equation (2) from \citet{Arce11}:
\begin{equation}
\dot{m}_{\mathrm{w}} = \frac{P_{\mathrm{bubble}}}{v_{\mathrm{w}} \tau_{\mathrm{w}} },
\label{equ:WindMassLoss}
\end{equation}
where $P_{\mathrm{bubble}}$ is the total momentum of bubbles. The wind velocity, $v_{\mathrm{w}}$, is the same as that in Eq. (\ref{equ:WindEnergyInjection}). $\tau_{\mathrm{w}}$ is the wind timescale, which is assumed $\sim 1$ Myr \citep{Arce11}.
From Eq. (\ref{equ:WindMassLoss}) we obtained the wind mass loss rates of each bubble ($\dot{m}_{\mathrm{w}}$), which are listed in Table \ref{bubble}. Summing $\dot{m}_{\mathrm{w}}$ of all bubbles we find $\dot{M}_{\mathrm{w}}$ to be $1.89 \times 10^{-5}$ M$_{\sun}$~yr$^{-1}$. Using Eq. (\ref{equ:WindEnergyInjection}) we obtained the wind energy injection rate ($\dot{E}_{\mathrm{w}}$) to be $\bf \sim 2 \times 10^{33}$ \textbf{erg s}$\bf ^{-1}$, 31\% of the total energy injection rate from bubbles in Taurus, which is comparable to the turbulent dissipation rate in Taurus. Therefore, the protostellar winds can drive bubbles to sustain turbulence in Taurus.

\subsection{\bf Potential Sources of Turbulent Motions in Taurus}
\textbf{The origin of turbulence in the molecular cloud has been intensely debated over the past three decades \citep[e.g.][]{Larson81, Heyer04}. \citet{Hennebelle12} suggests that for a large fraction of clouds the turbulent driving is external. Numerical simulations shows that the external sources of turbulence are likely to be large-scale H{\sc i} streams \citep{Ballesteros99}, shocks \citep{McKee07}, Alfv$\bf \acute{e}$n waves \citep{Nakamura07, Wang10}, supernovae explosion and galactic differential rotation \citep{Klessen10, Hennebelle12}. It is unclear which one is the source of turbulence in Taurus.}

\section{Conclusions}
\label{Conclusions}
We have studied the dynamic structures including outflows and bubbles within the Taurus molecular cloud using the 100 deg$^2$ FCRAO large-scale $^{12}$CO(1-0) and $^{13}$CO(1-0) maps and the Spitzer protostellar catalog. The high sensitivity and large spatial dynamic range of the maps provide us an excellent opportunity to undertake an unbiased search for outflows and bubbles in this region. We also analyzed the energy injection of these dynamic structures into the entire cloud. Our conclusions regarding the dynamic structures in Taurus and their properties are as follows.

1. We identified 55 outflows around the Spitzer YSOs in the main 44 deg$^2$ area of Taurus. In total, 31 of the detected outflows were previously unknown, increasing the number of outflows by a factor of 1.3.

2. We classified the outflows into 5 categories according to the morphology of contour maps and P-V diagrams. The classifications indicate the confidence level of the outflows. 76.3\% of the outflows are in the ``most probable'' category in our study.

3. Most of the outflows are driven by Class I, Flat and Class II YSOs while few outflows were found around Class III YSOs, which indicates that the outflow activity likely occurred in the earlier stage rather than the late phase of the star formation.

4. More bipolar and monopolar redshifted outflows were identified while few monopolar blueshifted ones were detected in our study.

5. We detected 37 bubbles in the 100 deg$^2$ region of Taurus. All the bubbles were previously unknown. The bubbles were identified by the integrated intensity maps, P-V diagrams, Gaussian fitting profiles and channel maps.

6. The gravitational binding energy of the Taurus molecular cloud is $\bf \sim 1.5 \times 10^{48}$ {\bf erg}. The total kinetic energy of outflows and bubbles in Taurus are $\bf \sim 3.9 \times 10^{45}$ erg and $\sim 9.2 \times 10^{46}$ erg, respectively. Neither outflows nor bubbles can balance the overall gravitational binding energy of Taurus.

7. The turbulent energy of the Taurus molecular cloud is $\sim 3.2 \times 10^{47}$ erg. The energy of all detected outflows and bubbles cannot have generated the observed turbulence in Taurus.

8. The rate of turbulent dissipation in Taurus is \textbf{between $\bf \sim 6.6 \times 10^{32}$ to $\bf \sim 3.1 \times 10^{33}$ erg s$^{-1}$}. The energy injection rates of outflows and bubbles are $\bf \sim 1.3 \times 10^{33}$ erg s$^{-1}$ and $\sim 6.4 \times 10^{33}$ erg s$^{-1}$, respectively. \textbf{Both outflows and bubbles can sustain the turbulence in Taurus at the current epoch.}

9. The stellar winds can drive bubbles to sustain turbulence in the Taurus molecular cloud.

\label{summary}

\clearpage

\begin{figure*}%[htbp]
\centering %sub_figures centered
%\includegraphics[width=0.40\textwidth]{041159.7+294236.eps}\hspace{-18mm}
%\includegraphics[width=0.40\textwidth]{pv_041159.7+294236_12CO_pixdown14arcmin_pixup18arcmin_midpoint_70.eps}
%
%\vspace{-5mm}
%
%\includegraphics[width=0.40\textwidth]{041159.7+294236_Left6.75000Right4.50000Down9Up16_Aver_red.eps}

\caption{TMO\_01 (SST 041159.7+294236): \textbf{upper left} and \textbf{upper right}, the same as upper two panels of Fig. \ref{sample-figure01} except for the following. In the upper left panel the integrated intervals for red lobe and grey-scale image are 8.75 to 10 km s$^{-1}$ and 6.5 to 7.5 km s$^{-1}$, respectively. In the upper right panel the position angle is $20^{\circ}$ and contour levels are 0.4 to 1.6 K by 0.3 K, 2 to 4 K by 0.5 K. \textbf{lower}, the same as the lower right panel of Fig. \ref{sample-figure01}.}
\label{sample-figure02}
\end{figure*}

\clearpage
\begin{figure*}%[htbp]
\centering %sub_figures centered
%\includegraphics[width=0.40\textwidth]{041414.5+282758.eps}\hspace{-8mm}
%\includegraphics[width=0.40\textwidth]{pv_041414.5+282758_12CO_pixdown10arcmin_pixup5arcmin_yso_45.eps}
%
%\vspace{-5mm}
%
%\includegraphics[width=0.40\textwidth]{041414.5+282758_Left3.25000Right2.25000Down6.80000Up1.80000_Aver_blue.eps}\hspace{-8mm}
%\includegraphics[width=0.40\textwidth]{041414.5+282758_Left5.50000Right5.50000Down1.80000Up0.500000_Aver_red.eps}

\caption{TMO\_03 (SST 041414.5+282758): The same as Fig. \ref{sample-figure01} except for the following. In the upper left panel the integrated intervals for red lobe and grey-scale image are 8.5 to 9.5 km s$^{-1}$ and 6 to 7 km s$^{-1}$, respectively. The blue and red contour levels are 30\%, 40\%,..., 90\% and 50\%, 60\%,..., 90\% of their peak value, respectively. In the upper right panel the position angle is $45^{\circ}$ and contour levels are 0.4 to 1.6 K by 0.3 K, 2 to 6 K by 0.5 K.}
\label{sample-figure03}
\end{figure*}

\clearpage
\begin{figure*}%[htbp]
\centering %sub_figures centered
%\includegraphics[width=0.40\textwidth]{041832.0+283115.eps}\hspace{-12mm}
%\includegraphics[width=0.40\textwidth]{pv_041832.0+283115_12CO_pixdown2arcmin_pixup8arcmin_yso_90.eps}
%
%\vspace{-5mm}
%
%\includegraphics[width=0.40\textwidth]{041832.0+283115_Left3.52500Right2.10000Down5.60000Up2.80000_Aver_blue.eps}\hspace{-12mm}
%\includegraphics[width=0.40\textwidth]{041832.0+283115_Left1.35000Right1.25000Down3.60000Up4.10000_Aver_red.eps}

\caption{TMO\_04 (SST 041832.0+283115): The same as Fig. \ref{sample-figure01} except for the following. In the upper left panel the integrated intervals for blue lobe, red lobe and grey-scale image are 4.4 to 5.3 km s$^{-1}$, 9.5 to 10.3 km s$^{-1}$ and 6.5 to 8 km s$^{-1}$, respectively. The red contour levels are 35\%, 50\%,..., \%90 of the peak value. In the upper right panel the position angle is $0^{\circ}$ and contour levels are 0.5 to 2 K by 0.3 K, 2.5 to 5.5 K by 0.5 K.}
\label{sample-figure04}
\end{figure*}

\clearpage
\begin{figure*}%[htbp]
\centering %sub_figures centered
%\includegraphics[width=0.40\textwidth]{041941.4+271607.eps}\hspace{-10mm}
%\includegraphics[width=0.40\textwidth]{pv_041941.4+271607_12CO_pixdown7arcmin_pixup10arcmin_midpoint_50.eps}
%
%\vspace{-5mm}
%
%\includegraphics[width=0.40\textwidth]{041941.4+271607_Left5.07500Right4.05000Down8.70000Up9.50000_Aver_blue.eps}

\caption{TMO\_05 (SST 041941.4+271607): \textbf{upper left} and \textbf{upper right}, the same as upper two panels of Fig. \ref{sample-figure01} except for the following. In the upper left panel the integrated intervals for the blue lobe and grey-scale image are 2 to 4.3 km s$^{-1}$ and 6 to 7 km s$^{-1}$, respectively. The blue levels are 30\%, 45\%,..., 90\% of the peak value. In the upper right panel the position angle is $40^{\circ}$ and contour levels are 0.3 to 0.9 K by 0.15 K, 1.1 to 1.9 K by 0.2 K. \textbf{lower}, the same as the lower left panel of Fig. \ref{sample-figure01}.}
\label{sample-figure05}
\end{figure*}

\clearpage
\begin{figure*}%[htbp]
\centering %sub_figures centered
%\includegraphics[width=0.40\textwidth]{041958.4+270957.eps}\hspace{-14mm}
%\includegraphics[width=0.40\textwidth]{pv_041958.4+270957_12CO_pixdown8arcmin_pixup8arcmin_midpoint_118.eps}
%
%\vspace{-5mm}
%
%\includegraphics[width=0.40\textwidth]{041958.4+270957_Left3.00000Right2.00000Down2.90000Up2.30000_Aver_blue.eps}\hspace{-8mm}
%\includegraphics[width=0.40\textwidth]{041958.4+270957_Left1.25000Right1.25000Down2.50000Up2_Aver_red.eps}

\caption{TMO\_06 (SST 041958.4+270957): The same as Fig. \ref{sample-figure01} except for the following. In the upper left panel the integrated intervals for the blue lobe, red lobe and grey-scale image are 2 to 4 km s$^{-1}$, 9 to 12 km s$^{-1}$ and 5 to 7 km s$^{-1}$, respectively. The blue and red contour levels are both 30\%, 40\%,..., 90\% of their peak value. In the upper right panel the position angle is $152^{\circ}$ and contour levels are 0.5 to 2 K by 0.3 K, 2.5 to 5.5 K by 0.5 K.}
\label{sample-figure06}
\end{figure*}

\clearpage
\begin{figure*}%[htbp]
\centering %sub_figures centered
%\includegraphics[width=0.40\textwidth]{042107.9+270220.eps}\hspace{-10mm}
%\includegraphics[width=0.40\textwidth]{pv_042107.9+270220_12CO_pixdown6arcmin_pixup6arcmin_midpoint_140.eps}
%
%\vspace{-5mm}
%
%\includegraphics[width=0.40\textwidth]{042107.9+270220_Left3.35000Right3.35000Down3.30000Up3.30000_Aver_blue.eps}\hspace{-10mm}
%\includegraphics[width=0.40\textwidth]{042107.9+270220_Left1.12500Right1.12500Down1.10000Up1.10000_Aver_red.eps}

\caption{TMO\_07 (SST 042107.9+270220): The same as Fig. \ref{sample-figure01} except for the following. In the upper left panel the integrated intervals for the blue lobe and red lobe are 3 to 4.5 km s$^{-1}$ and 8 to 9 km s$^{-1}$, respectively. The blue and red contour levels are 50\%, 60\%,..., 90\% and 40\%, 50\%,..., 90\% of their peak value, respectively. In the upper right panel the position angle is $130^{\circ}$ and contour levels are 0.5 to 2 K by 0.3 K, 2.5 to 4.5 K by 0.5 K.}
\label{sample-figure07}
\end{figure*}

\clearpage
\begin{figure*}%[htbp]
\centering %sub_figures centered
%\hspace{6mm}\includegraphics[width=0.40\textwidth]{042215.6+265706.eps}\hspace{-10mm}
%\includegraphics[width=0.40\textwidth]{pv_042215.6+265706_12CO_pixdown6arcmin_pixup12arcmin_midpoint_148.eps}
%
%\vspace{-5mm}
%
%\includegraphics[width=0.40\textwidth]{042215.6+265706_Left1.87500Right2.82500Down2.80000Up1.40000_Aver_blue.eps}\hspace{-6mm}
%\includegraphics[width=0.40\textwidth]{042215.6+265706_Left0.950000Right1.12500Down1Up1.10000_Aver_red.eps}

\caption{TMO\_08 (SST 042215.6+265706): The same as Fig. \ref{sample-figure01} except for the following. In the upper left panel the integrated intervals for the blue lobe,red lobe and grey-scale image are 2 to 4.5 km s$^{-1}$, 8.3 to 9.5 km s$^{-1}$ and 5.5 to 7 km s$^{-1}$, respectively. The blue and red contour levels are both 30\%, 40\%,..., 90\% of their peak value. In the upper right panel the position angle is $122^{\circ}$ and contour levels are 0.5 to 2 K by 0.3 K, 2.5 to 4 K by 0.5 K.}
\label{sample-figure08}
\end{figure*}

\clearpage
\begin{figure*}%[htbp]
\centering %sub_figures centered
%\includegraphics[width=0.40\textwidth]{042325.9+250354.eps}\hspace{-6mm}
%\includegraphics[width=0.40\textwidth]{pv_042325.9+250354_12CO_pixdown4arcmin_pixup21arcmin_yso_38.eps}
%
%\vspace{-5mm}
%
%\includegraphics[width=0.40\textwidth]{042325.9+250354_Left0.250000Right17.0000Down5Up13_Aver_red.eps}

\caption{TMO\_09 (SST 042325.9+250354): The same as Fig. \ref{sample-figure02} except for the following. In the upper left panel the integrated interval for the red lobe is 8.25 to 9.7 km s$^{-1}$. The red contour levels are 50\%, 60\%,..., 90\% of the peak value. In the upper right panel the position angle is $52^{\circ}$ and contour levels are 0.4 to 1.6 K by 0.3 K, 2 to 4.5 by 0.5 K.}
\label{sample-figure09}
\end{figure*}

\clearpage
\begin{figure*}%[htbp]
\centering %sub_figures centered
%\hspace{6mm}\includegraphics[width=0.40\textwidth]{042420.9+263051.eps}\hspace{-8mm}
%\includegraphics[width=0.40\textwidth]{pv_042420.9+263051_12CO_pixdown6arcmin_pixup10arcmin_midpoint_60.eps}
%
%\vspace{-5mm}
%
%\includegraphics[width=0.40\textwidth]{042420.9+263051_Left1.75000Right1.25000Down2.50000Up4_Aver_blue.eps}\hspace{-4mm}
%\includegraphics[width=0.40\textwidth]{042420.9+263051_Left4.25000Right2.50000Down1.30000Up1.70000_Aver_red.eps}

\caption{TMO\_10 (SST 042420.9+263051): The same as Fig. \ref{sample-figure01} except for the following. In the upper left panel the integrated intervals for the blue lobe,red lobe and grey-scale image are 4 to 4.75 km s$^{-1}$, 7.5 to 10 km s$^{-1}$ and 5.5 to 7.25 km s$^{-1}$, respectively. The blue and red contour levels are 50\%, 60\%,..., 90\% and 60\%, 70\%,..., 90\% of their peak value, respectively. In the upper right panel the position angle is $30^{\circ}$ and contour levels are 0.4 to 1.6 K by 0.3 K, 2 to 7 K by 0.5 K.}
\label{sample-figure10}
\end{figure*}

\clearpage
\begin{figure*}%[htbp]
\centering %sub_figures centered
%\includegraphics[width=0.40\textwidth]{042445.0+270144.eps}\hspace{-8mm}
%\includegraphics[width=0.40\textwidth]{pv_042445.0+270144_12CO_pixdown7arcmin_pixup7arcmin_midpoint_90.eps}
%
%\vspace{-5mm}
%
%\includegraphics[width=0.40\textwidth]{042445.0+270144_Left2.25000Right4.50000Down2Up5_Aver_red.eps}

\caption{TMO\_11 (SST 042445.0+270144): The same as Fig. \ref{sample-figure02} except for the following. In the upper left panel the integrated intervals for the red lobe and grey-scale image are 8 to 10.5 km s$^{-1}$ and 6 to 7 km s$^{-1}$, respectively. The red contour levels are 20\%, 30\%,..., 90\% of the peak value. In the upper right panel the position angle is $0^{\circ}$.}
\label{sample-figure11}
\end{figure*}

\clearpage
\begin{figure*}%[htbp]
\centering %sub_figures centered
%\includegraphics[width=0.40\textwidth]{042930.0+243955.eps}\hspace{-10mm}
%\includegraphics[width=0.40\textwidth]{pv_042930.0+243955_12CO_pixdown6arcmin_pixup6arcmin_midpoint_37.eps}
%
%\vspace{-5mm}
%
%\includegraphics[width=0.40\textwidth]{042930.0+243955_Left2.95000Right1.47500Down1.50000Up1.50000_Aver_red.eps}

\caption{TMO\_12 (SST 042930.0+243955): The same as Fig. \ref{sample-figure02} except for the following. In the upper left panel the integrated intervals for the red lobe and grey-scale image are 8 to 9.2 km s$^{-1}$ and 5.5 to 7.5 km s$^{-1}$, respectively. The red contour levels are 50\%, 60\%,..., 90\% of the peak value. In the upper right panel the position angle is $53^{\circ}$ and Contour levels are 0.4 to 4.8 K by 0.3 K.}
\label{sample-figure12}
\end{figure*}

\clearpage
\begin{figure*}%[htbp]
\centering %sub_figures centered
%\includegraphics[width=0.40\textwidth]{043110.4+254129.eps}\hspace{-4mm}
%\includegraphics[width=0.40\textwidth]{pv_043110.4+254129_12CO_pixdown10arcmin_pixup10arcmin_midpoint_35.eps}
%
%\vspace{-5mm}
%
%\includegraphics[width=0.40\textwidth]{043110.4+254129_Left5.00000Right2.25000Down2Up2_Aver_red.eps}\hspace{-2mm}
%\includegraphics[width=0.40\textwidth]{043110.4+254129_13CO.eps}

\caption{TMO\_13: \textbf{upper left} and \textbf{upper right}, the same as upper two panels of Fig. \ref{sample-figure02} except for the following. In the upper left panel the integrated intervals for the red lobe and grey-scale image are 7 to 9.5 km s$^{-1}$ and 5.5 to 6.5 km s$^{-1}$, respectively. The red contour levels are 40\%, 50\%,..., 90\% of the peak value. In the upper right panel the position angle is $55^{\circ}$ and contour levels are 0.4 to 1.6 K by 0.3 K, 2 to 4.5 K by 0.5 K. \textbf{lower left}, the same as the lower panel of Fig. \ref{sample-figure02}. \textbf{lower right}, $\rm {}^{13}CO$ integrated intensity map overlaid on the $\rm {}^{13}CO$ grey-scale image, the integrated intervals for the red lobe and grey-scale image and the red contour levels are the same with those in the upper left panel of this figure.}
\label{sample-figure13}
\end{figure*}

\clearpage
\begin{figure*}%[htbp]
\centering %sub_figures centered
%\includegraphics[width=0.40\textwidth]{043158.4+254329.eps}\hspace{-16mm}
%\includegraphics[width=0.40\textwidth]{pv_043158.4+254329_12CO_pixdown6arcmin_pixup6arcmin_midpoint_125.eps}
%
%\vspace{-5mm}
%
%\includegraphics[width=0.40\textwidth]{043158.4+254329_Left1.75000Right3.25000Down1Up4_Aver_blue.eps}\hspace{-12mm}
%\includegraphics[width=0.40\textwidth]{043158.4+254329_Left4.00000Right2.50000Down2Up2_Aver_red.eps}

\caption{TMO\_14 (SST 043158.4+254329): The same as Fig. \ref{sample-figure01} except for the following. In the upper left panel the integrated intervals for the blue lobe,red lobe and grey-scale image are 3 to 4.5 km s$^{-1}$, 7 to 10 km s$^{-1}$ and 5 to 6.5 km s$^{-1}$, respectively. The blue and red contour levels are both 50\%, 60\%,..., 90\% of their peak value. In the upper right panel the position angle is $145^{\circ}$ and contour levels are 0.4 to 1.6 K by 0.3 K, 2 to 4.5 K by 0.5 K.}
\label{sample-figure14}
\end{figure*}

\clearpage
\begin{figure*}%[htbp]
\centering %sub_figures centered
%\includegraphics[width=0.40\textwidth]{043214.6+223742.eps}\hspace{-10mm}
%\includegraphics[width=0.40\textwidth]{pv_043214.6+223742_12CO_pixdown16arcmin_pixup18arcmin_yso_35.eps}
%
%\vspace{-5mm}
%
%\includegraphics[width=0.40\textwidth]{043214.6+223742_Left2.12500Right1.07500Down1.10000Up2.10000_Aver_blue.eps}\hspace{-10mm}
%\includegraphics[width=0.40\textwidth]{043214.6+223742_Left2.12500Right1.07500Down1.70000Up2.10000_Aver_red.eps}

\caption{TMO\_15 (SST 043214.6+223742): The same as Fig. \ref{sample-figure01} except for the following. In the upper left panel the integrated intervals for the blue lobe,red lobe and grey-scale image are 2.8 to 4 km s$^{-1}$, 6.8 to 9 km s$^{-1}$ and 5 to 6 km s$^{-1}$, respectively. The blue and red contour levels are 60\%, 70\%,..., 90\% and 40\%, 50\%,..., 90\% of their peak value, respectively. In the upper right panel the position angle is $55^{\circ}$ and contour levels are 0.5 to 2 K by 0.3 K, 2.5 to 4 K by 0.5 K.}
\label{sample-figure15}
\end{figure*}

\clearpage
\begin{figure*}%[htbp]
\centering %sub_figures centered
%\includegraphics[width=0.40\textwidth]{043231.7+242002.eps}\hspace{-14mm}
%\includegraphics[width=0.40\textwidth]{pv_043231.7+242002_12CO_pixdown8arcmin_pixup8arcmin_midpoint_70.eps}
%
%\vspace{-5mm}
%
%\includegraphics[width=0.40\textwidth]{043231.7+242002_Left0.775000Right0.775000Down0.800000Up1.30000_Aver_blue.eps}\hspace{-12mm}
%\includegraphics[width=0.40\textwidth]{043231.7+242002_Left1.77500Right2.05000Down2.80000Up1.50000_Aver_red.eps}

\caption{TMO\_16 (SST 043231.7+242002): The same as Fig. \ref{sample-figure01} except for the following. In the upper left panel the integrated intervals for the blue lobe,red lobe and grey-scale image are 3 to 4.5 km s$^{-1}$, 7.5 to 8.7 km s$^{-1}$ and 5 to 7 km s$^{-1}$, respectively. The blue and red contour levels are 80\%, 90\% and 50\%, 60\%,..., 90\% of their peak value, respectively. In the upper right panel the position angle is $20^{\circ}$ and contour levels are 0.4 to 1.6 K by 0.3 K, 2 to 7 K by 0.5 K.}
\label{sample-figure16}
\end{figure*}

\clearpage
\begin{figure*}%[htbp]
\centering %sub_figures centered
%\includegraphics[width=0.40\textwidth]{043232.0+225726.eps}\hspace{-24mm}
%\includegraphics[width=0.40\textwidth]{pv_043232.0+225726_12CO_pixdown10arcmin_pixup12arcmin_midpoint_100.eps}
%
%\vspace{-5mm}
%
%\includegraphics[width=0.40\textwidth]{043232.0+225726_Left3.25000Right1.75000Down6.20000Up7.80000_Aver_red.eps}

\caption{TMO\_17 (SST 043232.0+225726): The same as Fig. \ref{sample-figure02} except for the following. In the upper left panel the integrated intervals for the red lobe and grey-scale image are 6.8 to 7.8 km s$^{-1}$ and 4.8 to 6.3 km s$^{-1}$, respectively. The red contour levels are 40\%, 50\%,..., 90\% of the peak value. In the upper right panel the position angle is $170^{\circ}$ and contour levels are 0.4 to 1.6 K by 0.3 K, 2 to 7 K by 0.5 K.}
\label{sample-figure17}
\end{figure*}

\clearpage
\begin{figure*}%[htbp]
\centering %sub_figures centered
%\includegraphics[width=0.40\textwidth]{043243.0+255231.eps}\hspace{-10mm}
%\includegraphics[width=0.40\textwidth]{pv_043243.0+255231_12CO_pixdown7arcmin_pixup5arcmin_yso_90.eps}
%
%\vspace{-5mm}
%
%\includegraphics[width=0.40\textwidth]{043243.0+255231_Left7.57500Right2.52500Down1.30000Up1.30000_Aver_blue.eps}\hspace{-10mm}
%\includegraphics[width=0.40\textwidth]{043243.0+255231_Left1.27500Right1.90000Down1.90000Up1.90000_Aver_red.eps}

\caption{TMO\_18 (SST 043243.0+255231): The same as Fig. \ref{sample-figure01} except for the following. In the upper left panel the integrated intervals for the blue lobe,red lobe and grey-scale image are 3.2 to 4 km s$^{-1}$, 7 to 8 km s$^{-1}$ and 5 to 6.5 km s$^{-1}$, respectively. The blue and red contour levels are 50\%, 60\%,..., 90\% and 60\%, 70\%,..., 90\% of their peak value, respectively. In the upper right panel the position angle is $0^{\circ}$ and contour levels are 0.4 to 1.6 K by 0.3 K, 2 to 4 K by 0.5 K.}
\label{sample-figure18}
\end{figure*}

\clearpage
\begin{figure*}%[htbp]
\centering %sub_figures centered
%\includegraphics[width=0.40\textwidth]{043415.2+225030.eps}\hspace{-8mm}
%\includegraphics[width=0.40\textwidth]{pv_043415.2+225030_12CO_pixdown4arcmin_pixup4arcmin_yso_145.eps}
%
%\vspace{-5mm}
%
%\includegraphics[width=0.40\textwidth]{043415.2+225030_Left2.00000Right2.00000Down2Up2_Aver_red.eps}

\caption{TMO\_19 (SST 043415.2+225030): The same as Fig. \ref{sample-figure02} except for the following. In the upper left panel the integrated intervals for the red lobe and grey-scale image are 7 to 8.5 km s$^{-1}$ and 6 to 6.5 km s$^{-1}$, respectively. The red contour levels are 40\%, 50\%,..., 90\% of the peak value. In the upper right panel the position angle is $125^{\circ}$.}
\label{sample-figure19}
\end{figure*}

\clearpage
\begin{figure*}%[htbp]
\centering %sub_figures centered
%\includegraphics[width=0.40\textwidth]{043724.8+270919.eps}\hspace{-10mm}
%\includegraphics[width=0.40\textwidth]{pv_043724.8+270919_12CO_pixdown4arcmin_pixup8arcmin_midpoint_0.eps}
%
%\vspace{-5mm}
%
%\includegraphics[width=0.40\textwidth]{043724.8+270919_Left1.25000Right5.75000Down2Up1_Aver_red.eps}

\caption{TMO\_20 (SST 043724.8+270919): The same as Fig. \ref{sample-figure02} except for the following. In the upper left panel the integrated intervals for the red lobe and grey-scale image are 8 to 10.5 km s$^{-1}$ and 5 to 7 km s$^{-1}$, respectively. The red contour levels are 30\%, 40\%,..., 90\% of the peak value. In the upper right panel the position angle is $90^{\circ}$ and contour levels are 0.5 to 2 K by 0.3 K, 2.5 to 4 K by 0.5 K.}
\label{sample-figure20}
\end{figure*}

\clearpage
\begin{figure*}%[htbp]
\centering %sub_figures centered
%\includegraphics[width=0.40\textwidth]{043953.9+260309.eps}\hspace{-10mm}
%\includegraphics[width=0.40\textwidth]{pv_043953.9+260309_12CO_pixdown5arcmin_pixup4arcmin_yso_0.eps}
%
%\vspace{-5mm}
%
%\includegraphics[width=0.40\textwidth]{043953.9+260309_Left0.850000Right2.97500Down0.850000Up0.850000_Aver_blue.eps}\hspace{-10mm}
%\includegraphics[width=0.40\textwidth]{043953.9+260309_Left2.37500Right2.12500Down1.02000Up1.86000_Aver_red.eps}

\caption{TMO\_21 (SST 043953.9+260309): The same as Fig. \ref{sample-figure01} except for the following. In the upper left panel the integrated intervals for the blue lobe,red lobe and grey-scale image are 2 to 4 km s$^{-1}$, 8 to 10 km s$^{-1}$ and 4.5 to 7.5 km s$^{-1}$, respectively. The blue and red contour levels are 30\%, 45\%,..., 90\% and 40\%, 50\%,..., 90\% of their peak value, respectively. In the upper right panel the position angle is $90^{\circ}$ and contour levels are 2 K by 0.3 K, 2.5 to 6 K by 0.5 K.}
\label{sample-figure21}
\end{figure*}

\clearpage
\begin{figure*}%[htbp]
\centering %sub_figures centered
%\includegraphics[width=0.40\textwidth]{044108.2+255607.eps}\hspace{-14mm}
%\includegraphics[width=0.40\textwidth]{pv_044108.2+255607_12CO_pixdown15arcmin_pixup15arcmin_midpoint_50.eps}
%
%\vspace{-5mm}
%
%\includegraphics[width=0.40\textwidth]{044108.2+255607_Left4.60000Right5.52500Down9.70000Up12_Aver_blue.eps}

\caption{TMO\_22 (SST 044108.2+255607): The same as Fig. \ref{sample-figure05} except for the following. In the upper left panel the integrated intervals for the blue lobe and grey-scale image are 3 to 4.5 km s$^{-1}$ and 4.6 to 7 km s$^{-1}$, respectively. The blue contour levels are 35\%, 50\%,..., 95\% of the peak value. In the upper right panel the contour levels are 0.4 to 1.6 K by 0.3 K, 2 to 4.5 K by 0.5 K.}
\label{sample-figure22}
\end{figure*}

\clearpage
\begin{figure*}%[htbp]
\centering %sub_figures centered
%\includegraphics[width=0.40\textwidth]{044112.6+254635.eps}\hspace{-18mm}
%\includegraphics[width=0.40\textwidth]{pv_044112.6+254635_12CO_pixdown6arcmin_pixup6arcmin_yso_90.eps}
%
%\vspace{-5mm}
%
%\includegraphics[width=0.40\textwidth]{044112.6+254635_Left1.00000Right1.75000Down4Up2_Aver_red.eps}

\caption{TMO\_23 (SST 044112.6+254635): The same as Fig. \ref{sample-figure02} except for the following. In the upper left panel the integrated intervals for the red lobe and grey-scale image are 7.5 to 10.5 km s$^{-1}$ and 4.5 to 7 km s$^{-1}$, respectively. The red contour levels are 40\%, 50\%,..., 90\% of the peak value. In the upper right panel the position angle is $0^{\circ}$.}
\label{sample-figure23}
\end{figure*}

\clearpage
\begin{figure*}%[htbp]
\centering %sub_figures centered
%\includegraphics[width=0.40\textwidth]{044207.7+252311.eps}\hspace{-8mm}
%\includegraphics[width=0.40\textwidth]{pv_044207.7+252311_12CO_pixdown3arcmin_pixup7arcmin_yso_75.eps}
%
%\vspace{-5mm}
%
%\includegraphics[width=0.40\textwidth]{044207.7+252311_Left1.25000Right2.75000Down1Up5_Aver_red.eps}

\caption{TMO\_24 (SST 044207.7+252311): The same as Fig. \ref{sample-figure02} except for the following. In the upper left panel the integrated intervals for the red lobe and grey-scale image are 8 to 9.5 km s$^{-1}$ and 5 to 7 km s$^{-1}$, respectively. The red contour levels are 30\%, 40\%,..., 90\% of the peak value. In the upper right panel the position angle is $15^{\circ}$ and contour levels are 0.4 to 1.6 K by 0.3 K, 2 to 4.5 K by 0.5 K.}
\label{sample-figure24}
\end{figure*}

\clearpage
\begin{figure*}%[htbp]
\centering %sub_figures centered
%\includegraphics[width=0.40\textwidth]{041858.1+281223.eps}\hspace{-10mm}
%\includegraphics[width=0.40\textwidth]{pv_041858.1+281223_12CO_pixdown6arcmin_pixup5arcmin_midpoint_90.eps}
%
%\vspace{-5mm}
%
%\includegraphics[width=0.40\textwidth]{041858.1+281223_Left3.05000Right2.27500Down2.30000Up3_Aver_blue.eps}
%
%\vspace{-5mm}
%
%\includegraphics[width=0.40\textwidth]{041858.1+281223_13CO.eps}\hspace{-10mm}
%\includegraphics[width=0.40\textwidth]{pv_041858.1+281223_13CO_pixdown6arcmin_pixup5arcmin_midpoint_90.eps}

\caption{TMO\_25 (SST 041858.1+281223): \textbf{upper left} and \textbf{upper right}, the same as the upper two panels of Fig. \ref{sample-figure05} except for the following. In the upper left panel the integrated intervals for the blue lobe and grey-scale image are 3.5 to 5.5 km s$^{-1}$ and 6.5 to 7.5 km s$^{-1}$, respectively. The blue contour levels are 50\%, 60\%,..., 90\% of the peak value. In the upper right panel the position angle is $0^{\circ}$ and contour levels are 0.5 to 2 K by 0.3 K, 2.5 to 6.5 K by 0.5 K. \textbf{middle}, the same as the lower panel of Fig. \ref{sample-figure05}. \textbf{lower left}, $\rm {}^{13}CO$ integrated intensity map overlaid on the $\rm {}^{13}CO$ grey-scale image, the integrated intervals for the blue lobe and the grey-scale image and the blue contour levels are the same with those in the upper left panel of this figure. \textbf{lower right}, P-V diagram of $\rm {}^{13}CO$, through the same position angle in the upper left panel of this figure. Contour levels are 0.15 to 0.9 K by 0.15 K, 1.1 to 1.9 K by 0.2 K and 2.1 to 3.9 K by 0.3 K.}
\label{sample-figure25}
\end{figure*}

\clearpage
\begin{figure*}%[htbp]
\centering %sub_figures centered
%\includegraphics[width=0.40\textwidth]{042318.2+264115.eps}\hspace{-12mm}
%\includegraphics[width=0.40\textwidth]{pv_042318.2+264115_12CO_pixdown15arcmin_pixup10arcmin_midpoint_40.eps}
%
%\vspace{-5mm}
%
%\includegraphics[width=0.40\textwidth]{042318.2+264115_Left8.25000Right4.25000Down4Up4_Aver_red.eps}
%
%\vspace{-5mm}
%
%\includegraphics[width=0.40\textwidth]{042318.2+264115_13CO.eps}\hspace{-12mm}
%\includegraphics[width=0.40\textwidth]{pv_042318.2+264115_13CO_pixdown12arcmin_pixup10arcmin_midpoint_125.eps}

\caption{TMO\_26 (SST 042318.2+264115): \textbf{upper left} and \textbf{upper right}, the same as the upper two panels of Fig. \ref{sample-figure02} except for the following. In the upper left panel the integrated intervals for the red lobe and grey-scale image are 7.5 to 10 km s$^{-1}$ and 6 to 7 km s$^{-1}$, respectively. The red contour levels are 40\%, 50\%,..., 90\% of the peak value. In the upper right panel the position angle is $50^{\circ}$. \textbf{middle}, the same as the lower panel of Fig. \ref{sample-figure02}. \textbf{lower left}, $\rm {}^{13}CO$ integrated intensity map overlaid on the $\rm {}^{13}CO$ grey-scale image, the integrated intervals for the red lobe and grey-scale image are the same with those in the upper left panel of this figure. The red contour levels are 30\%, 40\%,..., 90\% of the peak value. \textbf{lower right}, the same as the lower right panel of Fig. \ref{sample-figure25} except that the contour levels are 0.15 to 0.9 K by 0.15 K, 1.1 to 2.3 K by 0.2 K.}
\label{sample-figure26}
\end{figure*}

\clearpage
\begin{figure*}%[htbp]
\centering %sub_figures centered
%\includegraphics[width=0.40\textwidth]{042656.2+244335.eps}\hspace{-10mm}
%\includegraphics[width=0.40\textwidth]{pv_042656.2+244335_12CO_pixdown7arcmin_pixup4arcmin_midpoint_50.eps}
%
%\vspace{-5mm}
%
%\includegraphics[width=0.40\textwidth]{042656.2+244335_Left3.55000Right2.02500Down5.10000Up2_Aver_red.eps}
%
%\vspace{-5mm}
%
%\includegraphics[width=0.40\textwidth]{042656.2+244335_13CO.eps}\hspace{-10mm}
%\includegraphics[width=0.40\textwidth]{pv_042656.2+244335_13CO_pixdown9arcmin_pixup9arcmin_midpoint_50.eps}

\caption{TMO\_27 (SST 042656.2+244335): The same as Fig. \ref{sample-figure26} except for the following. In the upper left panel the integrated intervals for the red lobe and grey-scale image are 8.3 to 9.7 km s$^{-1}$ and 6 to 7.5 km s$^{-1}$, respectively. In the upper right panel the position angle is $40^{\circ}$. In the lower left panel the integrated intervals for the red lobe and grey-scale image and the red contour levels are the same with those in the upper left panel of this figure. In the lower right panel the position angle is the same with the upper right panel of this figure and the contour levels are 0.15 to 0.9 K by 0.15 K, 1.1 to 2.1 K by 0.2 K.}
\label{sample-figure27}
\end{figure*}

\clearpage
\begin{figure*}%[htbp]
\centering %sub_figures centered
%\includegraphics[width=0.40\textwidth]{042702.6+260530.eps}\hspace{-10mm}
%\includegraphics[width=0.40\textwidth]{pv_042702.6+260530_12CO_pixdown6arcmin_pixup6arcmin_midpoint_0.eps}
%
%\vspace{-5mm}
%
%\includegraphics[width=0.40\textwidth]{042702.6+260530_Left2.12500Right1.42500Down4.60000Up1.80000_Aver_blue.eps}\hspace{-12mm}
%\includegraphics[width=0.40\textwidth]{042702.6+260530_Left1.05000Right2.12500Down0.700000Up1.40000_Aver_red.eps}
%
%\vspace{-5mm}
%\includegraphics[width=0.40\textwidth]{042702.6+260530_13CO.eps}\hspace{-10mm}
%\includegraphics[width=0.40\textwidth]{pv_042702.6+260530_13CO_pixdown6arcmin_pixup6arcmin_midpoint_0.eps}

\caption{TMO\_28 (SST 042702.6+260530): \textbf{upper left} and \textbf{upper right}, the same as the upper two panels of Fig. \ref{sample-figure01} except for the following. In the upper left panel the integrated intervals for the blue and red lobe are 3.4 to 5 km s$^{-1}$ and 8 to 13 km s$^{-1}$, respectively. The blue and red contour levels are both 50\%, 60\%,..., 90\% of their peak value. In the upper right panel the position angle is $90^{\circ}$ and contour levels are 0.7 to 1.6 K by 0.3 K, 2 to 4.5 K by 0.5 K. \textbf{middle left} and \textbf{middle right}, the same as the lower two panels of Fig. \ref{sample-figure01}. \textbf{lower left}, $\rm {}^{13}CO$ integrated intensity map overlaid on the $\rm {}^{13}CO$ grey-scale image, the integrated intervals for the blue lobe, red lobe and grey-scale image and the blue and red contour levels are the same with those in the upper left panel of this figure. \textbf{lower right}, the same as the lower right panel of Fig. \ref{sample-figure25} except for that the position angle is $90^{\circ}$ and the contour levels are 0.2 to 2 K by 0.3 K, 2.5 to 4.5 K by 0.5 K.}
\label{sample-figure28}
\end{figure*}

\clearpage
\begin{figure*}%[htbp]
\centering %sub_figures centered
%\includegraphics[width=0.40\textwidth]{042702.8+254222.eps}\hspace{-8mm}
%\includegraphics[width=0.40\textwidth]{pv_042702.8+254222_12CO_pixdown10arcmin_pixup10arcmin_midpoint_110.eps}
%
%\vspace{-5mm}
%
%\includegraphics[width=0.40\textwidth]{042702.8+254222_Left6.25000Right2.75000Down2Up2_Aver_blue.eps}\hspace{-8mm}
%\includegraphics[width=0.40\textwidth]{042702.8+254222_Left4.00000Right2.00000Down4Up1_Aver_red.eps}
%
%\vspace{-5mm}
%
%\includegraphics[width=0.40\textwidth]{042702.8+254222_13CO.eps}\hspace{-12mm}
%\includegraphics[width=0.40\textwidth]{pv_042702.8+254222_13CO_pixdown10arcmin_pixup10arcmin_midpoint_15.eps}

\caption{TMO\_29 (SST 042702.8+254222): The same as Fig. \ref{sample-figure28} except for the following. In the upper left panel the integrated intervals for the blue lobe, red lobe and grey-scale image are 3.5 to 5.5 km s$^{-1}$, 8 to 9.5 km s$^{-1}$ and 6 to 7 km s$^{-1}$, respectively. The blue and red contour levels are 60\%, 70\%,..., 90\% and 30\%, 40\%,..., 90\% of their peak value, respectively. In the upper right panel the position angle is $160^{\circ}$ and the contour levels are 0.4 to 1.6 K by 0.3 K, 2 to 4.5 K by 0.5 K. In the lower left panel the integrated interval for the blue lobe is 4.5 to 5.9 km s$^{-1}$. The blue levels are 40\%, 50\%,..., 90\% of the peak value. In the lower right panel the position angle is $75^{\circ}$ and the contour levels are 0.15 to 0.9 K by 0.15 K, 1.1 to 1.9 K by 0.2 K and 2.2 to 2.8 K by 0.3 K.}
\label{sample-figure29}
\end{figure*}

\clearpage
\begin{figure*}%[htbp]
\centering %sub_figures centered
%\includegraphics[width=0.40\textwidth]{042757.3+261918.eps}\hspace{-12mm}
%\includegraphics[width=0.40\textwidth]{pv_042757.3+261918_12CO_pixdown4arcmin_pixup7arcmin_midpoint_0.eps}
%
%\vspace{-5mm}
%
%\includegraphics[width=0.40\textwidth]{042757.3+261918_Left2.00000Right6.50000Down7.30000Up1.37500_Aver_red.eps}\hspace{-2mm}
%\includegraphics[width=0.40\textwidth]{042757.3+261918_13CO.eps}

\caption{TMO\_30 (SST 042757.3+261918): The same as Fig. \ref{sample-figure13} except for the following. In the upper left panel the integrated intervals for the red lobe and grey-scale image are 8 to 10.5 km s$^{-1}$ and 6 to 7.5 km s$^{-1}$, respectively. The red contour levels are 30\%, 40\%,..., 90\% of the peak value. In the upper right panel the position angle is $90^{\circ}$ and the contour levels are 0.4 to 1.6 K by 0.3 K, 2 to 7 K by 0.5 K.}
\label{sample-figure30}
\end{figure*}

\clearpage
\begin{figure*}%[htbp]
\centering %sub_figures centered
%\includegraphics[width=0.40\textwidth]{042810.4+243553.eps}\hspace{-22mm}
%\includegraphics[width=0.40\textwidth]{pv_042810.4+243553_12CO_pixdown11arcmin_pixup8arcmin_midpoint_110.eps}
%
%\vspace{-5mm}
%
%\includegraphics[width=0.40\textwidth]{042810.4+243553_Left2.62500Right2.62500Down10.5000Up3.90000_Aver_red.eps}\hspace{-6mm}
%\includegraphics[width=0.40\textwidth]{042810.4+243553_13CO.eps}

\caption{TMO\_31 (SST 042810.4+243553): The same as Fig. \ref{sample-figure13} except for the following. In the upper left panel the integrated intervals for the red lobe and grey-scale image are 8 to 9.5 km s$^{-1}$ and 6 to 7 km s$^{-1}$, respectively. The red contour levels are 30\%, 45\%,..., 90\% of the peak value. In the upper right panel the position angle is $160^{\circ}$ and the contour levels are 0.4 to 1.6 K by 0.3 K, 2 to 5 K by 0.5 K.}
\label{sample-figure31}
\end{figure*}

\clearpage
\begin{figure*}%[htbp]
\centering %sub_figures centered
%\includegraphics[width=0.40\textwidth]{043051.7+244147.eps}\hspace{-10mm}
%\includegraphics[width=0.40\textwidth]{pv_043051.7+244147_12CO_pixdown5arcmin_pixup8arcmin_midpoint_30.eps}
%
%\vspace{-5mm}
%
%\includegraphics[width=0.40\textwidth]{043051.7+244147_Left2.72500Right5.82500Down1.10000Up4.50000_Aver_red.eps}\hspace{-8mm}
%\includegraphics[width=0.40\textwidth]{043051.7+244147_13CO.eps}

\caption{TMO\_32 (SST 043051.7+244147): The same as Fig. \ref{sample-figure13} except for the following. In the upper left panel the integrated intervals for the red lobe and grey-scale image are 8 to 11.3 km s$^{-1}$ and 5.5 to 7 km s$^{-1}$, respectively. The red contour levels are 30\%, 45\%,..., 90\% of the peak value. In the upper right panel the position angle is $60^{\circ}$ and the contour levels are 0.4 to 1.6 K by 0.3 K, 2 to 7 K by 0.5 K.}
\label{sample-figure32}
\end{figure*}

\clearpage
\begin{figure*}%[htbp]
\centering %sub_figures centered
%\includegraphics[width=0.40\textwidth]{043215.4+242859.eps}\hspace{-8mm}
%\includegraphics[width=0.40\textwidth]{pv_043215.4+242859_12CO_pixdown8arcmin_pixup14arcmin_midpoint_60.eps}
%
%\vspace{-5mm}
%
%\includegraphics[width=0.40\textwidth]{043215.4+242859_Left4.55000Right3.62500Down1.80000Up2.70000_Aver_blue.eps}\hspace{-8mm}
%\includegraphics[width=0.40\textwidth]{043215.4+242859_Left6.37500Right4.55000Down8.20000Up4.50000_Aver_red.eps}
%
%\vspace{-5mm}
%
%\includegraphics[width=0.40\textwidth]{043215.4+242859_13CO.eps}\hspace{-8mm}
%\includegraphics[width=0.40\textwidth]{pv_043215.4+242859_13CO_pixdown10arcmin_pixup12arcmin_midpoint_60.eps}

\caption{TMO\_33 (SST 043215.4+242859): The same as Fig. \ref{sample-figure28} except for the following. In the upper left panel the integrated intervals for the blue lobe, red lobe and grey-scale image are 3 to 4 km s$^{-1}$, 8 to 10.5 km s$^{-1}$ and 5 to 7 km s$^{-1}$, respectively. In the upper right panel the position angle is $30^{\circ}$ and the contour levels are 0.4 to 1.6 K by 0.3 K, 2 to 5.5 K by 0.5 K. In the lower right panel the contour levels are 0.15 to 0.9 K by 0.15 K, 1.1 to 3.2 K by 0.2 K.}
\label{sample-figure33}
\end{figure*}

\clearpage
\begin{figure*}%[htbp]
\centering %sub_figures centered
%\includegraphics[width=0.40\textwidth]{043307.8+261606.eps}\hspace{-8mm}
%\includegraphics[width=0.40\textwidth]{pv_043307.8+261606_12CO_pixdown8arcmin_pixup12arcmin_midpoint_140.eps}
%
%\vspace{-5mm}
%
%\includegraphics[width=0.40\textwidth]{043307.8+261606_Left3.12500Right1.57500Down1.60000Up1.60000_Aver_blue.eps}\hspace{-6mm}
%\includegraphics[width=0.40\textwidth]{043307.8+261606_13CO.eps}

\caption{TMO\_34 (SST 043307.8+261606): \textbf{upper left} and \textbf{upper right}, the same as the upper two panels of Fig. \ref{sample-figure05} except for the following. In the upper left panel the integrated intervals for the blue lobe and grey-scale image are 2.3 to 3.5 km s$^{-1}$ and 5.5 to 7 km s$^{-1}$, respectively. The blue contour levels are 30\%, 40\%,..., 90\% of the peak value. In the upper right panel the position angle is $130^{\circ}$ and contour levels are 0.4 to 1.6 K by 0.3 K, 2 to 4 K by 0.5 K. \textbf{lower left},  the same as the lower panel of Fig. \ref{sample-figure05}. \textbf{lower right}, the same as the lower left panel of Fig. \ref{sample-figure25}.}
\label{sample-figure34}
\end{figure*}

\clearpage
\begin{figure*}%[htbp]
\centering %sub_figures centered
%\includegraphics[width=0.40\textwidth]{043310.0+243343.eps}\hspace{-10mm}
%\includegraphics[width=0.40\textwidth]{pv_043310.0+243343_12CO_pixdown10arcmin_pixup7arcmin_midpoint_130.eps}
%
%\vspace{-5mm}
%
%\includegraphics[width=0.40\textwidth]{043310.0+243343_Left2.50000Right3.25000Down2Up2_Aver_red.eps}\hspace{-2mm}
%\includegraphics[width=0.40\textwidth]{043310.0+243343_13CO.eps}

\caption{TMO\_35 (SST 043310.0+243343): The same as Fig. \ref{sample-figure13} except for the following. In the upper left panel the integrated intervals for the red lobe and grey-scale image are 8 to 9.5 km s$^{-1}$ and 6 to 7 km s$^{-1}$, respectively. The red contour levels are 50\%, 60\%,..., 90\% of the peak value. In the upper right panel the position angle is $140^{\circ}$ and the contour levels are 0.4 to 1.6 K by 0.3 K, 2 to 5.5 K by 0.5 K.}
\label{sample-figure35}
\end{figure*}

\clearpage
\begin{figure*}%[htbp]
\centering %sub_figures centered
%\includegraphics[width=0.40\textwidth]{043316.5+225320.eps}\hspace{-18mm}
%\includegraphics[width=0.40\textwidth]{pv_043316.5+225320_12CO_pixdown8arcmin_pixup15arcmin_midpoint_90.eps}
%
%\vspace{-5mm}
%
%\includegraphics[width=0.40\textwidth]{043316.5+225320_Left4.00000Right4.75000Down5.50000Up8_Aver_blue.eps}\hspace{-12mm}
%\includegraphics[width=0.40\textwidth]{043316.5+225320_Left5.50000Right2.50000Down3.20000Up3.20000_Aver_red.eps}
%
%\vspace{-5mm}
%
%\includegraphics[width=0.40\textwidth]{043316.5+225320_13CO.eps}\hspace{-18mm}
%\includegraphics[width=0.40\textwidth]{pv_043316.5+225320_13CO_pixdown8arcmin_pixup15arcmin_midpoint_90.eps}

\caption{TMO\_36 (SST 043316.5+225320): The same as Fig. \ref{sample-figure28} except for the following. In the upper left panel the integrated intervals for the blue lobe, red lobe and grey-scale image are 2 to 4 km s$^{-1}$, 7 to 8 km s$^{-1}$ and 4.5 to 6.5 km s$^{-1}$, respectively. In the upper right panel the position angle is $0^{\circ}$ and the contour levels are 0.7 to 1.6 K by 0.3 K, 2 to 4 K by 0.5 K. In the lower right panel the contour levels are 0.3 to 1.2 K by 0.3 K, 1.5 to 6.5 K by 0.5 K.}
\label{sample-figure36}
\end{figure*}

\clearpage
\begin{figure*}%[htbp]
\centering %sub_figures centered
%\includegraphics[width=0.40\textwidth]{043334.0+242117.eps}\hspace{-8mm}
%\includegraphics[width=0.40\textwidth]{pv_043334.0+242117_12CO_pixdown10arcmin_pixup12arcmin_yso_55.eps}
%
%\vspace{-5mm}
%
%\includegraphics[width=0.40\textwidth]{043334.0+242117_Left6.75000Right8.50000Down9Up10_Aver_red.eps}\hspace{-4mm}
%\includegraphics[width=0.40\textwidth]{043334.0+242117_13CO.eps}

\caption{TMO\_37 (SST 043334.0+242117): The same as Fig. \ref{sample-figure13} except for the following. In the upper left panel the integrated intervals for the red lobe and grey-scale image are 8 to 9.5 km s$^{-1}$ and 6 to 7 km s$^{-1}$, respectively. The red contour levels are 30\%, 45\%,..., 90\% of the peak value. In the upper right panel the position angle is $35^{\circ}$ and the contour levels are 0.4 to 1.6 K by 0.3 K, 2 to 5 K by 0.5 K.}
\label{sample-figure37}
\end{figure*}

\clearpage
\begin{figure*}%[htbp]
\centering %sub_figures centered
%\includegraphics[width=0.40\textwidth]{043336.7+260949.eps}\hspace{-10mm}
%\includegraphics[width=0.40\textwidth]{pv_043336.7+260949_12CO_pixdown10arcmin_pixup10arcmin_midpoint_20.eps}
%
%\vspace{-5mm}
%
%\includegraphics[width=0.40\textwidth]{043336.7+260949_Left3.75000Right9.25000Down3.10000Up4.60000_Aver_blue.eps}\hspace{-6mm}
%\includegraphics[width=0.40\textwidth]{043336.7+260949_13CO.eps}

\caption{TMO\_38 (SST 043336.7+260949): The same as Fig. \ref{sample-figure34} except for the following. In the upper left panel the integrated intervals for the blue lobe and grey-scale image are 3.0 to 4.1 km s$^{-1}$ and 5.5 to 6.5 km s$^{-1}$, respectively. The blue contour levels are 30\%, 45\%,..., 90\% of the peak value. In the upper right panel the position angle is $70^{\circ}$.}
\label{sample-figure38}
\end{figure*}

\clearpage
\begin{figure*}%[htbp]
\centering %sub_figures centered
%\includegraphics[width=0.40\textwidth]{043557.6+225357.eps}\hspace{-10mm}
%\includegraphics[width=0.40\textwidth]{pv_043557.6+225357_12CO_pixdown5arcmin_pixup5arcmin_yso_0.eps}
%
%\vspace{-5mm}
%
%\includegraphics[width=0.40\textwidth]{043557.6+225357_Left3.67500Right3.67500Down1.80000Up2.80000_Aver_blue.eps}\hspace{-10mm}
%\includegraphics[width=0.40\textwidth]{043557.6+225357_Left2.77500Right3.67500Down1.80000Up2.80000_Aver_red.eps}
%
%\vspace{-5mm}
%
%\includegraphics[width=0.40\textwidth]{043557.6+225357_13CO.eps}\hspace{-10mm}
%\includegraphics[width=0.40\textwidth]{pv_043557.6+225357_13CO_pixdown5arcmin_pixup5arcmin_yso_0.eps}

\caption{TMO\_39 (SST 043557.6+225357): The same as Fig. \ref{sample-figure28} except for the following. In the upper left panel the integrated intervals for the blue lobe, red lobe and grey-scale image are 2 to 4 km s$^{-1}$, 7 to 8 km s$^{-1}$ and 5 to 6 km s$^{-1}$, respectively. The blue and red contour levels are 35\%, 50\%,..., 95\% and 60\%, 70\%,..., 90\% of their peak value, respectively. In the upper right panel the contour levels are 0.7 to 1.6 K by 0.3 K, 2 to 7 K by 0.5 K. In the lower right panel the contour levels are 0.15 to 0.9 K by 0.15 K, 1.1 to 1.9 K by 0.2 K and 2.2 to 4 K by 0.3 K.}
\label{sample-figure39}
\end{figure*}

\clearpage
\begin{figure*}%[htbp]
\centering %sub_figures centered
%\includegraphics[width=0.40\textwidth]{043911.2+252710.eps}\hspace{-12mm}
%\includegraphics[width=0.40\textwidth]{pv_043911.2+252710_12CO_pixdown6arcmin_pixup6arcmin_yso_0.eps}
%
%\vspace{-5mm}
%
%\includegraphics[width=0.40\textwidth]{043911.2+252710_Left2.50000Right1.75000Down2.50000Up1.80000_Aver_red.eps}\hspace{-4mm}
%\includegraphics[width=0.40\textwidth]{043911.2+252710_13CO.eps}\hspace{-18mm}

\caption{TMO\_40: The same as Fig. \ref{sample-figure13} except for the following. In the upper left panel the integrated intervals for the red lobe and grey-scale image are 8.5 to 11 km s$^{-1}$ and 4.5 to 7 km s$^{-1}$, respectively. The red contour levels are 30\%, 40\%,..., 90\% of the peak value. In the upper right panel the position angle is $90^{\circ}$ and the contour levels are 0.7 to 1.6 K by 0.3 K, 2 to 5 K by 0.5 K.}
\label{sample-figure40}
\end{figure*}

\clearpage
\begin{figure*}%[htbp]
\centering %sub_figures centered
%\includegraphics[width=0.40\textwidth]{043913.8+255320.eps}\hspace{-12mm}
%\includegraphics[width=0.40\textwidth]{pv_043913.8+255320_12CO_pixdown12arcmin_pixup10arcmin_midpoint_38.eps}
%
%\vspace{-5mm}
%
%\includegraphics[width=0.40\textwidth]{043913.8+255320_Left1.72500Right1.72500Down2.60000Up1.80000_Aver_blue.eps}\hspace{-12mm}
%\includegraphics[width=0.40\textwidth]{043913.8+255320_Left2.57500Right2.57500Down6Up8.60000_Aver_red.eps}
%
%\vspace{-5mm}
%
%\includegraphics[width=0.40\textwidth]{043913.8+255320_13CO.eps}\hspace{-12mm}
%\includegraphics[width=0.40\textwidth]{pv_043913.8+255320_13CO_pixdown12arcmin_pixup10arcmin_midpoint_38.eps}

\caption{TMO\_41 (SST 043913.8+255320): The same as Fig. \ref{sample-figure28} except for the following. In the upper left panel the integrated intervals for the blue lobe and grey-scale image are 1.3 to 4.5 km s$^{-1}$ and 5 to 6 km s$^{-1}$, respectively. The blue and red contour levels are 45\%, 60\%,..., 90\% and 30\%, 45\%,..., 90\% of their peak value, respectively. In the upper right panel the position angle is $52^{\circ}$ and the contour levels are 0.4 to 1.6 K by 0.3 K, 2 to 5 K by 0.5 K. In the lower left panel the blue and red contour levels are 60\%, 75\%, 90\% and 45\%, 60\%,..., 90\% of their peak value, respectively. In the lower right panel the contour levels are 0.1 to 1.6 K by 0.3 K, 2 to 5 K by 0.5 K.}
\label{sample-figure41}
\end{figure*}

\clearpage
\begin{figure*}%[htbp]
\centering %sub_figures centered
%\includegraphics[width=0.40\textwidth]{044802.3+253359.eps}\hspace{-18mm}
%\includegraphics[width=0.40\textwidth]{pv_044802.3+253359_12CO_pixdown8arcmin_pixup8arcmin_midpoint_170.eps}
%
%\vspace{-5mm}
%
%\includegraphics[width=0.40\textwidth]{044802.3+253359_Left3.32500Right2.00000Down8.60000Up6.60000_Aver_blue.eps}\hspace{-12mm}
%\includegraphics[width=0.40\textwidth]{044802.3+253359_Left1.32500Right2.00000Down1.30000Up1.30000_Aver_red.eps}
%
%\vspace{-5mm}
%
%\includegraphics[width=0.40\textwidth]{044802.3+253359_13CO.eps}\hspace{-18mm}
%\includegraphics[width=0.40\textwidth]{pv_044802.3+253359_13CO_pixdown8arcmin_pixup8arcmin_midpoint_170.eps}

\caption{TMO\_42 (SST 044802.3+253359): The same as Fig. \ref{sample-figure28} except for the following. In the upper left panel the integrated intervals for the blue lobe, red lobe and grey-scale image are 2.8 to 3.5 km s$^{-1}$, 8 to 9.5 km s$^{-1}$ and 5 to 7 km s$^{-1}$, respectively. The blue and red contour levels are both 30\%, 45\%,..., 90\% of their peak value. In the upper right panel the position angle is $100^{\circ}$ and the contour levels are 0.4 to 1.6 K by 0.3 K, 2 to 5 K by 0.5 K. In the lower right panel the contour levels are 0.1 to 1.6 K by 0.3 K, 2 to 6.5 K by 0.5 K.}
\label{sample-figure42}
\end{figure*}

\clearpage
\begin{figure*}%[htbp]
\centering %sub_figures centered
%\includegraphics[width=0.40\textwidth]{041851.4+282026.eps}\hspace{-10mm}
%\includegraphics[width=0.40\textwidth]{pv_041851.4+282026_12CO_pixdown8arcmin_pixup8arcmin_yso_135.eps}
%
%\vspace{-5mm}
%
%\includegraphics[width=0.40\textwidth]{041851.4+282026_Left4.72500Right4.15000Down4.10000Up4.70000_Aver_blue.eps}

\caption{TMO\_43 (SST 041851.4+282026): The same as Fig. \ref{sample-figure05} except for the following. In the upper left panel the integrated intervals for the blue lobe and grey-scale image are 4 to 5 km s$^{-1}$ and 6.5 to 8 km s$^{-1}$, respectively. The blue contour levels are 40\%, 50\%,..., 90\% of the peak value. In the upper right panel the position angle is $135^{\circ}$ and contour levels are 0.4 to 1.6 K by 0.3 K, 2 to 6.5 K by 0.5 K.}
\label{sample-figure43}
\end{figure*}

\clearpage
\begin{figure*}%[htbp]
\centering %sub_figures centered
%\includegraphics[width=0.40\textwidth]{042021.4+281349.eps}\hspace{-8mm}
%\includegraphics[width=0.40\textwidth]{pv_042021.4+281349_12CO_pixdown19arcmin_pixup25arcmin_midpoint_105.eps}
%
%\vspace{-5mm}
%
%\includegraphics[width=0.40\textwidth]{042021.4+281349_Left9.27500Right5.72500Down6.40000Up7.90000_Aver_blue.eps}\hspace{-12mm}
%\includegraphics[width=0.40\textwidth]{042021.4+281349_Left2.85000Right5.72500Down4.30000Up8.60000_Aver_red.eps}

\caption{TMO\_44 (SST 042021.4+281349): The same as Fig. \ref{sample-figure01} except for the following. In the upper left panel the integrated intervals for the blue lobe,red lobe and grey-scale image are 4.5 to 5.5 km s$^{-1}$, 8.3 to 8.8 km s$^{-1}$ and 6.5 to 7.5 km s$^{-1}$, respectively. The blue and red contour levels are both 30\%, 40\%,..., 90\% of their peak value. In the upper right panel the position angle is $165^{\circ}$ and contour levels are 0.4 to 5.5 K by 0.3 K.}
\label{sample-figure44}
\end{figure*}

\clearpage
\begin{figure*}%[htbp]
\centering %sub_figures centered
%\includegraphics[width=0.40\textwidth]{042653.3+255858.eps}\hspace{-8mm}
%\includegraphics[width=0.40\textwidth]{pv_042653.3+255858_12CO_pixdown10arcmin_pixup10arcmin_midpoint_30.eps}
%
%\vspace{-5mm}
%
%\includegraphics[width=0.40\textwidth]{042653.3+255858_Left1.75000Right0.500000Down2Up2_Aver_blue.eps}\hspace{-10mm}
%\includegraphics[width=0.40\textwidth]{042653.3+255858_Left0.250000Right2.50000Down1Up1_Aver_red.eps}

\caption{TMO\_45 (SST 042653.3+255858):  The same as Fig. \ref{sample-figure01} except for the following. In the upper left panel the integrated intervals for the blue lobe,red lobe and grey-scale image are 2 to 5.5 km s$^{-1}$, 7.5 to 10 km s$^{-1}$ and 6 to 7 km s$^{-1}$, respectively. The blue and red contour levels are both 60\%, 70\%,..., 90\% of their peak value. In the upper right panel the position angle is $60^{\circ}$ and contour levels are 0.4 to 1.6 K by 0.3 K, 2 to 3.5 K by 0.5 K.}
\label{sample-figure45}
\end{figure*}

\clearpage
\begin{figure*}%[htbp]
\centering %sub_figures centered
%\includegraphics[width=0.40\textwidth]{043956.1+262802.eps}\hspace{-25mm}
%\includegraphics[width=0.40\textwidth]{pv_043956.1+262802_12CO_pixdown12arcmin_pixup6arcmin_midpoint_70.eps}
%
%\vspace{-5mm}
%
%\includegraphics[width=0.40\textwidth]{043956.1+262802_Left2.45000Right3.27500Down9.80000Up3.30000_Aver_blue.eps}\hspace{-12mm}
%\includegraphics[width=0.40\textwidth]{043956.1+262802_Left0.825000Right4.07500Down13.1000Up4.90000_Aver_red.eps}

\caption{TMO\_46: The same as Fig. \ref{sample-figure01} except for the following. In the upper left panel the integrated intervals for the blue lobe,red lobe and grey-scale image are 1 to 4 km s$^{-1}$, 7.5 to 9 km s$^{-1}$ and 4.5 to 7 km s$^{-1}$, respectively. The blue and red contour levels are 30\%, 45\%,..., 90\% and 60\%, 70\%,..., 90\% of their peak value, respectively. In the upper right panel the position angle is $20^{\circ}$ and contour levels are 0.4 to 1.6 K by 0.3 K, 2 to 4.5 K by 0.5 K.}
\label{sample-figure46}
\end{figure*}

\clearpage
\begin{figure*}%[htbp]
\centering %sub_figures centered
%\includegraphics[width=0.40\textwidth]{043535.3+240819.eps}\hspace{-18mm}
%\includegraphics[width=0.40\textwidth]{pv_043535.3+240819_12CO_pixdown15arcmin_pixup10arcmin_yso_90.eps}
%
%\vspace{-5mm}
%
%\includegraphics[width=0.40\textwidth]{043535.3+240819_Left2.50000Right5.75000Down2.60000Up0.700000_Aver_blue.eps}\hspace{-14mm}
%\includegraphics[width=0.40\textwidth]{043535.3+240819_Left2.00000Right3.25000Down8.50000Up1.30000_Aver_red.eps}
%
%\vspace{-5mm}
%
%\includegraphics[width=0.40\textwidth]{043535.3+240819_13CO.eps}\hspace{-18mm}
%\includegraphics[width=0.40\textwidth]{pv_043535.3+240819_13CO_pixdown15arcmin_pixup10arcmin_yso_90.eps}

\caption{TMO\_47 (SST 043535.3+240819): The same as Fig. \ref{sample-figure28} except for the following. In the upper left panel the integrated intervals for the blue lobe, red lobe and grey-scale image are 4 to 4.8 km s$^{-1}$, 7 to 8.5 km s$^{-1}$ and 5 to 7 km s$^{-1}$, respectively. The red contour levels are 30\%, 45\%,..., 90\% of the peak value. In the upper right panel the position angle is $0^{\circ}$ and the contour levels are 0.7 to 1.6 K by 0.3 K, 2 to 6 K by 0.5 K. In the lower right panel the contour levels are 0.1 to 1.6 K by 0.3 K, 2 to 4 K by 0.5 K.}
\label{sample-figure47}
\end{figure*}

\clearpage
\begin{figure*}%[htbp]
\centering %sub_figures centered
%\includegraphics[width=0.40\textwidth]{041535.6+284741.eps}\hspace{-10mm}
%\includegraphics[width=0.40\textwidth]{pv_041535.6+284741_12CO_pixdown6arcmin_pixup6arcmin_midpoint_90_amarath.eps}
%
%\vspace{-5mm}
%
%\includegraphics[width=0.40\textwidth]{041535.6+284741_Left0.750000Right1.25000Down1Up2_Aver_blue.eps}\hspace{-10mm}
%\includegraphics[width=0.40\textwidth]{041535.6+284741_Left0.500000Right1.25000Down1Up1_Aver_red.eps}

\caption{TMO\_48 (SST 041535.6+284741): The same as Fig. \ref{sample-figure01} except for the following. In the upper left panel the integrated intervals for the blue lobe,red lobe and grey-scale image are -1 to 4 km s$^{-1}$, 9.25 to 13 km s$^{-1}$ and 6.5 to 8.5 km s$^{-1}$, respectively. The red contour levels are 40\%, 50\%,..., 90\% of the peak value. In the upper right panel the position angle is $0^{\circ}$ and contour levels are 0.4 to 1.6 K by 0.3 K, 2 to 6.5 K by 0.5 K.}
\label{sample-figure48}
\end{figure*}

\clearpage
\begin{figure*}%[htbp]
\centering %sub_figures centered
%\includegraphics[width=0.40\textwidth]{041733.7+282046.eps}\hspace{-14mm}
%\includegraphics[width=0.40\textwidth]{pv_041733.7+282046_12CO_pixdown8arcmin_pixup7arcmin_midpoint_67.eps}
%
%\vspace{-5mm}
%
%\includegraphics[width=0.40\textwidth]{041733.7+282046_Left4.27500Right2.85000Down9.30000Up7.90000_Aver_red.eps}

\caption{TMO\_49 (SST 041733.7+282046): The same as Fig. \ref{sample-figure02} except for the following. In the upper left panel the integrated interval for the red lobe is 9.5 to 13 km s$^{-1}$. In the upper right panel the position angle is $23^{\circ}$ and contour levels are 0.4 to 1.6 K by 0.3 K, 2 to 7 K by 0.5 K.}
\label{sample-figure49}
\end{figure*}

\clearpage
\begin{figure*}%[htbp]
\centering %sub_figures centered
%\includegraphics[width=0.40\textwidth]{041810.5+284447.eps}\hspace{-18mm}
%\includegraphics[width=0.40\textwidth]{pv_041810.5+284447_12CO_pixdown6arcmin_pixup10arcmin_yso_60.eps}
%
%\vspace{-5mm}
%
%\includegraphics[width=0.40\textwidth]{041810.5+284447_Left2.02500Right2.02500Down7.50000Up5.40000_Aver_red.eps}

\caption{TMO\_50 (SST 041810.5+284447): The same as Fig. \ref{sample-figure02} except for the following. In the upper left panel the integrated intervals for the red lobe and grey-scale image are 9.3 to 10 km s$^{-1}$ and 6.5 to 8.5 km s$^{-1}$, respectively. The red contour levels are 50\%, 60\%,..., 90\% of the peak value. In the upper right panel the position angle is $30^{\circ}$ and contour levels are 0.4 to 1.6 K by 0.3 K, 2 to 5 K by 0.5 K.}
\label{sample-figure50}
\end{figure*}

\clearpage
\begin{figure*}%[htbp]
\centering %sub_figures centered
%\includegraphics[width=0.40\textwidth]{041831.1+281629.eps}\hspace{-10mm}
%\includegraphics[width=0.40\textwidth]{pv_041831.1+281629_12CO_pixdown3arcmin_pixup4arcmin_yso_10.eps}
%
%\vspace{-5mm}
%
%\includegraphics[width=0.40\textwidth]{041831.1+281629_Left1.00000Right0.700000Down1.10000Up0.800000_Aver_red.eps}

\caption{TMO\_51 (SST 041831.1+281629): The same as Fig. \ref{sample-figure02} except for the following. In the upper left panel the integrated intervals for the red lobe and grey-scale image are 9.3 to 10.2 km s$^{-1}$ and 6.5 to 8 km s$^{-1}$, respectively. The red contour levels are 30\%, 40\%,..., 90\% of the peak value. In the upper right panel the position angle is $80^{\circ}$ and contour levels are 0.5 to 2 K by 0.3 K, 2.5 to 6 K by 0.5 K.}
\label{sample-figure51}
\end{figure*}

\clearpage
\begin{figure*}%[htbp]
\centering %sub_figures centered
%\includegraphics[width=0.40\textwidth]{041831.2+282617.eps}\hspace{-14mm}
%\includegraphics[width=0.40\textwidth]{pv_041831.2+282617_12CO_pixdown5arcmin_pixup2arcmin_yso_30.eps}
%
%\vspace{-5mm}
%
%\includegraphics[width=0.40\textwidth]{041831.2+282617_Left1.90000Right0.250000Down3.25000Up0.0500000_Aver_red.eps}

\caption{TMO\_52 (SST 041831.2+282617): The same as Fig. \ref{sample-figure02} except for the following. In the upper left panel the integrated intervals for the red lobe and grey-scale image are 9.3 to 9.8 km s$^{-1}$ and 6 to 8 km s$^{-1}$, respectively. The red contour levels are 30\%, 40\%,..., 90\% of the peak value. In the upper right panel the position angle is $60^{\circ}$ and contour levels are 0.4 to 1.6 K by 0.3 K, 2 to 6 K by 0.5 K.}
\label{sample-figure52}
\end{figure*}

\clearpage
\begin{figure*}%[htbp]
\centering %sub_figures centered
%\includegraphics[width=0.40\textwidth]{041841.3+282725.eps}\hspace{-32mm}
%\includegraphics[width=0.40\textwidth]{pv_041841.3+282725_12CO_pixdown5arcmin_pixup13arcmin_yso_106.eps}
%
%\vspace{-5mm}
%
%\includegraphics[width=0.40\textwidth]{041841.3+282725_Left0.800000Right0.800000Down6Up4_Aver_red.eps}

\caption{TMO\_53 (SST 041841.3+282725): The same as Fig. \ref{sample-figure02} except for the following. In the upper left panel the integrated intervals for the red lobe and grey-scale image are 9.3 to 9.9 km s$^{-1}$ and 6 to 8 km s$^{-1}$, respectively. The red contour levels are 60\%, 70\%,..., 90\% of the peak value. In the upper right panel the position angle is $164^{\circ}$ and contour levels are 0.4 to 1.6 K by 0.3 K, 2 to 6 K by 0.5 K.}
\label{sample-figure53}
\end{figure*}

\clearpage
\begin{figure*}%[htbp]
\centering %sub_figures centered
%\includegraphics[width=0.40\textwidth]{042154.5+265231.eps}\hspace{-14mm}
%\includegraphics[width=0.40\textwidth]{pv_042154.5+265231_12CO_pixdown9arcmin_pixup5arcmin_midpoint_90.eps}
%
%\vspace{-5mm}
%
%\includegraphics[width=0.40\textwidth]{042154.5+265231_Left2.50000Right2.50000Down2Up3.40000_Aver_blue.eps}\hspace{-12mm}
%\includegraphics[width=0.40\textwidth]{042154.5+265231_Left3.50000Right4.25000Down2Up3.40000_Aver_red.eps}

\caption{TMO\_54 (SST 042154.5+265231): The same as Fig. \ref{sample-figure01} except for the following. In the upper left panel the integrated intervals for the blue lobe,red lobe and grey-scale image are 3.8 to 4.5 km s$^{-1}$, 7.5 to 8.3 km s$^{-1}$ and 5.5 to 7 km s$^{-1}$, respectively. The blue and red contour levels are 30\%, 45\%,..., 90\% and 50\%, 60\%,..., 90\% of their peak value, respectively. In the upper right panel the position angle is $0^{\circ}$ and contour levels are 0.5 to 2 K by 0.3 K, 2.5 to 4 K by 0.5 K.}
\label{sample-figure54}
\end{figure*}

\clearpage
\begin{figure*}%[htbp]
\centering %sub_figures centered
%\includegraphics[width=0.40\textwidth]{042904.9+264907.eps}\hspace{-10mm}
%\includegraphics[width=0.40\textwidth]{pv_042904.9+264907_12CO_pixdown8arcmin_pixup4arcmin_midpoint_40.eps}
%
%\vspace{-5mm}
%
%\includegraphics[width=0.40\textwidth]{042904.9+264907_Left2.40000Right1.80000Down9.60000Up1.80000_Aver_red.eps}

\caption{TMO\_55 (SST 042904.9+264907): The same as Fig. \ref{sample-figure02} except for the following. In the upper left panel the integrated intervals for the red lobe and grey-scale image are 8.3 to 9 km s$^{-1}$ and 5.5 to 7 km s$^{-1}$, respectively. In the upper right panel the position angle is $50^{\circ}$ and contour levels are 0.4 to 4.2 K by 0.3 K.}
\label{sample-figure55}
\end{figure*}
%
%
%
%
%%%%%%%%%%%%%%%%%%%%%%%%%%%%%%%%%
%%%%%%%%%%%%%%%%%%%%%%%%%%%%%%%%%
%%%%%%%%%%%%%%%%%%%%%%%%%%%%%%%%%
%%%%%%%%%%%%%%%%%%%%%%%%%%%%%%%%%
%%%%%%%%%%%%%%%%%%%%%%%%%%%%%%%%%
%%%%%%%%%%%%%%%%%%%%%%%%%%%%%%%%%
%
%
%
%
\begin{figure*}%[htbp]
\centering %sub_figures centered
\includegraphics[width=0.325\textwidth]{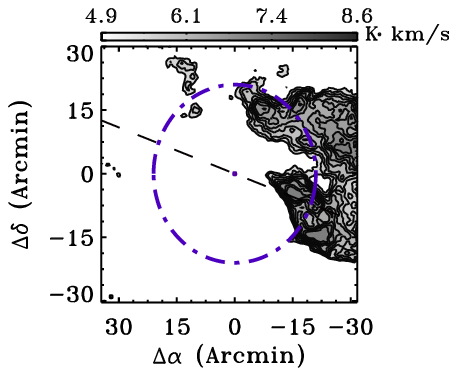}\hspace{-6mm}
\includegraphics[width=0.261\textwidth]{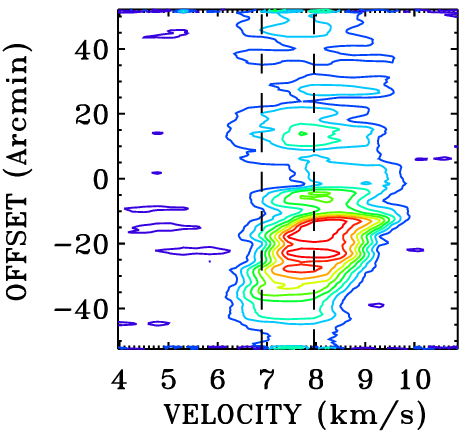}\hspace{-6mm}
\includegraphics[width=0.275\textwidth]{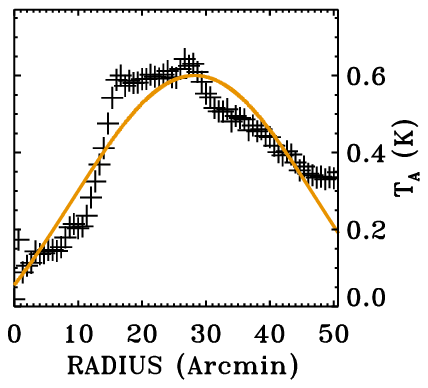}

\vspace{-10mm}

\includegraphics[width=.9\textwidth]{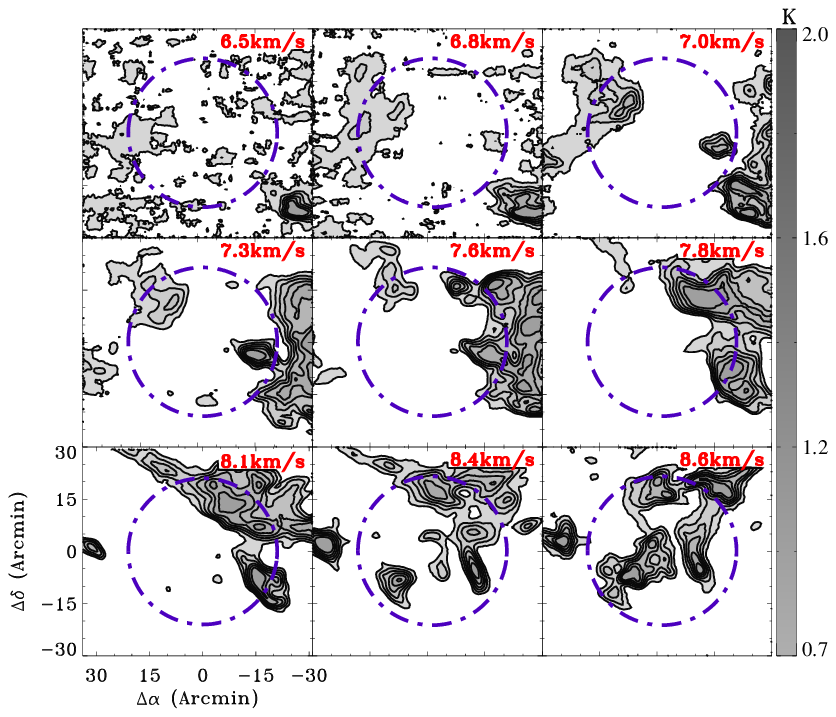}\hspace{15mm}

\vspace{-20mm}

\caption{TMB\_01: \textbf{upper left}, $\rm {}^{13}CO$ integrated intensity map, integrated over 7.3 to 8.4 km s$^{-1}$. The starting contour and the contour step are 3.3 K km s$^{-1}$ and 0.6 K km s$^{-1}$, respectively. The black dashed line represents a cut for the P-V diagram shown in the upper middle panel, and the amaranth open circle indicates the origin on the Y-axis of the upper middle panel. The purple dashed circle shows the approximate extent of the CO emission of the bubble. The black open circle on the lower left corner marks the beam. \textbf{upper middle}, P-V diagram of $\rm {}^{13}CO$, through the slice shown by the black dashed line in the upper left panel at a position angle of $110^{\circ}$. The contour levels are 0.1 to 1.2 K by 0.1 K. The two black dashed vertical lines show the expansion velocity interval determined by visual inspection. \textbf{upper right}, gaussian fit to the azimuthally averaged profile of $\rm {}^{13}CO$ intensity. \textbf{lower}, Channel maps of $^{13}$CO emission, the number on the upper right corner of each panel indicates the central velocity of the channel map. The purple dashed circles are the same as that in the upper left panel.
\label{Bubble_fig1.}}
\end{figure*}

\begin{figure*}%[htbp]
\centering %sub_figures centered
%\includegraphics[width=0.325\textwidth]{contour041428+274553_13CO.eps}\hspace{-6mm}
%\includegraphics[width=0.261\textwidth]{1764_1098_10_13CO.eps}\hspace{-6mm}
%\includegraphics[width=0.275\textwidth]{1764_1098_47_42_45_13CO.eps}
%\vspace{-10mm}
%
%\includegraphics[width=.9\textwidth]{channel041428+274553_13CO.eps}\hspace{15mm}
\caption{TMB\_02: The same as Fig. \ref{Bubble_fig1.} except for the following. In the upper left panel the integrated velocity interval is from 5.4 to 6.5 km s$^{-1}$, the starting contour and the contour step are 2.7 K km s$^{-1}$ and 0.5 K km s$^{-1}$, respectively. In the upper middle panel the position angle is $80^{\circ}$ and the contour levels are 0.1 to 1.0 K by 0.1 K.
\label{Bubble_fig2.}}
\end{figure*}

\begin{figure*}%[htbp]
\centering %sub_figures centered
%\includegraphics[width=0.325\textwidth]{contour041620+282853_13CO.eps}\hspace{-6mm}
%\includegraphics[width=0.261\textwidth]{1685_1227_110_13CO.eps}\hspace{-6mm}
%\includegraphics[width=0.275\textwidth]{1685_1227_59_40_57_13CO.eps}
%\vspace{-10mm}
%
%\includegraphics[width=.9\textwidth]{channel041620+282853_13CO.eps}\hspace{15mm}
\caption{TMB\_03: The same as Fig. \ref{Bubble_fig1.} except for the following. In the upper left panel the integrated velocity interval is from 4.6 to 6.2 km s$^{-1}$ and the contour step is 0.9 K km s$^{-1}$. In the upper middle panel the position angle is $160^{\circ}$ and the contour levels are 0.1 to 1.7 K by 0.2 K.
\label{Bubble_fig3.}}
\end{figure*}

\clearpage
\begin{figure*}%[htbp]
\centering %sub_figures centered
%\includegraphics[width=0.325\textwidth]{contour041905+273333_13CO.eps}\hspace{-6mm}
%\includegraphics[width=0.261\textwidth]{1581_1061_70_13CO.eps}\hspace{-6mm}
%\includegraphics[width=0.275\textwidth]{1581_1061_48_44_46_13CO.eps}
%\vspace{-10mm}
%
%\includegraphics[width=.9\textwidth]{channel041905+273333_13CO.eps}\hspace{15mm}
\caption{TMB\_04: The same as Fig. \ref{Bubble_fig1.} except for the following. In the upper left panel the integrated velocity interval is from 6.0 to 7.0 km s$^{-1}$ and the contour step is 0.8 K km s$^{-1}$. In the upper middle panel the position angle is $20^{\circ}$ and the contour levels are 0.1 to 2.2 K by 0.2 K.
\label{Bubble_fig4.}}
\end{figure*}

\begin{figure*}%[htbp]
\centering %sub_figures centered
%\includegraphics[width=0.325\textwidth]{contour042112+265533_13CO.eps}\hspace{-6mm}
%\includegraphics[width=0.261\textwidth]{1499_947_80_13CO.eps}\hspace{-6mm}
%\includegraphics[width=0.275\textwidth]{1499_947_44_40_42_13CO.eps}
%\vspace{-10mm}
%
%\includegraphics[width=.9\textwidth]{channel042112+265533_13CO.eps}\hspace{15mm}
\caption{TMB\_05: The same as Fig. \ref{Bubble_fig1.} except for the following. In the upper left panel the integrated velocity interval is from 4.4 to 6.0 km s$^{-1}$, the starting contour and the contour step is 2.7 K km s$^{-1}$ and 1 K km s$^{-1}$, respectively. In the upper middle panel the position angle is $10^{\circ}$ and the contour levels are 0.1 to 1.9 K by 0.2 K.
\label{Bubble_fig5.}}
\end{figure*}

\begin{figure*}%[htbp]
\centering %sub_figures centered
%\includegraphics[width=0.325\textwidth]{contour042517+253213_13CO.eps}\hspace{-6mm}
%\includegraphics[width=0.261\textwidth]{1339_697_160_13CO.eps}\hspace{-6mm}
%\includegraphics[width=0.275\textwidth]{1339_697_61_41_59_13CO.eps}
%\vspace{-10mm}
%
%\includegraphics[width=.9\textwidth]{channel042517+253213_13CO.eps}\hspace{15mm}
\caption{TMB\_06: The same as Fig. \ref{Bubble_fig1.} except for the following. In the upper left panel the integrated velocity interval is from 5.2 to 6.2 km s$^{-1}$, the starting contour and the contour step are 2.1 K km s$^{-1}$ and 0.3 K km s$^{-1}$, respectively.
\label{Bubble_fig6.}}
\end{figure*}

\begin{figure*}%[htbp]
\centering %sub_figures centered
%\includegraphics[width=0.325\textwidth]{contour042529+261013_13CO.eps}\hspace{-6mm}
%\includegraphics[width=0.261\textwidth]{1329_811_80_13CO.eps}\hspace{-6mm}
%\includegraphics[width=0.275\textwidth]{1329_811_88_44_86_13CO.eps}
%\vspace{-10mm}
%
%\includegraphics[width=.9\textwidth]{channel042529+261013_13CO.eps}\hspace{15mm}
\caption{TMB\_07: The same as Fig. \ref{Bubble_fig1.} except for the following. In the upper left panel the integrated velocity interval is from 5.7 to 7.8 km s$^{-1}$, the starting contour and the contour step are 4.7 K km s$^{-1}$ and 0.9 K km s$^{-1}$, respectively. In the upper middle panel the position angle is $10^{\circ}$ and the contour levels are 0.1 to 1.1 K by 0.1 K.
\label{Bubble_fig7.}}
\end{figure*}

\begin{figure*}%[htbp]
\centering %sub_figures centered
%\includegraphics[width=0.325\textwidth]{contour042707+242013_13CO.eps}\hspace{-6mm}
%\includegraphics[width=0.261\textwidth]{1267_481_80_13CO.eps}\hspace{-6mm}
%\includegraphics[width=0.275\textwidth]{1267_481_55_44_53_13CO.eps}
%\vspace{-10mm}
%
%\includegraphics[width=.9\textwidth]{channel042707+242013_13CO.eps}\hspace{15mm}
\caption{TMB\_08: The same as Fig. \ref{Bubble_fig1.} except for the following. In the upper left panel the integrated velocity interval is from 5.7 to 7.8 km s$^{-1}$ and the contour step is 1 K km s$^{-1}$. In the upper middle panel the position angle is $10^{\circ}$ and the contour levels are 0.1 to 1.1 K by 0.1 K.
\label{Bubble_fig8.}}
\end{figure*}

\begin{figure*}%[htbp]
\centering %sub_figures centered
%\includegraphics[width=0.325\textwidth]{contour042731+261653_13CO.eps}\hspace{-6mm}
%\includegraphics[width=0.261\textwidth]{1247_831_10_13CO.eps}\hspace{-6mm}
%\includegraphics[width=0.275\textwidth]{1247_831_39_44_37_13CO.eps}
%\vspace{-10mm}
%
%\includegraphics[width=.9\textwidth]{channel042731+261653_13CO.eps}\hspace{15mm}
\caption{TMB\_09: The same as Fig. \ref{Bubble_fig1.} except for the following. In the upper left panel the integrated velocity interval is from 6.2 to 7.0 km s$^{-1}$, the starting contour and the contour step are 3.4 K km s$^{-1}$ and 0.7 K km s$^{-1}$, respectively. In the upper middle panel the position angle is $80^{\circ}$.
\label{Bubble_fig9.}}
\end{figure*}

\begin{figure*}%[htbp]
\centering %sub_figures centered
%\includegraphics[width=0.325\textwidth]{contour042852+241433_13CO.eps}\hspace{-6mm}
%\includegraphics[width=0.261\textwidth]{1195_464_160_13CO.eps}\hspace{-6mm}
%\includegraphics[width=0.275\textwidth]{1195_464_124_39_122_13CO.eps}
%\vspace{-10mm}
%
%\includegraphics[width=.9\textwidth]{channel042852+241433_13CO.eps}\hspace{15mm}
\caption{TMB\_10: The same as Fig. \ref{Bubble_fig1.} except for the following. In the upper left panel the integrated velocity interval is from 4.4 to 5.7 km s$^{-1}$, the starting contour and the contour step are 2.3 K km s$^{-1}$ and 0.8 K km s$^{-1}$, respectively. In the upper middle panel the contour levels are 0.1 to 1.8 K by 0.2 K.
\label{Bubble_fig10.}}
\end{figure*}

\begin{figure*}%[htbp]
\centering %sub_figures centered
%\includegraphics[width=0.325\textwidth]{contour042932+263233_13CO.eps}\hspace{-6mm}
%\includegraphics[width=0.261\textwidth]{1165_878_50_13CO.eps}\hspace{-6mm}
%\includegraphics[width=0.275\textwidth]{1165_878_66_47_64_13CO.eps}
%\vspace{-10mm}
%
%\includegraphics[width=.9\textwidth]{channel042932+263233_13CO.eps}\hspace{15mm}
\caption{TMB\_11: The same as Fig. \ref{Bubble_fig1.} except for the following. In the upper left panel the integrated velocity interval is from 6.2 to 8.1 km s$^{-1}$, the starting contour and the contour step are 4.8 K km s$^{-1}$ and 0.9 K km s$^{-1}$, respectively. In the upper middle panel the position angle is $40^{\circ}$ and the contour levels are 0.1 to 1.6 K by 0.2 K.
\label{Bubble_fig11.}}
\end{figure*}

\begin{figure*}%[htbp]
\centering %sub_figures centered
%\includegraphics[width=0.325\textwidth]{contour042944+263253_13CO.eps}\hspace{-6mm}
%\includegraphics[width=0.261\textwidth]{1157_879_110_13CO.eps}\hspace{-6mm}
%\includegraphics[width=0.275\textwidth]{1157_879_55_45_41_13CO.eps}
%\vspace{-10mm}
%
%\includegraphics[width=.9\textwidth]{channel042944+263253_13CO.eps}\hspace{15mm}
\caption{TMB\_12: The same as Fig. \ref{Bubble_fig1.} except for the following. In the upper left panel the integrated velocity interval is from 5.4 to 7.8 km s$^{-1}$, the starting contour and the contour step are 6.2 K km s$^{-1}$ and 0.9 K km s$^{-1}$, respectively. In the upper middle panel the position angle is $160^{\circ}$ and the contour levels are 0.1 to 1.5 K by 0.2 K.
\label{Bubble_fig12.}}
\end{figure*}

\begin{figure*}%[htbp]
\centering %sub_figures centered
%\includegraphics[width=0.325\textwidth]{contour043031+242613_13CO.eps}\hspace{-6mm}
%\includegraphics[width=0.261\textwidth]{1127_499_50_13CO.eps}\hspace{-6mm}
%\includegraphics[width=0.275\textwidth]{1127_499_57_41_44_13CO.eps}
%\vspace{-10mm}
%
%\includegraphics[width=.9\textwidth]{channel043031+242613_13CO.eps}\hspace{15mm}
\caption{TMB\_13: The same as Fig. \ref{Bubble_fig1.} except for the following. In the upper left panel the integrated velocity interval is from 4.4 to 6.5 km s$^{-1}$, the starting contour and the contour step are 5.0 K km s$^{-1}$ and 0.9 K km s$^{-1}$, respectively. In the upper middle panel the position angle is $40^{\circ}$ and the contour levels are 0.1 to 1.7 K by 0.2 K.
\label{Bubble_fig13.}}
\end{figure*}

\begin{figure*}%[htbp]
\centering %sub_figures centered
%\includegraphics[width=0.325\textwidth]{contour043114+292553_13CO.eps}\hspace{-6mm}
%\includegraphics[width=0.261\textwidth]{1095_1398_140_13CO.eps}\hspace{-6mm}
%\includegraphics[width=0.275\textwidth]{1095_1398_66_25_64_13CO.eps}
%\vspace{-10mm}
%
%\includegraphics[width=.9\textwidth]{channel043114+292553_13CO.eps}\hspace{15mm}
\caption{TMB\_14: The same as Fig. \ref{Bubble_fig1.} except for the following. In the upper left panel the integrated velocity interval is from 0.7 to 2.5 km s$^{-1}$ and the starting contour is 1.3 K km s$^{-1}$. In the upper middle panel the position angle is $130^{\circ}$ and the contour levels are 0.1 to 0.8 K by 0.1 K.
\label{Bubble_fig14.}}
\end{figure*}

\begin{figure*}%[htbp]
\centering %sub_figures centered
%\includegraphics[width=0.325\textwidth]{contour043130+241433_13CO.eps}\hspace{-6mm}
%\includegraphics[width=0.261\textwidth]{1087_464_70_13CO.eps}\hspace{-6mm}
%\includegraphics[width=0.275\textwidth]{1087_464_66_47_34_13CO.eps}
%\vspace{-10mm}
%
%\includegraphics[width=.9\textwidth]{channel043130+241433_13CO.eps}\hspace{15mm}
\caption{TMB\_15: The same as Fig. \ref{Bubble_fig1.} except for the following. In the upper left panel the integrated velocity interval is from 6.8 to 7.8 km s$^{-1}$, the starting contour and the contour step are 2.9 K km s$^{-1}$ and 0.8 K km s$^{-1}$, respectively. In the upper middle panel the position angle is $20^{\circ}$ and the contour levels are 0.1 to 1.4 K by 0.1 K.
\label{Bubble_fig15.}}
\end{figure*}

\begin{figure*}%[htbp]
\centering %sub_figures centered
%\includegraphics[width=0.325\textwidth]{contour043132+240953_13CO.eps}\hspace{-6mm}
%\includegraphics[width=0.261\textwidth]{1086_450_80_13CO.eps}\hspace{-6mm}
%\includegraphics[width=0.275\textwidth]{1086_450_55_47_46_13CO.eps}
%\vspace{-10mm}
%
%\includegraphics[width=.9\textwidth]{channel043132+240953_13CO.eps}\hspace{15mm}
\caption{TMB\_16: The same as Fig. \ref{Bubble_fig1.} except for the following. In the upper left panel the integrated velocity interval is from 6.5 to 8.1 km s$^{-1}$, the starting contour and the contour step are 4.2 K km s$^{-1}$ and 0.9 K km s$^{-1}$, respectively. In the upper middle panel the position angle is $10^{\circ}$ and the contour levels are 0.1 to 1.8 K by 0.2 K.
\label{Bubble_fig16.}}
\end{figure*}

\begin{figure*}%[htbp]
\centering %sub_figures centered
%\includegraphics[width=0.325\textwidth]{contour043135+233513_13CO.eps}\hspace{-6mm}
%\includegraphics[width=0.261\textwidth]{1084_346_100_13CO.eps}\hspace{-6mm}
%\includegraphics[width=0.275\textwidth]{1084_346_33_40_31_13CO.eps}
%\vspace{-10mm}
%
%\includegraphics[width=.9\textwidth]{channel043135+233513_13CO.eps}\hspace{15mm}
\caption{TMB\_17: The same as Fig. \ref{Bubble_fig1.} except for the following. In the upper left panel the integrated velocity interval is from 5.2 to 6.0 km s$^{-1}$ and the starting contour is 1.6 K km s$^{-1}$. In the upper middle panel the position angle is $170^{\circ}$ and the contour levels are 0.1 to 1.3 K by 0.1 K.
\label{Bubble_fig17.}}
\end{figure*}

\begin{figure*}%[htbp]
\centering %sub_figures centered
%\includegraphics[width=0.325\textwidth]{contour043150+242213_13CO.eps}\hspace{-6mm}
%\includegraphics[width=0.261\textwidth]{1073_487_100_13CO.eps}\hspace{-6mm}
%\includegraphics[width=0.275\textwidth]{1073_487_44_42_35_13CO.eps}
%\vspace{-10mm}
%
%\includegraphics[width=.9\textwidth]{channel043150+242213_13CO.eps}\hspace{15mm}
\caption{TMB\_18: The same as Fig. \ref{Bubble_fig1.} except for the following. In the upper left panel the integrated velocity interval is from 5.2 to 7.0 km s$^{-1}$, the starting contour and the contour step are 7.2 K km s$^{-1}$ and 1 K km s$^{-1}$, respectively. In the upper middle panel the position angle is $170^{\circ}$ and the contour levels are 0.1 to 1.9 K by 0.2 K.
\label{Bubble_fig18.}}
\end{figure*}

\begin{figure*}%[htbp]
\centering %sub_figures centered
%\includegraphics[width=0.325\textwidth]{contour043159+254313_13CO.eps}\hspace{-6mm}
%\includegraphics[width=0.261\textwidth]{1067_730_160_13CO.eps}\hspace{-6mm}
%\includegraphics[width=0.275\textwidth]{1067_730_55_42_53_13CO.eps}
%\vspace{-10mm}
%
%\includegraphics[width=.9\textwidth]{channel043159+254313_13CO.eps}\hspace{15mm}
\caption{TMB\_19: The same as Fig. \ref{Bubble_fig1.} except for the following. In the upper left panel the integrated velocity interval is from 4.9 to 6.8 km s$^{-1}$, the starting contour and the contour step are 5.0 K km s$^{-1}$ and 0.8 K km s$^{-1}$, respectively. In the upper middle panel the contour levels are 0.1 to 1.5 K by 0.1 K.
\label{Bubble_fig19.}}
\end{figure*}

\begin{figure*}%[htbp]
\centering %sub_figures centered
%\includegraphics[width=0.325\textwidth]{contour043203+253653_13CO.eps}\hspace{-6mm}
%\includegraphics[width=0.261\textwidth]{1064_711_140_13CO.eps}\hspace{-6mm}
%\includegraphics[width=0.275\textwidth]{1064_711_33_42_31_13CO.eps}
%\vspace{-10mm}
%
%\includegraphics[width=.9\textwidth]{channel043203+253653_13CO.eps}\hspace{15mm}
\caption{TMB\_20: The same as Fig. \ref{Bubble_fig1.} except for the following. In the upper left panel the integrated velocity interval is from 5.4 to 6.5 km s$^{-1}$ and the starting contour is 3.6 K km s$^{-1}$. In the upper middle panel the position angle is $130^{\circ}$ and the contour levels are 0.1 to 1.8 K by 0.2 K.
\label{Bubble_fig20.}}
\end{figure*}

\clearpage
\begin{figure*}%[htbp]
\centering %sub_figures centered
%\includegraphics[width=0.325\textwidth]{contour043237+292913_13CO.eps}\hspace{-6mm}
%\includegraphics[width=0.261\textwidth]{1041_1408_130_13CO.eps}\hspace{-6mm}
%\includegraphics[width=0.275\textwidth]{1041_1408_48_28_46_13CO.eps}
%\vspace{-10mm}
%
%\includegraphics[width=.9\textwidth]{channel043237+292913_13CO.eps}\hspace{15mm}
\caption{TMB\_21: The same as Fig. \ref{Bubble_fig1.} except for the following. In the upper left panel the integrated velocity interval is from 1.4 to 3.6 km s$^{-1}$, the starting contour and the contour step are 2.2 K km s$^{-1}$ and 0.7 K km s$^{-1}$, respectively. In the upper middle panel the position angle is $140^{\circ}$ and the contour levels are 0.1 to 0.8 K by 0.1 K.
\label{Bubble_fig21.}}
\end{figure*}

\begin{figure*}%[htbp]
\centering %sub_figures centered
%\includegraphics[width=0.325\textwidth]{contour043239+244613_13CO.eps}\hspace{-6mm}
%\includegraphics[width=0.261\textwidth]{1040_559_150_13CO.eps}\hspace{-6mm}
%\includegraphics[width=0.275\textwidth]{1040_559_84_38_82_13CO.eps}
%\vspace{-10mm}
%
%\includegraphics[width=.9\textwidth]{channel043239+244613_13CO.eps}\hspace{15mm}
\caption{TMB\_22: The same as Fig. \ref{Bubble_fig1.} except for the following. In the upper left panel the integrated velocity interval is from 4.4 to 5.4 km s$^{-1}$, the starting contour and the contour step are 2.0 K km s$^{-1}$ and 0.7 K km s$^{-1}$, respectively. In the upper middle panel the position angle is $120^{\circ}$ and the contour levels are 0.1 to 1.5 K by 0.1 K.
\label{Bubble_fig22.}}
\end{figure*}

\begin{figure*}%[htbp]
\centering %sub_figures centered
%\includegraphics[width=0.325\textwidth]{contour043310+260853_13CO.eps}\hspace{-6mm}
%\includegraphics[width=0.261\textwidth]{1019_807_150_13CO.eps}\hspace{-6mm}
%\includegraphics[width=0.275\textwidth]{1019_807_22_40_20_13CO.eps}
%\vspace{-10mm}
%
%\includegraphics[width=.9\textwidth]{channel043310+260853_13CO.eps}\hspace{15mm}
\caption{TMB\_23: The same as Fig. \ref{Bubble_fig1.} except for the following. In the upper left panel the integrated velocity interval is from 4.9 to 6.0 km s$^{-1}$, the starting contour and the contour step are 3.6 K km s$^{-1}$ and 0.4 K km s$^{-1}$, respectively. In the upper middle panel the position angle is $120^{\circ}$ and the contour levels are 0.1 to 1.7 K by 0.2 K.
\label{Bubble_fig23.}}
\end{figure*}

\begin{figure*}%[htbp]
\centering %sub_figures centered
%\includegraphics[width=0.325\textwidth]{contour043313+252453_13CO.eps}\hspace{-6mm}
%\includegraphics[width=0.261\textwidth]{1017_675_20_13CO.eps}\hspace{-6mm}
%\includegraphics[width=0.275\textwidth]{1017_675_39_41_37_13CO.eps}
%\vspace{-10mm}
%
%\includegraphics[width=.9\textwidth]{channel043313+252453_13CO.eps}\hspace{15mm}
\caption{TMB\_24: The same as Fig. \ref{Bubble_fig1.} except for the following. In the upper left panel the integrated velocity interval is from 4.9 to 6.0 km s$^{-1}$, the starting contour and the contour step are 4.2 K km s$^{-1}$ and 0.5 K km s$^{-1}$, respectively. In the upper middle panel the position angle is $70^{\circ}$ and the contour levels are 0.1 to 0.9 K by 0.1 K.
\label{Bubble_fig24.}}
\end{figure*}

\begin{figure*}%[htbp]
\centering %sub_figures centered
%\includegraphics[width=0.325\textwidth]{contour043334+242053_13CO.eps}\hspace{-6mm}
%\includegraphics[width=0.261\textwidth]{1002_483_120_13CO.eps}\hspace{-6mm}
%\includegraphics[width=0.275\textwidth]{1002_483_22_40_15_13CO.eps}
%\vspace{-10mm}
%
%\includegraphics[width=.9\textwidth]{channel043334+242053_13CO.eps}\hspace{15mm}
\caption{TMB\_25: The same as Fig. \ref{Bubble_fig1.} except for the following. In the upper left panel the integrated velocity interval is from 4.9 to 7.8 km s$^{-1}$, the starting contour and the contour step are 8.2 K km s$^{-1}$ and 1.2 K km s$^{-1}$, respectively. In the upper middle panel the position angle is $150^{\circ}$ and the contour levels are 0.1 to 1.0 K by 0.1 K.
\label{Bubble_fig25.}}
\end{figure*}

\begin{figure*}%[htbp]
\centering %sub_figures centered
%\includegraphics[width=0.325\textwidth]{contour043447+293713_13CO.eps}\hspace{-6mm}
%\includegraphics[width=0.261\textwidth]{956_1432_10_13CO.eps}\hspace{-6mm}
%\includegraphics[width=0.275\textwidth]{956_1432_36_32_34_13CO.eps}
%\vspace{-10mm}
%
%\includegraphics[width=.9\textwidth]{channel043447+293713_13CO.eps}\hspace{15mm}
\caption{TMB\_26: The same as Fig. \ref{Bubble_fig1.} except for the following. In the upper left panel the integrated velocity interval is from 2.0 to 4.1 km s$^{-1}$, the starting contour and the contour step are 1.5 K km s$^{-1}$ and 0.8 K km s$^{-1}$, respectively. In the upper middle panel the position angle is $80^{\circ}$ and the contour levels are 0.1 to 0.6K by 0.1 K.
\label{Bubble_fig26.}}
\end{figure*}

\begin{figure*}%[htbp]
\centering %sub_figures centered
%\includegraphics[width=0.325\textwidth]{contour043602+282313_13CO.eps}\hspace{-6mm}
%\includegraphics[width=0.261\textwidth]{906_1210_130_13CO.eps}\hspace{-6mm}
%\includegraphics[width=0.275\textwidth]{906_1210_110_34_108_13CO.eps}
%\vspace{-10mm}
%
%\includegraphics[width=.9\textwidth]{channel043602+282313_13CO.eps}\hspace{15mm}
\caption{TMB\_27: The same as Fig. \ref{Bubble_fig1.} except for the following. In the upper left panel the integrated velocity interval is from 2.5 to 4.9 km s$^{-1}$ and the starting contour is 1.8 K km s$^{-1}$. In the upper middle panel the position angle is $140^{\circ}$.
\label{Bubble_fig27.}}
\end{figure*}

\begin{figure*}%[htbp]
\centering %sub_figures centered
%\includegraphics[width=0.325\textwidth]{contour043623+253633_13CO.eps}\hspace{-6mm}
%\includegraphics[width=0.261\textwidth]{888_710_50_13CO.eps}\hspace{-6mm}
%\includegraphics[width=0.275\textwidth]{888_710_110_43_108_13CO.eps}
%\vspace{-10mm}
%
%\includegraphics[width=.9\textwidth]{channel043623+253633_13CO.eps}\hspace{15mm}
\caption{TMB\_28: The same as Fig. \ref{Bubble_fig1.} except for the following. In the upper left panel the integrated velocity interval is from 4.4 to 7.6 km s$^{-1}$, the starting contour and the contour step are 7.5 K km s$^{-1}$ and 1.2 K km s$^{-1}$, respectively. In the upper middle panel the position angle is $40^{\circ}$.
\label{Bubble_fig28.}}
\end{figure*}

\clearpage
\begin{figure*}%[htbp]
\centering %sub_figures centered
%\includegraphics[width=0.325\textwidth]{contour043704+254633_13CO.eps}\hspace{-6mm}
%\includegraphics[width=0.261\textwidth]{861_740_70_13CO.eps}\hspace{-6mm}
%\includegraphics[width=0.275\textwidth]{861_740_66_38_41_13CO.eps}
%\vspace{-10mm}
%
%\includegraphics[width=.9\textwidth]{channel043704+254633_13CO.eps}\hspace{15mm}
\caption{TMB\_29: The same as Fig. \ref{Bubble_fig1.} except for the following. In the upper left panel the integrated velocity interval is from 4.4 to 5.7 km s$^{-1}$, the starting contour and the contour step are 4.0 K km s$^{-1}$ and 0.4 K km s$^{-1}$, respectively. In the upper middle panel the position angle is $20^{\circ}$ and the contour levels are 0.1 to 1.3 K by 0.1 K.
\label{Bubble_fig29.}}
\end{figure*}

\begin{figure*}%[htbp]
\centering %sub_figures centered
%\includegraphics[width=0.325\textwidth]{contour043811+260553_13CO.eps}\hspace{-6mm}
%\includegraphics[width=0.261\textwidth]{816_798_140_13CO.eps}\hspace{-6mm}
%\includegraphics[width=0.275\textwidth]{816_798_150_39_148_13CO.eps}
%\vspace{-10mm}
%
%\includegraphics[width=.9\textwidth]{channel043811+260553_13CO.eps}\hspace{15mm}
\caption{TMB\_30: The same as Fig. \ref{Bubble_fig1.} except for the following. In the upper left panel the integrated velocity interval is from 4.4 to 6.0 km s$^{-1}$ and the starting contour is 5.1 K km s$^{-1}$. In the upper middle panel the position angle is $130^{\circ}$ and the contour levels are 0.1 to 1.4 K by 0.1 K.
\label{Bubble_fig30.}}
\end{figure*}

\begin{figure*}%[htbp]
\centering %sub_figures centered
%\includegraphics[width=0.325\textwidth]{contour043911+290513_13CO.eps}\hspace{-6mm}
%\includegraphics[width=0.261\textwidth]{783_1336_150_13CO.eps}\hspace{-6mm}
%\includegraphics[width=0.275\textwidth]{783_1336_99_40_62_13CO.eps}
%\vspace{-10mm}
%
%\includegraphics[width=.9\textwidth]{channel043911+290513_13CO.eps}\hspace{15mm}
\caption{TMB\_31: The same as Fig. \ref{Bubble_fig1.} except for the following. In the upper left panel the integrated velocity interval is from 4.6 to 6.0 km s$^{-1}$ and the starting contour is 1.7 K km s$^{-1}$. In the upper middle panel the position angle is $120^{\circ}$ and the contour levels are 0.1 to 0.8 K by 0.1 K.
\label{Bubble_fig31.}}
\end{figure*}

\begin{figure*}%[htbp]
\centering %sub_figures centered
%\includegraphics[width=0.325\textwidth]{contour043948+283533_13CO.eps}\hspace{-6mm}
%\includegraphics[width=0.261\textwidth]{757_1247_40_13CO.eps}\hspace{-6mm}
%\includegraphics[width=0.275\textwidth]{757_1247_55_32_53_13CO.eps}
%\vspace{-10mm}
%
%\includegraphics[width=.9\textwidth]{channel043948+283533_13CO.eps}\hspace{15mm}
\caption{TMB\_32: The same as Fig. \ref{Bubble_fig1.} except for the following. In the upper left panel the integrated velocity interval is from 2.2 to 4.1 km s$^{-1}$, the starting contour and the contour step are 1.4 K km s$^{-1}$ and 0.4 K km s$^{-1}$, respectively. In the upper middle panel the position angle is $50^{\circ}$ and the contour levels are 0.1 to 0.5 K by 0.1 K.
\label{Bubble_fig32.}}
\end{figure*}

\begin{figure*}%[htbp]
\centering %sub_figures centered
%\includegraphics[width=0.325\textwidth]{contour044110+253113_13CO.eps}\hspace{-6mm}
%\includegraphics[width=0.261\textwidth]{694_694_40_13CO.eps}\hspace{-6mm}
%\includegraphics[width=0.275\textwidth]{694_694_55_47_46_13CO.eps}
%\vspace{-10mm}
%
%\includegraphics[width=.9\textwidth]{channel044110+253113_13CO.eps}\hspace{15mm}
\caption{TMB\_33: The same as Fig. \ref{Bubble_fig1.} except for the following. In the upper left panel the integrated velocity interval is from 7.0 to 7.8 km s$^{-1}$, the starting contour and the contour step are 2.0 K km s$^{-1}$ and 0.4 K km s$^{-1}$, respectively. In the upper middle panel the position angle is $50^{\circ}$ and the contour levels are 0.1 to 2.2 K by 0.2 K.
\label{Bubble_fig33.}}
\end{figure*}

\begin{figure*}%[htbp]
\centering %sub_figures centered
%\includegraphics[width=0.325\textwidth]{contour044420+283653_13CO.eps}\hspace{-6mm}
%\includegraphics[width=0.261\textwidth]{578_1251_50_13CO.eps}\hspace{-6mm}
%\includegraphics[width=0.275\textwidth]{578_1251_99_43_97_13CO.eps}
%\vspace{-10mm}
%
%\includegraphics[width=.9\textwidth]{channel044420+283653_13CO.eps}\hspace{15mm}
\caption{TMB\_34: The same as Fig. \ref{Bubble_fig1.} except for the following. In the upper left panel the integrated velocity interval is from 5.2 to 7.8 km s$^{-1}$, the starting contour and the contour step are 3.8 K km s$^{-1}$ and 0.7 K km s$^{-1}$, respectively. In the upper middle panel the position angle is $40^{\circ}$.
\label{Bubble_fig34.}}
\end{figure*}

\begin{figure*}%[htbp]
\centering %sub_figures centered
%\includegraphics[width=0.325\textwidth]{contour044612+250733_13CO.eps}\hspace{-6mm}
%\includegraphics[width=0.261\textwidth]{488_623_40_13CO.eps}\hspace{-6mm}
%\includegraphics[width=0.275\textwidth]{488_623_48_41_46_13CO.eps}
%\vspace{-10mm}
%
%\includegraphics[width=.9\textwidth]{channel044612+250733_13CO.eps}\hspace{15mm}
\caption{TMS\_35: The same as Fig. \ref{Bubble_fig1.} except for the following. In the upper left panel the integrated velocity interval is from 4.9 to 6.2 km s$^{-1}$ and the starting contour is 4.0 K km s$^{-1}$. In the upper middle panel the position angle is $50^{\circ}$ and the contour levels are 0.1 to 0.9 K by 0.1 K.
\label{Bubble_fig35.}}
\end{figure*}

\begin{figure*}%[htbp]
\centering %sub_figures centered
%\includegraphics[width=0.325\textwidth]{contour044643+245913_13CO.eps}\hspace{-6mm}
%\includegraphics[width=0.261\textwidth]{466_598_60_13CO.eps}\hspace{-6mm}
%\includegraphics[width=0.275\textwidth]{466_598_44_44_42_13CO.eps}
%\vspace{-10mm}
%
%\includegraphics[width=.9\textwidth]{channel044643+245913_13CO.eps}\hspace{15mm}
\caption{TMB\_36: The same as Fig. \ref{Bubble_fig1.} except for the following. In the upper left panel the integrated velocity interval is from 5.7 to 7.3 km s$^{-1}$, the starting contour and the contour step are 4.0 K km s$^{-1}$ and 0.4 K km s$^{-1}$, respectively. In the upper middle panel the position angle is $30^{\circ}$ and the contour levels are 0.1 to 0.8 K by 0.1 K.
\label{Bubble_fig36.}}
\end{figure*}

\begin{figure*}%[htbp]
\centering %sub_figures centered
%\includegraphics[width=0.325\textwidth]{contour044812+245033_13CO.eps}\hspace{-6mm}
%\includegraphics[width=0.261\textwidth]{405_572_160_13CO.eps}\hspace{-6mm}
%\includegraphics[width=0.275\textwidth]{405_572_50_37_48_13CO.eps}
%\vspace{-10mm}
%
%\includegraphics[width=.9\textwidth]{channel044812+245033_13CO.eps}\hspace{15mm}
\caption{TMB\_37: The same as Fig. \ref{Bubble_fig1.} except for the following. In the upper left panel the integrated velocity interval is from 3.3 to 5.4 km s$^{-1}$, the starting contour and the contour step are 2.2 K km s$^{-1}$ and 0.7 K km s$^{-1}$, respectively. In the upper middle panel the contour levels are 0.1 to 1.1 K by 0.1 K.
\label{Bubble_fig37.}}
\end{figure*}

\section*{Acknowledgments}
We are grateful to Dr. Y.L. Yue, Dr. Z.Y. Zhang, Dr. T. Liu, Dr. X.Y. Gao and Dr. Z.Y. Ren for their kind and valuable advice and support. We would like to thank the anonymous referee for the careful inspection of the manuscript and constructive comments particularly the important suggestions to examine the turbulent dissipation issue to improve the quality of this study. We also thank Prof. W. Butler Burton for help in the review process. This work is partly supported by the China Ministry of Science and Technology under State Key Development Program for Basic Research (2012CB821802), and the National Natural Science Foundation of China (11373038, 11373045), the Hundred Talents Program of the Chinese Academy of Sciences and the Young Researcher Grant of National Astronomical Observatories, Chinese Academy of Sciences.

\appendix

\section{Derivations of Outflow Parameters}
\label{AppendixOutflowParameters}
To calculate molecular outflow parameters we need first to obtain the column density first.
A simple solution of the equation of radiative transfer is
\begin{equation}
T_{\rm s}=[T_{\rm ex}-T_{\rm bg}](1-e^{-\tau_\nu}),
\label{equ:RadiativeTransfer}
\end{equation}
where $T_{\rm s}$ is the source temperature, $T_{\rm ex}$ is the excitation temperature and $T_{\rm bg}$ is the background temperature, and $\nu$ is the frequency of the transition. The modified Planck function $J$ is defined as
\begin{equation}
J(T)=\frac{h\nu/k}{e^{\frac{h\nu}{kT}}-1},
\label{equ:PlanckFunction}
\end{equation}
where $k$ is Boltzmann's constant and $h$ is Planck's constant. The definition of optical depth in terms of upper level column density is expressed as \citep{Wilson13}
\begin{equation}
\int \tau_\nu\ d\nu=\frac{A_{\rm ul}c^2N_{\rm u}}{8\pi\nu^2}(e^{h\nu/kT_{\rm ex}}-1).
\label{equ:OpticalDepth}
\end{equation}
Assuming $\tau$ $\ll$ 1, $h\nu/k$ $\ll$ $T_{\rm ex}$ and $T_{\rm bg}$ $\ll$ $T_{\rm ex}$, we get the column density of the rotational upper level of the transition in the outflow by combining Eq. (\ref{equ:RadiativeTransfer}) with Eq. (\ref{equ:OpticalDepth}).
\begin{equation}
N_{\rm u}(^{12}{\rm CO})=\frac{8\pi k\nu^2}{hc^3 A_{\rm ul}}\int T_{\rm s} dv,
\label{equ:ColumnDensityUpperLevel}
\end{equation}
where $T_{\rm s}$ is the observed source antenna temperature with proper correction for antenna efficiency, $c$ is the speed of light, and $A_{\rm ul}$ is the spontaneous transition rate from the upper level($J+1$) to the lower level($J$), which can be expressed as
\begin{equation}
A_{\rm ul}=\frac{64\pi^4\nu^3\mu^2_{\rm d}}{3hc^3} \frac{J+1}{2J+3},
\label{equ:Aul}
\end{equation}
where $\mu_d$ is the permanent electric dipole moment of a molecule and $J$ = 0.\\
The total column density of the outflow is
\begin{equation}
N_{\rm tot}(^{12}{\rm CO})=\frac{N_{\rm u}(^{12}\rm CO)}{f_{\rm u}},
\label{equ:ColumnDensityTotal}
\end{equation}
where $f_{\rm u}$ is the fraction of the $\rm {}^{12}CO$ in the upper level of the transition. Under local thermal equilibrium(LTE) conditions, $f_u$ is given by
\begin{equation}
f_{\rm u}=\frac{g_{\rm u} \exp(-h\nu/kT_{\rm ex})}{Q(T_{\rm ex})},
\label{equ:CorrectionFactor}
\end{equation}
where the statistical weight of the upper level $g_{\rm u}=2(J+1)+1$, the LTE partition function (for $kT_{\rm ex}$ $\gg$ $hB$) $Q(T_{\rm ex})=kT_{\rm ex}/hB$ and the rotational constant $B=\nu/[2(J+1)]$ for the $J+1 \rightarrow J$ transition \citep{Tennyson05}.
Then we can derive the total column density of outflow from Eq. (\ref{equ:ColumnDensityUpperLevel}), Eq. (\ref{equ:Aul}), Eq. (\ref{equ:ColumnDensityTotal}) and Eq. (\ref{equ:CorrectionFactor}),
\begin{equation}
N_{\rm tot}(^{12}{\rm CO})=\frac{3k^2T_{\rm ex}}{4\pi^3\mu^2_{\rm d}h\nu^2 \exp(-h\nu/kT_{\rm ex})}\int T_{\rm s} dv
\label{equ:DerivedTotalColumnDensity}
\end{equation}
If there is a high velocity wing in $\rm {}^{12}CO$ but not in $\rm {}^{13}CO$ profile, we assume $\rm {}^{12}CO$ is optically thin. Then we can calculate the column density of outflow from Eq. (\ref{equ:DerivedTotalColumnDensity}). If there is high velocity wing both in $\rm {}^{12}CO$ and $\rm {}^{13}CO$ profile, we can correct the optical depth of $\rm {}^{12}CO$ using the following equation
\begin{equation}
\frac{T^*_{\rm a}(^{12}{\rm CO})}{T^*_{\rm a}(^{13}{\rm CO})}=\frac{1-e^{-\tau(^{12}{\rm CO})}}{1-e^{-\tau(^{13}{\rm CO})}}.
\label{equ:TaTaoRatio}
\end{equation}
Here $T^*_{\rm a}(^{12}{\rm CO})$ and $T^*_{\rm a}(^{13}{\rm CO})$ are the antenna temperatures of $\rm{}^{12}{CO}$ and $\rm{}^{13}{CO}$ (with proper correction for antenna efficiency), respectively. $\tau(^{12}{\rm CO})$ and $\tau(^{13}{\rm CO})$ are the optical depths of $\rm {}^{12}CO$ and $\rm {}^{13}CO$, respectively. We assume the abundance ratio of $\rm {}^{12}CO$ to $\rm {}^{13}CO$ is 65 \citep{Langer93}. The correction factor for opacity is defined as
\begin{equation}
f_\tau=\frac{\tau(^{12}{\rm CO})}{1-e^{-\tau(^{12}{\rm CO})}}.
\label{equ:CorrectionFactorTao}
\end{equation}
Then we get the corrected total column density of outflow as
\begin{equation}
N_{\rm ctot}(^{12}{\rm CO})=f_{\tau}N_{\rm tot}(^{12}{\rm CO}).
\label{equ:CorrectedTotalColumnDensity}
\end{equation}
After obtaining the column density we can calculate other parameters of outflow. The mass of outflow can be calculated from
\begin{equation}
M_{\rm gas}=N_{\rm tot}(^{12}{\rm CO})[{\rm H_2/CO}]\mu_{\rm g}m({\rm H})S,
\label{equ:OutflowMass}
\end{equation}
where $\mu_{\rm g}$=2.72 is the mean molecular weight \citep{Brunt10}, $m(\rm H)$=$1.67\times10^{-24}$ g is the mass of a hydrogen atom, and $[\rm {H_2}]/[\rm{CO}]$ is assumed to be $10^4$, and $S$ is the area of the outflow.\\
The momentum ($P$) and energy ($E$) of the outflow can be calculated from
\begin{equation}
P=M_{\rm gas}|\overline{v}|,
%P=\int M_{channel}(v-v_0) dv
\label{equ:momentum}
\end{equation}
\begin{equation}
E=\frac{1}{2}M_{\rm gas}\overline{v}^2,
\label{equ:energy}
\end{equation}
where $\overline{v}$ is the average velocity of the outflow relative to the cloud systemic velocity and $M_{\rm gas}$ is obtained from Eq.(\ref{equ:OutflowMass}).\\
The dynamical timescale $t_{\rm dyn}$ can be estimated from
\begin{equation}
t_{\rm dyn}=\frac{L}{|\overline{v}|},
\label{equ:TimeScale}
\end{equation}
where $L$ is the typical linear scale of the outflow lobe.
The outflow luminosity, $L_{\rm flow}$, can be estimated by dividing the kinetic energy by the dynamical timescale. It can be expressed as
\begin{equation}
L_{\rm flow}=\frac{E}{t_{\rm dyn}}.
\label{equ:OutflowLuminosity}
\end{equation}

%\bibliographystyle{apj}
%\bibliography{journals,taurus_feedback}

\end{document}